\pdfoutput=1

\documentclass[aps,pra,superscriptaddress,floatfix,altaffilletter,showpacs,showkeys]{revtex4-1}

\usepackage[utf8]{inputenc}
\usepackage{amsmath,amsfonts,amssymb,amsthm}
\usepackage{graphicx}
\usepackage{subcaption}
\usepackage{comment}

\usepackage{xcolor}

\usepackage{siunitx}

\newcommand{\imu}{\mathrm{i}}
\newcommand{\dint}[2][]{\!\mathrm{d}^{#1}#2\,}                 

\renewcommand{\Re}{\mathrm{Re}\,}
\renewcommand{\Im}{\mathrm{Im}\,}
\renewcommand{\vec}[1]{\mathbf{{#1}}}                         
\newcommand{\pvec}[1]{\mathbf{{#1}}_\parallel }
                         
\newcommand{\vecUnit}[1]{\mathbf{\hat{#1}}}                         
\newcommand{\pvecUnit}[1]{\mathbf{\hat{#1}}_\parallel }

\def\mathclap#1{\text{\hbox to 0pt{\hss$\mathsurround=0pt#1$\hss}}}

\newcommand{\etal}{\textit{et~al.}}

\graphicspath{.,{Figures/}}         

\begin{document}



\title{Non-parametric reconstruction of the statistical properties of penetrable, isotropic randomly rough surfaces from in-plane, co-polarized light scattering data: Application to computer generated and experimental scattering data}



\author{Ver\'{o}nica P. Simonsen}
\affiliation{PoreLab, NTNU -- Norwegian University of Science and Technology, NO-7491 Trondheim, Norway}

\author{Dick Bedeaux}
\affiliation{PoreLab, NTNU -- Norwegian University of Science and Technology, NO-7491 Trondheim, Norway}

\author{Ingve Simonsen}
\email{ingve.simonsen@ntnu.no}
\affiliation{Department of Physics, NTNU -- Norwegian University of Science and Technology, NO-7491 Trondheim, Norway}
\affiliation{Surface du Verre et Interfaces, UMR 125 CNRS/Saint-Gobain, F-93303 Aubervilliers, France}



\date{\today}


\begin{abstract}
   An approach is introduced for the non-parametric reconstruction of the statistical properties of penetrable, isotropic randomly rough surfaces from in-plane, co-polarized light scattering data.
   Starting from expressions within the Kirchhoff approximation for the light  scattered diffusely by a two-dimensional randomly rough surface, an analytic expression for the normalized surface height correlation function is obtained as an integral over the in-plane and co-polarized scattering data with the introduction of only a couple of additional approximations. The inversion approach consists of two main steps.  In the first step the surface roughness is estimated. Next, this value is used to obtain the functional form of the surface height correlation function without initially assuming any particular form for this function (non-parametric inversion). The input data used in validating this inversion approach consist of in-plane and co-polarized scattering data obtained for different forms of the correlation function by either computer simulations or by experiments for two-dimensional randomly rough dielectric or metallic surfaces. Good agreement was obtained between the correlation function and surface roughness obtained during the reconstruction and the corresponding quantities assumed when generating the input scattering data; this was the case for both dielectric and metallic surfaces, for both p- and s-polarized light, and for different polar angles of incidence.
  The proposed inversion approach provides an accurate, efficient, robust and contact-less  method based on in-plane and co-polarized scattering data for the non-parametric characterization of the statistical properties of isotropic two-dimensional randomly rough dielectric and metallic surface.
\end{abstract}


\maketitle



\section{Introduction}
The ability to characterize quantitatively the roughness of a solid surface is important for both basic science reasons and applications.  Among the basic science reasons are the in-situ monitoring of the growth of crystals to elucidate the mechanisms responsible for them and the fact that adhesion properties of a rough surface can be significantly different from a planar surface.  Applications of such characterization techniques include determining the quality of optical elements such as mirrors and lenses, and even the quality of the paint coating an automobile body.  Other applications are, in the oil and gas industry, the corrosion testing for integrity management and the optimized life extension of production facilities, and the analysis of the formation of scale at a surface.  Such characterizations currently are very difficult to carry out, in particular over large surface areas.

The most direct way of obtaining the statistical properties of two dimensional randomly rough surfaces is first to measure their surface topographies and to infer their statistical properties from such data.  Various scanning probe microscopy techniques are often used for obtaining such maps for which atomic force microscopy and contact profilometry~\cite{Book:Kaupp2006,Book:Thomas1999} represent classic examples.  However, such contact methods have several disadvantages.  First, to directly measure the surface topography over larger areas of the mean-plane of the surface is often time-consuming and, therefore, not very practical.  Second, such measurements can display significant probe artifacts since what is measured is the surface morphology convoluted by the shape of the tip of the probe.  Finally, there are serious challenges for how to integrate such measurements into a production line setup for which the surface one wants to characterize is moving.

For all the above reasons and others, there have been significant efforts invested into obtaining the statistical properties of surfaces by \emph{indirect} means that do not require prior knowledge of the surface topography.  One such approach is based on recording the intensity of the acoustic or electromagnetic waves scattered from the rough surface and to use \emph{inverse scattering theory} to obtain information about the surface that produced them.  For instance, when a monochromatic incident wave is scattered from a rough surface, a speckle pattern is produced that encodes information about the surface topography~\cite{Book:Goodman1985}.  The inversion of light scattering data to reconstruct the surface profile itself is a very difficult problem in general that has been solved only in a few special cases; some examples are given in Refs.~\onlinecite{Greffet1995,Arhab2011}. However, it is more efficient to directly reconstruct the statistical properties of the randomly rough surface from such scattering data.  For instance, the surface root mean square~(rms) roughness and the surface height auto-correlation function~\cite{Ogilvy1987,Book:Stout2000,Simonsen2004-3}.  The availability of a relatively simple, accurate, and flexible method for obtaining these statistical properties, and even the dielectric constant of the scattering medium if it is not known in advance, will represent a significant contribution to the characterization of surface roughness by contactless methods.  The advantage of such inverse scattering methods are that they typically are fast, relatively cheap to implement, and can cover large surface areas, which is in particular important when dealing with randomly rough surfaces. 

This problem was initially studied for two dimensional randomly rough surfaces by Chandley~\cite{Chandley1976} in the mid 1970s.  He modeled the scattered light by scalar diffraction theory in combination with a random phase screen model.  Under some additional assumptions he was able to invert the scattering data with respect to the statistical properties of the surface by the use of a two-dimensional Fourier transform.  One of the drawbacks of this is that one needs to measure the full angular distribution of the scattered waves, something that is both time-consuming and requires more sophisticated experimental equipment.  About one and a half decade after the initial paper of Chandley, Marx and Vorburger~\cite{Marx1990}  performed a similar study for the scattering from a perfectly conducting surface.  These authors based the calculation of the scattered intensity on the scalar Kirchhoff approximation.  By assuming a particular form for the normalized surface-height correlation function they were able to use the expressions for the intensity in combination with a least-square procedure to invert experimental scattering data.  Zhao and colleagues~\cite{Zhao1996} combined the work of Chandley and  Marx and Vorburger in the sense that they used a Fourier based technique for the inversion (like Chandley), but the scattering model that they used was obtained within the scalar Kirchhoff approximation, similarly to Marx and Vorburger.  The advantage of the approach of  Zhao~\etal\  was that the full angular distribution of the scattered light was not needed.

The assumption that the form of the correlation function is known in advance is restrictive.  This is in particular expected to be the case for many industrial surfaces were often multiscale correlation functions are seen.  A more flexible mathematical approach was recently presented by Zamani and coworkers~\cite{Zamani2016}.  Their non-parametric expression directly relates the surface-height correlation function to the diffusely scattered intensity along a linear path at fixed polar angle, but only for the case of non-penetrable randomly rough surfaces, illuminated by monochromatic scalar waves (also see Refs.~\onlinecite{Zamani2012,Zamani2012a}).  The approach of Zamani~\etal\  was applied to the reconstruction of scattering data collected for the scattering from a rough silicon surface for fixed polar angle of incidence.  Results for the height-height correlation function obtained by inversion were compared to what was obtained by analyzing the surface morphology directly, and favorable agreement was found.

In passing we mention that a set of inverse wave scattering approaches recently were presented for the determination of the Hurst exponent and the topothesy of self-affine surfaces~\cite{Simonsen2014-04,Simonsen1999-1}; such surfaces are examples of scale-free surfaces (if cut-offs can be neglected), and they display fractal properties  at sufficiently small length scales.

The inversion schemes presented in Refs.~\onlinecite{Chandley1976,Marx1990,Zhao1996,Zamani2016} all have one thing in common: they all assume scalar wave theory.  In a series of studies, Maradudin and coworkers~\cite{Chakrabarti2013a,Simonsen2014-03,Simonsen2014-05,Simonsen2016-06,Simonsen2018-06} recently presented a class of inversion approaches for penetrable surfaces that are based on electromagnetic theory.  In particular, second order phase perturbation theory~\cite{Shen1980,NavarreteAlcala2009}, which is a vector theory, was used to derive expressions for the polarization dependent intensity scattered from two-dimensional, penetrable randomly rough surfaces.  When such expressions for the polarization dependent scattered intensity were combined with an assumption of the form of the correlation function, in-plane and co-polarized scattering data were inverted with respect to the statistical properties of the underlying rough surfaces in a least-square procedure.  This approach was applied successfully to the reconstruction of both computer generated and experimental scattering data obtained for rough dielectric~\cite{Simonsen2014-05} and metallic~\cite{Simonsen2018-06} surfaces.  It should be mentioned that for dielectric surfaces, the inversion approach for in-plane p-to-p scattering has some known issues for scattering angles in the vicinity of the Brewster angle~\cite{Simonsen2016-06}.  Finally, we stress that these inversion approaches are \emph{parametric} since the form of the correlation function needs to be assumed in advance.

The purpose of this paper, is to extend the non-parametric inversion approach for scalar waves introduced by Zamani~\etal~\cite{Zamani2016} to include both penetrable media and electromagnetic waves.  To this end, we use the stationary phase approximation to the Kirchhoff integrals at a two-dimensional randomly rough penetrable surface, dielectric or metallic, and use them to calculate the intensity of the light scattered diffusely by the rough surface when it is illuminated by a linearly polarized electromagnetic wave.  By the introduction of an additional approximation, we are able to analytically invert co-polarized input scattering data in the plane of incidence with respect to the normalized height-height correlation function and the surface roughness.  In our approach, we first estimate the surface rms-roughness value by means of either an iterative method or an analytic expression, and then use it to obtain the functional form of the surface height correlation function without initially assuming any particular form of this function.  We demonstrate the accuracy of our method by applying it to scattering data calculated from the original height correlation function, or obtained either by computer simulations, or in experimental measurements of the diffuse component of the scattered light as a function of the angle of incidence for different surfaces.

\smallskip
The remaining part of this paper is organized as follows: In Sec.~\ref{sec:Theory} we present the scattering system and derive the expression for the surface height correlation function, on the basis of the Kirchhoff approximation.  Section~\ref{sec:Results} presents the numerical and experimental results by the use of this method for different correlation functions, materials and polarizations.  The conclusions reached on the basis of these results are summarized in Sec.~\ref{sec:Conclusions}.  The paper ends with an appendix detailing the derivation of the expressions, along with the computational details.

\section{Theory}
\label{sec:Theory}
\subsection{The scattering system}
%
%
%
\begin{figure}[tbph] 
  \centering
  \includegraphics*[width=0.4\columnwidth]{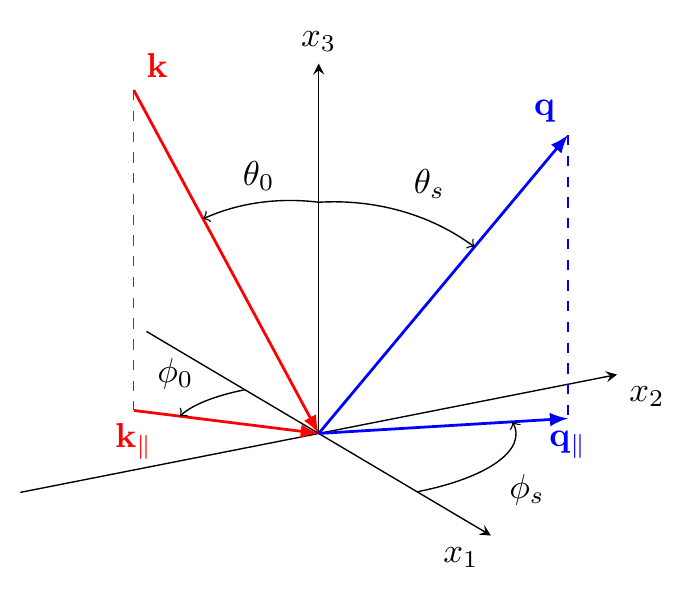}
  \caption{A schematic depiction of the scattering geometry considered in this work.}
  \label{fig:1}
\end{figure}
%
%
The physical system that we study consists of vacuum in the region $x_3 > \zeta(\pvec{x})$, and a medium whose dielectric constant is $\varepsilon$, in the region $x_3 < \zeta(\pvec{x})$~[Fig.~\ref{fig:1}].  Here $\pvec{x} = (x_1,x_2,0)$ is a position vector in the plane $x_3 = 0$.  The surface profile function $x_3=\zeta(\pvec{x})$ is assumed to be a single-valued  function of $\pvec{x}$ that is differentiable with respect to $x_1$ and $x_2$.  It is also assumed to constitute a  stationary, zero-mean, isotropic, Gaussian random process defined by
\begin{subequations}
\label{eq:1}
\begin{align}
  \left< \zeta^2(\pvec{x} ) \right>
  &=
    \delta^2
    \label{eq:1a}\\
  \left< \zeta(\pvec{x}) \zeta (\pvec{x} ') \right>
  &=
    \delta^2 W( | \pvec{x} - \pvec{x}' | ), 
    \label{eq:1b} 
\end{align}
\end{subequations}
where the angle brackets denote an average over the ensemble of realizations of $\zeta(\pvec{x})$. The function $W(|\pvec{x} |)$ is the {\it normalized surface height auto-correlation function}, normalized so that $W(0) = 1$, and $\delta$ denotes the rms height of the surface.

\subsection{The mean differential reflection coefficient}
The differential reflection coefficient $[ \partial R_{\alpha\beta}(\pvec{q} |\pvec{k} ) /\partial\Omega_s]$ is defined such that $[ \partial R_{\alpha\beta}(\pvec{q} |\pvec{k} ) /\partial\Omega_s] \mathrm{d}\Omega_s$ is the fraction of the total time-averaged flux in an incident field of polarization $\beta$, the projection of whose wave vector on the mean scattering plane is $\pvec{k}$ that is scattered into a field of $\alpha$ polarization the projection of whose wave vector on the mean scattering plane is $\pvec{q}$, within an element of solid angle $\textrm{d}\Omega_s$ about the scattering direction defined by the polar and azimuthal scattering angles $(\theta_s,\phi_s)$~[see Fig.~\ref{fig:1}].
The polarization indices $\alpha$ and $\beta$ take values in the set $\{p,s \}$.  As we here deal with scattering from a randomly rough surface, it is the average of this function over the ensemble of realizations of the surface profile function that has to be calculated.  The contribution to this average from the light scattered incoherently (diffusely) is~\cite{Simonsen2014-01} 
\begin{align}
  \left< \frac{\partial R_{\alpha\beta}(\pvec{q} |\pvec{k} )}{\partial\Omega_s}\right>_\textrm{incoh} 
    &= 
      \frac{1}{S} \left( \frac{\omega}{2\pi c}\right)^2 \frac{\cos^2\theta_s}{\cos\theta_0}
      \left[
      \left< \left| R_{\alpha\beta}(\pvec{q} |\pvec{k} ) \right|^2\right>
      -
      \Big| \Big< R_{\alpha\beta}(\pvec{q} |\pvec{k} ) \Big> \Big|^2
      \right], 
    \label{eq:mdrc-definition}
\end{align}
where $S$ is the area of the $x_1x_2$ plane covered by the randomly rough surface, and $\{ R_{\alpha\beta}(\pvec{q} |\pvec{k} )\}$ is a set of scattering amplitudes (relative to a plane wave basis).  The parallel (or in-plane) wave vectors $\pvec{k}$ and $\pvec{q}$ that appear in Eq.~\eqref{eq:mdrc-definition} can, in the propagating regime, be expressed in terms of the angles of incidence $(\theta_0,\phi_0)$ and scattering $(\theta_s,\phi_s)$ [see Fig.~\ref{fig:1}] as
\begin{subequations}
  \label{eq:kq_par_definition}
  \begin{align}
    \pvec{k} &= \frac{\omega}{c} \sin \theta_0 \left( \cos\phi_0, \sin\phi_0, 0 \right)
               \label{eq:kq_par_definition-k}
    \\
    \pvec{q} &= \frac{\omega}{c} \sin \theta_s \left( \cos\phi_s, \sin\phi_s, 0 \right),
               \label{eq:kq_par_definition-q}
  \end{align}
\end{subequations}
and the norms of these vectors are $k_\parallel=| \pvec{k}|$ and $q_\parallel=| \pvec{q}|$, respectively.  To calculate the mean differential reflection coefficients~(DRCs), one needs to know the scattering amplitudes $R_{\alpha\beta}(\pvec{q} |\pvec{k} )$ that define them, and to calculate these amplitudes is in general a far from trivial task. For instance, for a randomly rough surface a rigorous way of obtaining them requires the numerical solution of a coupled set of inhomogeneous integral equations~\cite{Simonsen2009-1,Simonsen2009-9}; such calculations consume a significant amount of resources to perform on large scale computer facilities.

Instead of relaying on purely numerical and time consuming calculations in order to obtain the intensity scattered from a rough surface, numerous approximate methods have been developed for the same purpose over the last decades~\cite{Book:Beckmann1987,Ogilvy1987,Book:Ishimaru1999,Book:Voronovich1999,Book:Maradudin2007,Simonsen2004-3}.  Chief among them is the so-called \textit{Kirchhoff approximation}~\cite{Sancer1969,Book:Ulaby1986,Voronovich2007,Book:Voronovich1999}.  The basic assumption underlying this approach is that the rough surface has gentle and small slopes, \textit{i.e.} the characteristic horizontal length-scale of the surface (its correlation length) is large compare with both its rms-roughness and the wavelength of the incident light~\cite{Thorsos1988,Voronovich2007}.  Therefore, the scattering of light from any point of the surface can be calculated as if the light is reflected from an infinite plane that is tangent to the surface at this point.  If the substrate is a (non-penetrable) perfect conductor, the Kirchhoff integrals defining the reflection amplitudes can be calculated analytically and the statistical average required to obtain the mean DRCs can also be performed analytically under the assumption that the surface roughness constitutes a Gaussian random process.  This situation is in sharp contrast to what happens when the substrate is penetrable.  In this case, the local reflection amplitudes depend on the local angles of incidence and the Kirchhoff integrals cannot be obtained analytically, and they have instead to be evaluated numerically for each surface realization on which basis the ensemble average is calculated~\cite{Voronovich2007}.  This process of calculating the ensemble average is somewhat time-consuming even if no large equations system has to be solved.  However, analytic approximations to the Kirchhoff integrals can be obtained if we apply to them the stationary-phase method.  The adaption of this approximation physically means that the scattering occurs only along directions for which there are specular points at the surface.  In this way, the expression for the scattering amplitudes $R_{\alpha\beta}(\pvec{q} | \pvec{k} )$ takes the form~\cite[Ch.~12]{Book:Ulaby1986}
\begin{subequations}
  \label{eq:R-Kichhoff}
  \begin{align}
    \label{eq:R-Kichhoff-a}
    R_{\alpha\beta}(\pvec{q} | \pvec{k} )
    &=
      \frac{
      1
      }{
      2 \alpha_0(q_\parallel)
      \left[ \alpha_0(q_\parallel) + \alpha_0(k_\parallel) \right]
      }
      \frac{
      {\mathcal S}_{\alpha\beta} (\pvec{q} | \pvec{k} )
      }{
      \Delta^2(\pvec{q} | \pvec{k} )
      }
      r( \pvec{q} | \pvec{k} ),
  \end{align}
  with
  \begin{align}
    \label{eq:R-Kichhoff-b}
    r( \pvec{q} | \pvec{k} )
    &=
      \int \dint[2]{x_\parallel}
      \exp \left[ -\imu ( \pvec{q} -\pvec{k} ) \cdot \pvec{x}  \right]
      \exp \left[ -\imu \left\{ \alpha_0(q_\parallel) + \alpha_0(k_\parallel) \right\} \zeta(\pvec{x})  \right].
  \end{align}
\end{subequations}
In writing the expressions in Eq.~\eqref{eq:R-Kichhoff} we have defined the local tangential wave vector of the scattered light as
\begin{align}
  \Delta ( \pvec{q} |  \pvec{k} )
  &=
    \frac{ 1 }{ \sqrt{2} }
    \left[
    \left( \frac{\omega}{c} \right)^2
    -
    \alpha_0(q_\parallel) \alpha_0(k_\parallel)
    +
    \pvec{q} \cdot \pvec{k}
    \right]^\frac{1}{2},
\end{align}
the expressions for the matrix elements ${\mathcal S}_{\alpha\beta} (\pvec{q} | \pvec{k} )$ are defined in Appendix~\ref{Appendix:A}, and the 3rd component of the scattered wave vector  $\vec{q}$, whose parallel component is $\pvec{q}$, reads~[$\vec{q}=\pvec{q}+\alpha_0(q_\parallel)\vecUnit{x}_3$]
\begin{align}
  \label{eq:alpha_0}
 \alpha_0 ( q_\parallel ) 
  &= 
  \left[ \left(\frac{\omega}{c} \right)^2 -  q_\parallel^2 \right]^{\frac{1}{2}} 
  \qquad \Re\alpha_0 ( q_\parallel ) > 0 \quad
    \Im \alpha_0 ( q_\parallel ) > 0,
\end{align}
with the expression for $\alpha_0(k_\parallel)$ defined in an analogous manner. In the propagating regime for which  $k_\parallel<\omega/c$ and $q_\parallel<\omega/c$, it follows from Eq.~\eqref{eq:kq_par_definition} that $\alpha_0(k_\parallel) = (\omega/c)\cos\theta_0$ and  $\alpha_0(q_\parallel) = (\omega/c)\cos\theta_s$.  It should be noticed from Eq.~\eqref{eq:R-Kichhoff} that the dependence of  $R_{\alpha\beta}(\pvec{q}|\pvec{k})$ on the surface profile function $\zeta(\pvec{x})$ only enters through $r(\pvec{q}|\pvec{k})$ via the argument of the second exponential factor of its integrand.  The results in Eq.~\eqref{eq:R-Kichhoff} constitute the scattering amplitudes obtained by applying  the stationary phase approximation to the integrals of the Kirchhoff approximation; some (but not all)  authors use the \textrm{tangent plane approximation} to refer to this combined approximation.  In the following, we will simply refer loosely to this combined approximation as the Kirchhoff approximation.

The expressions for the mean DRCs within the Kirchhoff approximation are obtained from Eq.~\eqref{eq:mdrc-definition} by the substitution of the expressions for the scattering amplitudes from Eq.~\eqref{eq:R-Kichhoff}.  The ensemble averages that appear in the expressions obtain in this way can be calculated analytically since it has been assumed that the surface profile function $\zeta(\pvec{x})$ constitutes an isotropic, Gaussian random process.  Under these assumptions, it can be demonstrated that the expression for the incoherent component of the mean DRCs can be expressed as (see Appendix~\ref{Appendix:A} for details)
\begin{subequations}
  \label{eq:MDRC-approx}
  \begin{align}
    \left< \frac{\partial R_{\alpha\beta}(\pvec{q}|\pvec{k})}{\partial\Omega_s}\right>_{\textrm{incoh}}
    &=
      \Phi_{\alpha\beta}\left(\pvec{q}|\pvec{k}\right)
      L( \pvec{q}| \pvec{k} )_{\textrm{incoh}},
      \label{eq:MDRC-approx-a}
  \end{align}
  with 
  \begin{align}
    \Phi_{\alpha\beta}\left(\pvec{q}|\pvec{k}\right)
    &=
      \frac{
      \omega/c
      }{
      4 \alpha_0(k_\parallel)
      \Lambda^2( \pvec{q} | \pvec{k} )      
      }
      \frac{
      \left| {\mathcal S}_{\alpha\beta} (\pvec{q} | \pvec{k} ) \right|^2
      }{
      \Delta^4(\pvec{q} | \pvec{k} )
      }
  \label{eq:Phi}
  \end{align}
  and 
  \begin{align}
    \label{eq:L}
    L( \pvec{q}| \pvec{k} )_{\textrm{incoh}}
    &=
      \frac{1}{2\pi}
      \exp
      \left[
      -\delta^2
      \Lambda^2( \pvec{q} | \pvec{k} )
      \right]
      \displaystyle \int_0^\infty \dint{u_\parallel}
      u_\parallel \textrm{J}_0\!\left( |\pvec{q} - \pvec{k}| u_\parallel \right)
       \left\{
      \exp
      \left[
      \delta^2
      \Lambda^2( \pvec{q} | \pvec{k} )
       W( u_\parallel )
      \right]      
      -
      1
      \right\}.
  \end{align}
\end{subequations}
In writing these expressions, we have introduced the \emph{wave vector transfer} for the scattering process 
\begin{align}
  \vec{Q}=\vec{q}-\vec{k}
  =
  \pvec{Q}
  +
   \Lambda( \pvec{q} | \pvec{k} )\vecUnit{x}_3  
\end{align}
which 3rd component can be written in the form  
\begin{align}
  \label{eq:Lambda}
  \Lambda( \pvec{q} | \pvec{k} )
  &=
    \alpha_0(|\pvec{q}|) + \alpha_0(|\pvec{k}| ),
\end{align}
and $\textrm{J}_0(\cdot)$ denotes the Bessel function of the first kind and order zero~\cite[Ch.~10]{Book:Olver2010}.  When evaluating the expressions in Eq.~\eqref{eq:MDRC-approx} it should be recalled that the parallel wave vectors of the incident and scattered light, $\pvec{k}$ and $\pvec{q}$ respectively, as well as  $\Lambda(\pvec{q} | \pvec{k})$,  are defined in terms of the angles of incidence and scattering through the use of the expressions in Eq.~\eqref{eq:kq_par_definition}.  Furthermore, care should be taken when evaluating the expressions in Eq.~\eqref{eq:MDRC-approx} for $\pvec{q}\approx-\pvec{k}$ since the $\Delta$-function vanishes in the backscattering direction, \textit{i.e.\!} $\Delta( -\pvec{k} | \pvec{k} )=0$.  However, the quantity ${\mathcal S}_{\alpha\beta} ( -\pvec{k} | \pvec{k} )$ also vanishes, and it can be shown that the ratio $|{\mathcal S}_{\alpha\beta} ( -\pvec{k} | \pvec{k} )|^2 / \Delta^4( -\pvec{k} | \pvec{k} )$ remains finite~\cite{NavarreteAlcala2009}.

\smallskip
The expressions in Eq.~\eqref{eq:MDRC-approx} are derived by applying the method of stationary phase to the expressions for the scattering amplitudes that are obtained under the assumption of the Kirchhoff approximation~\cite{Sancer1969,Book:Ulaby1986}.  Therefore, the validity of the expressions~\eqref{eq:MDRC-approx} that approximates the mean DRC relies on the validity of the Kirchhoff approximation; this approximation is valid when $2kr_c\cos^3\theta \gg 1$~\cite{Fung1981,Ogilvy1987}, where $k$ is the wavenumber of the incident light, $r_c$ is the integrated radius of curvature, and $\theta$ represents the local angle of incidence.  Under the assumption of Gaussianly correlated surface roughness of correlation length $a$~[see Eq.~\eqref{eq:gaussW}], it can be shown that the validity criterion becomes $(\pi/\sqrt{3})(a^2/\lambda \delta) \cos^3\theta \gg 1$~\cite{Millet2004}.


\subsection{Derivation of the surface height correlation function}

The expression for the incoherent component of the mean DRCs in Eq.~\eqref{eq:MDRC-approx} forms the starting point of the non-parametric inversion scheme for the surface height auto-correlation function that we develop in this work.  Let us start by noting that the factor $L( \pvec{q}| \pvec{k} )_{\textrm{incoh}}$ contains \emph{all} the dependence of the problem on the surface roughness while $\Phi_{\alpha\beta}\left(\pvec{q}|\pvec{k}\right)$ on the other hand, is a polarization dependent geometrical factor that is independent of the roughness parameters.  We will now demonstrate how this observation can be combined with angular resolved scattering data to uncover the statistical properties of the randomly rough surface assumed in obtaining the scattering data.

In the following, it will be assumed that a metallic or dielectric randomly rough surface is illuminated by a $p$- or $s$-polarized plane-wave.  Since the surface roughness is isotropic, without loss of generality we can choose a coordinate system so that $\phi_0=\ang{0}$; this means that the plane of incidence is the $x_1x_3$-plane, $\pvec{k}$ is a constant parallel wave vector with $\pvecUnit{k}=\vecUnit{x}_1$ where a caret over a vector means that it is a unit vector, and $k_\parallel=|\pvec{k}|=(\omega/c)\sin\theta_0$. 

Angular resolved scattering data $\left< \partial R_{\alpha\beta}(\pvec{q}|\pvec{k})/\partial\Omega_s\right>$ are collected for a given $\pvec{k}$ and a set of values of the parallel scattered wave vector $\pvec{q}$, and polarization indices $\alpha$ and $\beta$ of the scattered and incident light, respectively.  From such scattering data, the incoherent component can be extracted.  Here we will assume that such data are collected within the plane of incident so that $\pvecUnit{q}=\pm\vecUnit{x}_1$ [$\phi_s=\ang{0},\ang{180}$], and that the polar angle of scattering $\theta_s$ varies in the interval from \ang{0} to \ang{90}.  We feel that this angular dependence is the most intuitive from an experimental point of view.  It should be pointed out, however, that the inversion scheme that we introduce does not depend critically on this assumption; alternatively, one may have chosen to collect scattering data for an azimuthal angle of scattering in the range $-\ang{180}<\phi_s<\ang{180}$ and a fixed polar angle of scattering $\theta_s=\theta_0$; for instance, this was the configuration assumed in the approach presented in Ref.~(\onlinecite{Zamani2016}). Furthermore, we will primarily be interested in scattering data for which the polarizations of the incident and the scattered light are the same [$\alpha=\beta\in\{p,s\}$], that is, we deal with \textit{co-polarized} scattering data.  This choice is mainly motivated by the fact that within the plane of incidence, cross-polarized scattering [$\alpha\neq \beta$] is a multiple scattering effect and, therefore, the intensity of cross-polarized scattered light for the surfaces we are interested in will be significantly smaller than the intensity of co-polarized scattered light, unless the sample is so that multiple scattering is the dominating scattering mechanism.  Under these assumptions, the scattering data on which we will base our inversion scheme, consists of the angular dependence of the incoherent component of the co-polarized mean~DRC in the plane of incidence.  In what follows, this quantity will be denoted  $\left.\left[\left< \partial R_{\alpha\alpha}( \pvec{q}|\pvec{k})/\partial\Omega_s\right>_{\textrm{incoh}}\right] \right|_{q_2=0}$.

\smallskip
The derivation of the equations that our inversion scheme will be based upon, starts by developing an approximation for $L( \pvec{q}| \pvec{k} )_{\textrm{incoh}}$.  To this end, we expand the function $\Lambda( \pvec{q} | \pvec{k} )=\Lambda( \pvec{k}+\pvec{Q}| \pvec{k} )$ around the specular direction $\pvec{q}=\pvec{k}$, or vanishing the  parallel wave vector transfer $\pvec{Q}=\pvec{q}-\pvec{k}$, with the result that  
\begin{align}
  \label{eq:Lambda-expansion}
  \Lambda( \pvec{q} | \pvec{k} )
  \equiv
  \Lambda( \pvec{k}+\pvec{Q} | \pvec{k} )
  &=
  \Lambda( \pvec{k} | \pvec{k} )
    +
    \mathcal{O} ( Q_\parallel ).
    %
\end{align}
By retaining only the leading order term of this expansion in the integral defining $L( \pvec{q}| \pvec{k} )_{\textrm{incoh}}$, and substituting the resulting expression into Eq.~\eqref{eq:MDRC-approx-a} followed by a rearrangement of terms in the resulting equation, we get that
\begin{align}
  \label{eq:MDRC_approximated_and_rearranged}
  \displaystyle \int_0^\infty \dint{u_\parallel}
    u_\parallel \textrm{J}_0\!\left( |\pvec{q} - \pvec{k}| u_\parallel \right)
    \left\{
    \exp
    \left[
      \delta^2
      \Lambda^2( \pvec{k} | \pvec{k} )
      W( u_\parallel )
      \right]      
      -
      1
    \right\}
    %
  &
  \approx
      2 \pi
      \frac{
      \exp
      \left[
      \delta^2
      \Lambda^2( \pvec{q} | \pvec{k} )
      \right]
      }{
      \Phi_{\alpha\beta}\left(\pvec{q}|\pvec{k}\right)
      }
      \left< \frac{\partial R_{\alpha\beta}(\pvec{q}|\pvec{k})}{\partial\Omega_s}\right>_{\textrm{incoh}}.
\end{align}
The approximation~\eqref{eq:Lambda-expansion}~[$\Lambda( \pvec{q} | \pvec{k} )\approx \Lambda(\pvec{k} | \pvec{k} )$], and therefore the validity of the expression in Eq.~\eqref{eq:MDRC_approximated_and_rearranged}, is good when $\big(2\pvec{k}\cdot \pvec{Q}+Q_\parallel^2\big) / \alpha_0^2(k_\parallel) \ll 1$, that is, for sufficiently small angular regions around the specular direction and this approximation is best for small polar angles of incidence.  It should be remarked that the majority of the diffusely scattered light typically is concentrated around the specular direction for which expression~\eqref{eq:MDRC_approximated_and_rearranged} is valid.

The integral on the left-hand side of Eq.~\eqref{eq:MDRC_approximated_and_rearranged} is the Hankel (or Fourier-Bessel) transform of order zero~\cite[Ch.~7]{Book:Debnath2015} of the function $\{\exp\left[\delta^2 \Lambda^2( \pvec{k} | \pvec{k} ) W( u_\parallel ) \right] - 1\}$.  Notice that this is only the case after the approximation  $\Lambda( \pvec{q} | \pvec{k} ) \approx   \Lambda( \pvec{k} | \pvec{k} )$ has been applied.  Motivated by the definition of the inverse Hankel transform of order zero~\cite{Book:Debnath2015}, we multiply the expression in Eq.~\eqref{eq:MDRC_approximated_and_rearranged} by $|\pvec{q}-\pvec{k}| \textrm{J}_0\!\left( |\pvec{q} - \pvec{k}| x_\parallel \right)$, set $q_2=0$ to guarantee that we are in the plane of incidence, and  integrate the resulting expression over all propagating values $-\omega/c\leq q_1 \leq \omega/c$.  In this way, after interchanging the order of the $u_\parallel$- and $q_1$-integrals on the left-hand side of the resulting equation, and using $\Lambda(\pvec{k}|\pvec{k})=2\alpha_0(k_\parallel)$, we obtain
\begin{align}
  \label{eq:MDRC_approximated_and_rearranged-2}
  \displaystyle \int_0^\infty \dint{u_\parallel}
  u_\parallel 
  \left\{
    \exp
  \left[
    4
    \delta^2
    \alpha_0^2(k_\parallel)
    W( u_\parallel )
    \right]      
    -
    1
  \right\}
  \displaystyle \int_{-\omega/c}^{\omega/c} \dint{q_1}
  \! \! \left.
  \left[
  |\pvec{q} - \pvec{k}|
  \textrm{J}_0\!\left( |\pvec{q} - \pvec{k}| x_\parallel \right)
  \textrm{J}_0\!\left( |\pvec{q} - \pvec{k}| u_\parallel \right)
  \right]
  \right|_{q_2=0}
  %
  %
  %
  &
  \approx
    \widehat{D}_\alpha(x_\parallel | \pvec{k}, \delta ).
\end{align}
In writing this equation, we have defined the  in-plane co-polarized scattering data dependent function 
\begin{align}
  \widehat{D}_{\alpha}(x_\parallel | \pvec{k}, \delta )
  &=
    2 \pi
    \int^{\omega/c}_{-\omega/c} \dint{q_1}
    \left. \left[
    \left|\pvec{q}-\pvec{k}\right| \mathrm{J}_0\!\left( |\pvec{q}-\pvec{k}| x_\parallel\right)
    \frac{
    \exp
    \left[
    \delta^2
    \Lambda^2( \pvec{q} | \pvec{k} )
    \right]
    }{
    \Phi_{\alpha\alpha}\left(\pvec{q}|\pvec{k}\right)
    }
    \left< \frac{\partial R_{\alpha\alpha}(\pvec{q}|\pvec{k})}{\partial\Omega_s}\right>_{\textrm{incoh}}
    \right]\right|_{q_2=0}.
    \label{eq:W_estimate-eq_b}
\end{align}
The limits of the $q_1$-integration in Eq.\eqref{eq:MDRC_approximated_and_rearranged-2} are $\pm\omega/c$, and for optical frequencies as we are concerned about here, $\omega/c$ is of the order $\num{E6}$ $m^{-1}$.  Therefore, as an additional approximation, we will extend the limits of the $q_1$-integration in Eq.\eqref{eq:MDRC_approximated_and_rearranged-2} to $\pm\infty$ so that 
\begin{align}
  \label{eq:Bessel-othogonality}
  \displaystyle \int_{-\omega/c}^{\omega/c} \dint{q_1}
  \! \! \left.
  \left[
  |\pvec{q} - \pvec{k}| 
  \textrm{J}_0\!\left( |\pvec{q} - \pvec{k}| x_\parallel \right) 
  \textrm{J}_0\!\left( |\pvec{q} - \pvec{k}| u_\parallel \right)
  \right]
  \right|_{q_2=0}
  %
  %
  %
  %
  %
  &\approx
  \displaystyle \int_{-\infty}^{\infty} \dint{q_1}
  \! \! \left.
  \left[
  |\pvec{q} - \pvec{k}|
  \textrm{J}_0\!\left( |\pvec{q} - \pvec{k}| x_\parallel \right)
  \textrm{J}_0\!\left( |\pvec{q} - \pvec{k}| u_\parallel \right)
  \right]
    \right|_{q_2=0}
    \nonumber
  \\
  &=
  2                        
  \displaystyle \int_{0}^{\infty} \dint{Q_\parallel}
    Q_\parallel
  \textrm{J}_0\!\left( Q_\parallel x_\parallel \right)
  \textrm{J}_0\!\left( Q_\parallel u_\parallel \right)
    %
    \nonumber
  \\
  &=
    \frac{2}{u_\parallel}
    \delta\left( u_\parallel - x_\parallel \right).
\end{align}
Here in the second transition we have changed variable to $Q_\parallel=q_1-k_1$ and used the fact that the integrand is an even function of this variable, while the last transition follows from the orthogonality of the Bessel functions~\cite{Leon2014}.  When the approximation and the result from Eq.~\eqref{eq:Bessel-othogonality} are introduced into Eq.~\eqref{eq:MDRC_approximated_and_rearranged}, the $u_\parallel$-integral of the resulting equation can be readily performed.  After some straightforward algebra, one obtains the following non-parametric estimate for the surface height correlation function
\begin{subequations}
  \label{eq:statistical-property-estimation}
  \begin{align}
    W( x_\parallel)
    &\approx
      \frac{1}{4\delta^2\alpha_0^2(k_\parallel)}
      \ln
      \left[
      1
      +
      \frac{1}{2} \widehat{D}_{\alpha}( x_\parallel | \pvec{k}, \delta ) 
      \right].
      \label{eq:W_estimate-eq_a}
  \end{align}
  The expression in  Eq.~\eqref{eq:W_estimate-eq_a} depends on the surface roughness $\delta$ which we do not know \textit{a priori}.  However, an estimate for it is obtained from the normalization condition of the correlation function $W(0)=1$.  In this way, we find that an estimate $\delta=\hat{\delta}_\star$ for the surface roughness --- obtained on the basis of the in-plane scattering data --- is the solution (or fixed-point) of the transcendental equation 
  \begin{align} 
    \hat{\delta}_\star
    &= 
       \frac{
      \sqrt{
      \ln \left[
      1
      +
      \frac{1}{2} \widehat{D}_{\alpha}( 0 | \pvec{k}, \hat{\delta}_\star )
      \right] }
       }{
      2\alpha_0(k_\parallel)
      }.
    \label{eq:delta_estimate}
  \end{align}
\end{subequations}
 The non-parametric reconstruction approach that we propose in this work is based on the expressions in Eqs.~\eqref{eq:W_estimate-eq_b} and \eqref{eq:statistical-property-estimation}, and it consists of two main steps.  In the first step we estimate the surface rms-roughness $\hat{\delta}_\star$ from the input data on the basis of Eq.~\eqref{eq:delta_estimate}.  Below we detail how this equation can be solved numerically.  Next, the estimated rms-roughness value $\delta=\hat{\delta}_\star$ is introduced on the right-hand side of Eq.~\eqref{eq:W_estimate-eq_a}, and used to obtain the function form of the surface height correlation function $W(x_\parallel)$ without initially assuming any particular form for this function.   

\smallskip
The solution of the transcendental equation~\eqref{eq:delta_estimate} is conveniently obtained by iterations.  This is done by replacing $\hat{\delta}_\star$ on the right-hand and left-hand sides of this equation by $\hat{\delta}_{n}$ and $\hat{\delta}_{n+1}$, respectively, where $\hat{\delta}_{n}$ is the $n$th approximation to the fixed-point value $\hat{\delta}_\star$ and $n\geq 0$ is an integer.  For a sufficiently large value of $n$, it is hoped that convergence can be reached.  To obtain the first iterate $\hat{\delta}_{0}$ to be used in the iterative solution of this equation, we introduce an additional approximation into Eq.~\eqref{eq:statistical-property-estimation}.  As we will detail in the next paragraph, this will lead to an alternative and \emph{simplified set of reconstruction formulae} [see Eq.~\eqref{eq:statistical-property-estimation-simplified} below] that is similar to that of Eq.~\eqref{eq:statistical-property-estimation}, but which use is expected to result in less or equally accurate reconstruction of the surface height correlation function and the surface roughness.  The advantage of using this simplified formulation is that the surface roughness that it produces, $\delta_\star$, is available in a closed-form~[see Eq.~\eqref{eq:delta_estimate-simplified} below]  and can therefore readily be calculated.  We will often use the term \emph{non-iterative reconstruction} to refer to the reconstruction that is performed on the basis of the simplified approach.  For this reason, the iterative solution of Eq.~\eqref{eq:delta_estimate} can be started by setting $\hat{\delta}_0=\delta_\star$.

%
%
Now we will present the alternative set of simplified reconstruction equations that, for instance, is used to obtain the expression for  $\delta_\star$.  The simplified set of reconstruction equations is obtained by applying  the approximation $\Lambda(\pvec{q}|\pvec{k})\approx \Lambda(\pvec{k}|\pvec{k})$ to the integrand on the right-hand side of Eq.~\eqref{eq:W_estimate-eq_b}, with the consequence that \emph{all} surface roughness dependent factors of the integrand can be moved outside the integral in this equation.  In this way, we obtain the alternative way of estimating the surface height correlation function with the result that 
\begin{subequations}
  \label{eq:statistical-property-estimation-simplified}
  \begin{align}
    W( x_\parallel)
    &\approx
      \frac{1}{4 \delta^2_\star \alpha_0^2(k_\parallel)}
      \ln
      \left[
      1 + \frac{1}{2} D_{\alpha}( x_\parallel | \pvec{k} )
      \exp\left[ 4 \delta^{2}_\star\alpha^2_0(k_\parallel)\right]
      \right],
      \label{eq:W_estimate-eq_a-simplified}
  \end{align}
  where a surface roughness \textit{independent} function is defined by
  \begin{align}
    D_{\alpha}(x_\parallel | \pvec{k} )
    &=
      2 \pi
      \int^{\omega/c}_{-\omega/c} \dint{q_1}
      \left. \left[
      \left|\pvec{q}-\pvec{k}\right| \mathrm{J}_0\!\left( |\pvec{q}-\pvec{k}| x_\parallel\right)
      \Phi_{\alpha\alpha}^{-1}\left(\pvec{q}|\pvec{k}\right)
      \left< \frac{\partial R_{\alpha\alpha}(\pvec{q}|\pvec{k})}{\partial\Omega_s}\right>_{\textrm{incoh}}
      \right]\right|_{q_2=0}.
      \label{eq:W_estimate-eq_b-simplified}
  \end{align}
  It should be noted that this function and $\widehat{D}_{\alpha}(x_\parallel | \pvec{k}, \delta )$ as defined in Eq.~\eqref{eq:W_estimate-eq_b} are related by $\widehat{D}_{\alpha}(x_\parallel | \pvec{k}, \delta ) \approx D_{\alpha}(x_\parallel | \pvec{k} ) \exp[4\delta^{2}\alpha_0^2(k_\parallel)]$.  Still the estimate for the rms-roughness of the surface is obtained when the normalization condition $\widehat{W}( 0 )=1$ is applied to the expressions in Eq.~\eqref{eq:W_estimate-eq_a-simplified}.  In this way, we obtain an expression that can be solved explicitly for the surface roughness to produce
  \begin{align}
    \delta_\star
    &= 
       \frac{
          \sqrt{ - \ln \left[ 1 - \frac{1}{2} D_{\alpha}( 0 | \pvec{k} ) \right] }
       }{
      2\alpha_0(k_\parallel)
      }.
    \label{eq:delta_estimate-simplified}
  \end{align}
\end{subequations}
The expressions in Eq.~\eqref{eq:statistical-property-estimation-simplified} constitute the simplified set of reconstruction formulae for the surface height correlation function and the rms-roughness (non-iterative reconstruction).


\section{Results and discussion}
\label{sec:Results}

%
\begin{figure}[tbh]
  \centering
  \includegraphics[width=0.46\columnwidth]{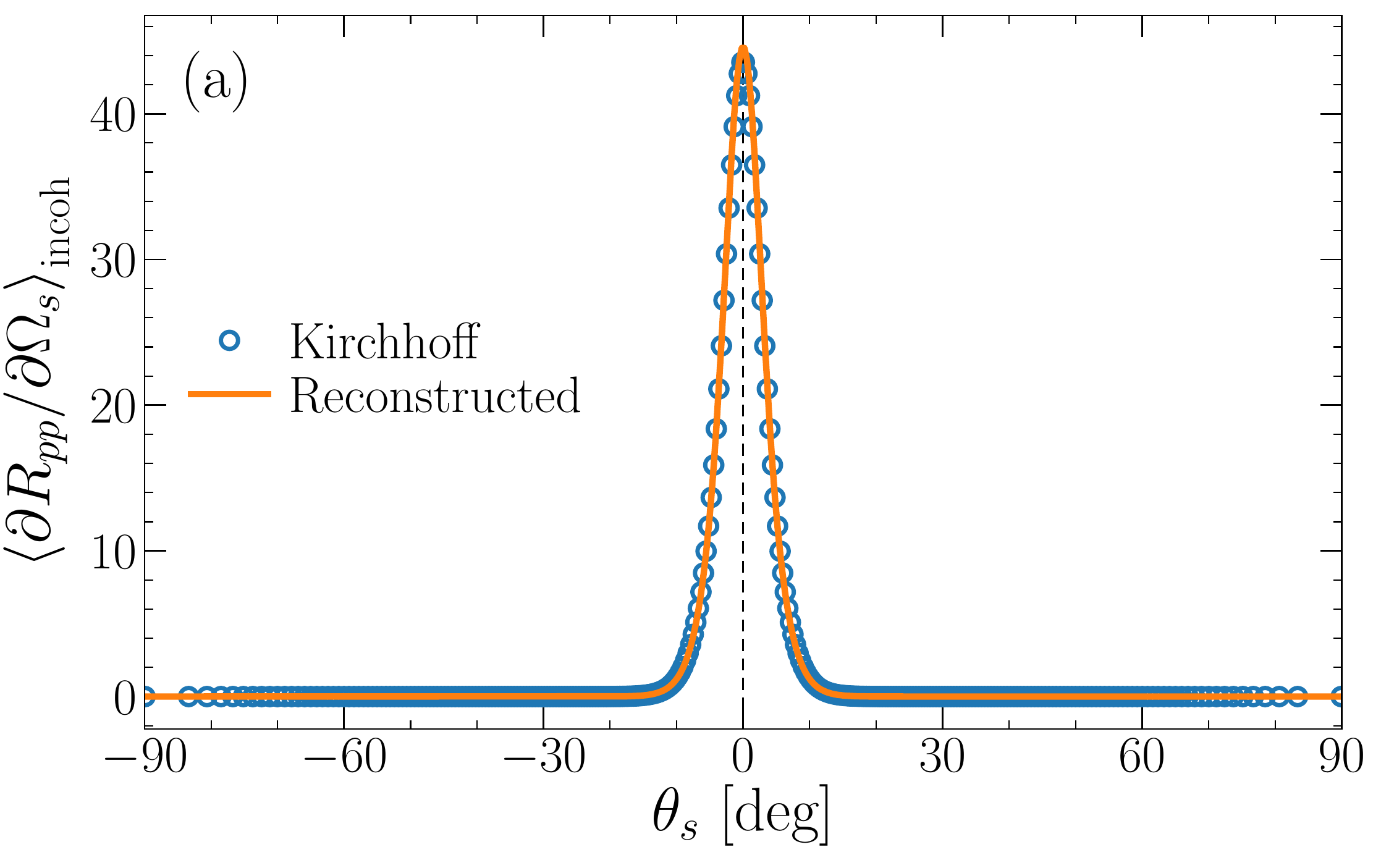}
  \qquad
  \includegraphics[width=0.46\columnwidth]{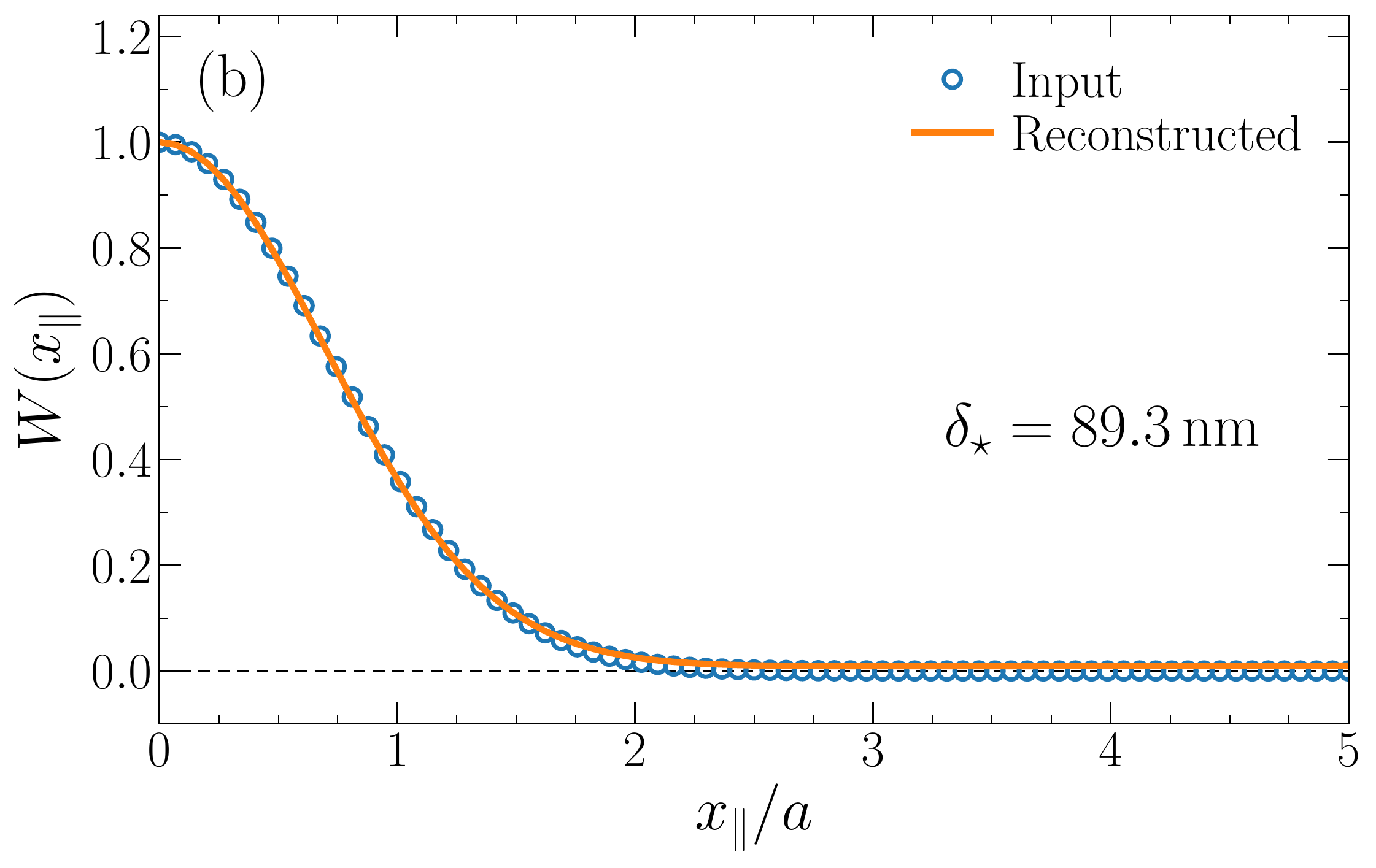}
  \\*[1ex]
  \includegraphics[width=0.46\columnwidth]{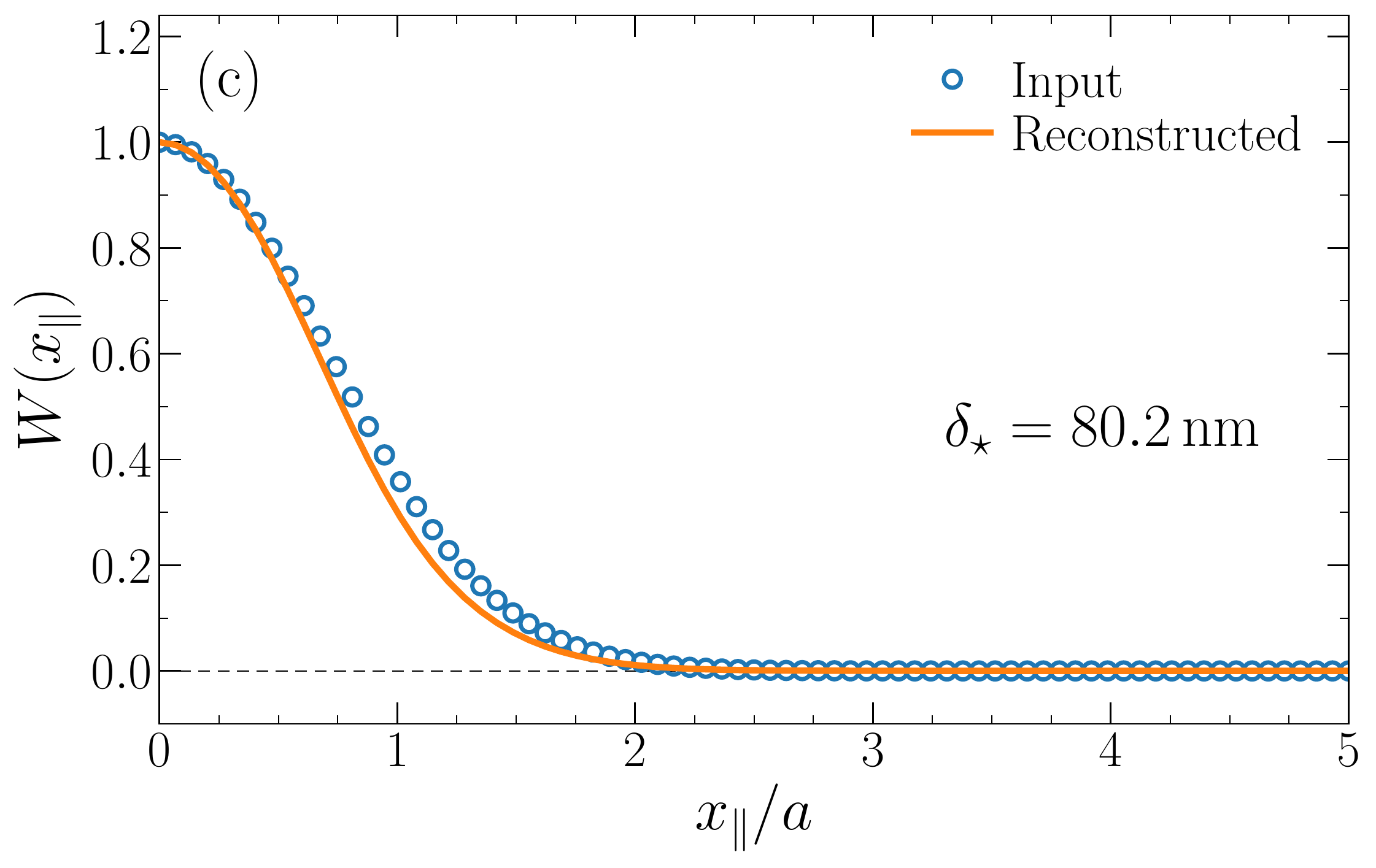}
  \qquad
  \includegraphics[width=0.46\columnwidth]{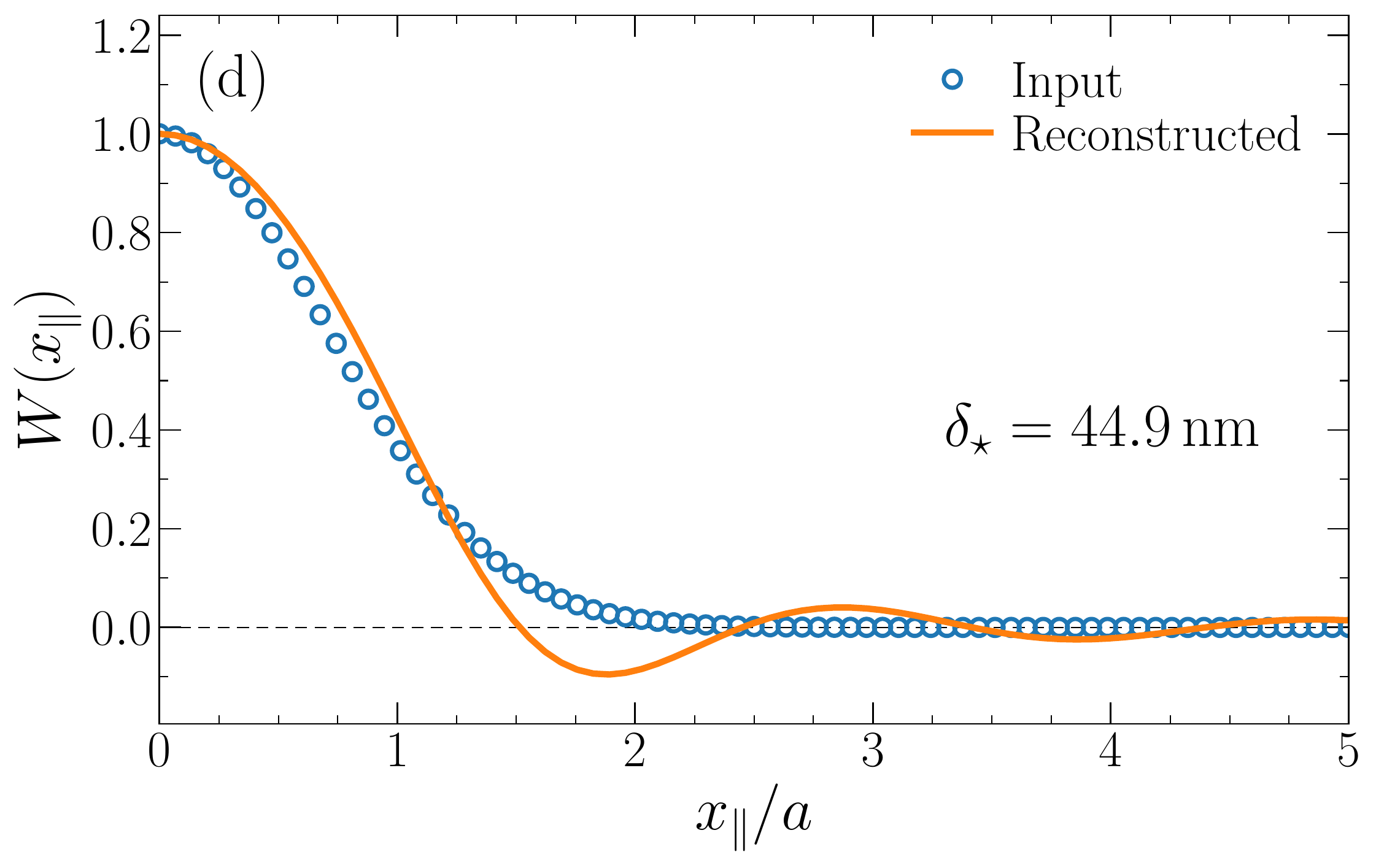}  
  \caption{Reconstruction of the normalized-surface-height auto-correlation function  $W(x_\parallel)$ and surface roughness $\delta$ based on in-plane and p-to-p scattering data corresponding to three different rough \textit{vacuum-silver} systems. In all cases, the polar angle of incidence is $\theta_0=\ang{0}$, the wavelength of the incident light is $\lambda=\SI{632.8}{nm}$, and the surface rms-roughness is  $\delta=\lambda/7=\SI{90.4}{nm}$. The dielectric function of silver at wavelength $\lambda$ is  $\varepsilon=-18.28+\mathrm{i}\,0.48$~\cite{Johnson1972}. The surface correlation function $W(x_\parallel)$ is assumed to have the Gaussian form~\eqref{eq:gaussW} and characterized by the a correlation length of (a,b)~$a=6\lambda$; (c)~$a=2\lambda$; and (d)~$a=0.5\lambda$. The functional forms of the resulting correlation functions are presented as open symbols in Panels~(b)--(d). The scattering data  $\left<\partial R_{pp}/\partial\Omega_s\right>_{\textrm{incoh}}$ in the plane of incidence that our reconstruction approach uses as input, were calculated on the basis of the Kirchhoff approximation from Eq.~\eqref{eq:MDRC-approx}; the open symbols in Panel~(a) represent such data for $a=6\lambda$ and the thin vertical dashed line indicates the polar angle of incidence. When non-iterative reconstruction of the input scattering data is performed on the basis of Eq.~\eqref{eq:statistical-property-estimation-simplified},  the reconstructed correlation functions are obtained and presented as solid lines in Panels~(b)--(d). The corresponding rms-roughness obtained during the reconstructions are
    (b)~$\delta_\star=\SI{89.3}{nm}$; (c)~$\delta_\star=\SI{80.2}{nm}$; and (d)~$\delta_\star=\SI{44.9}{nm}$. The solid line in Panel~(a) is the mean DRC for p-to-p scattering obtained from Eq.~\eqref{eq:MDRC-approx} by assuming the reconstructed correlation function  and $\delta=\delta_\star$.}
 \label{fig:Kirchhoff-Ex1}
\end{figure}
%
%

We will now illustrate the inversion approach developed in the previous section by applying it to in-plane and co-polarized scattering data (the input data), obtained either by computer simulations or in experimental measurements.  This will be done for scattering data obtained for both rough dielectric and metallic surfaces that are characterized by correlation functions $W(x_\parallel)$ of different functional forms.  The purpose of doing so, is to judge the performance and the reliability of the inversion approach that we propose.


%
\begin{figure}[!tbh]
  \centering
  \includegraphics[width=0.475\columnwidth]{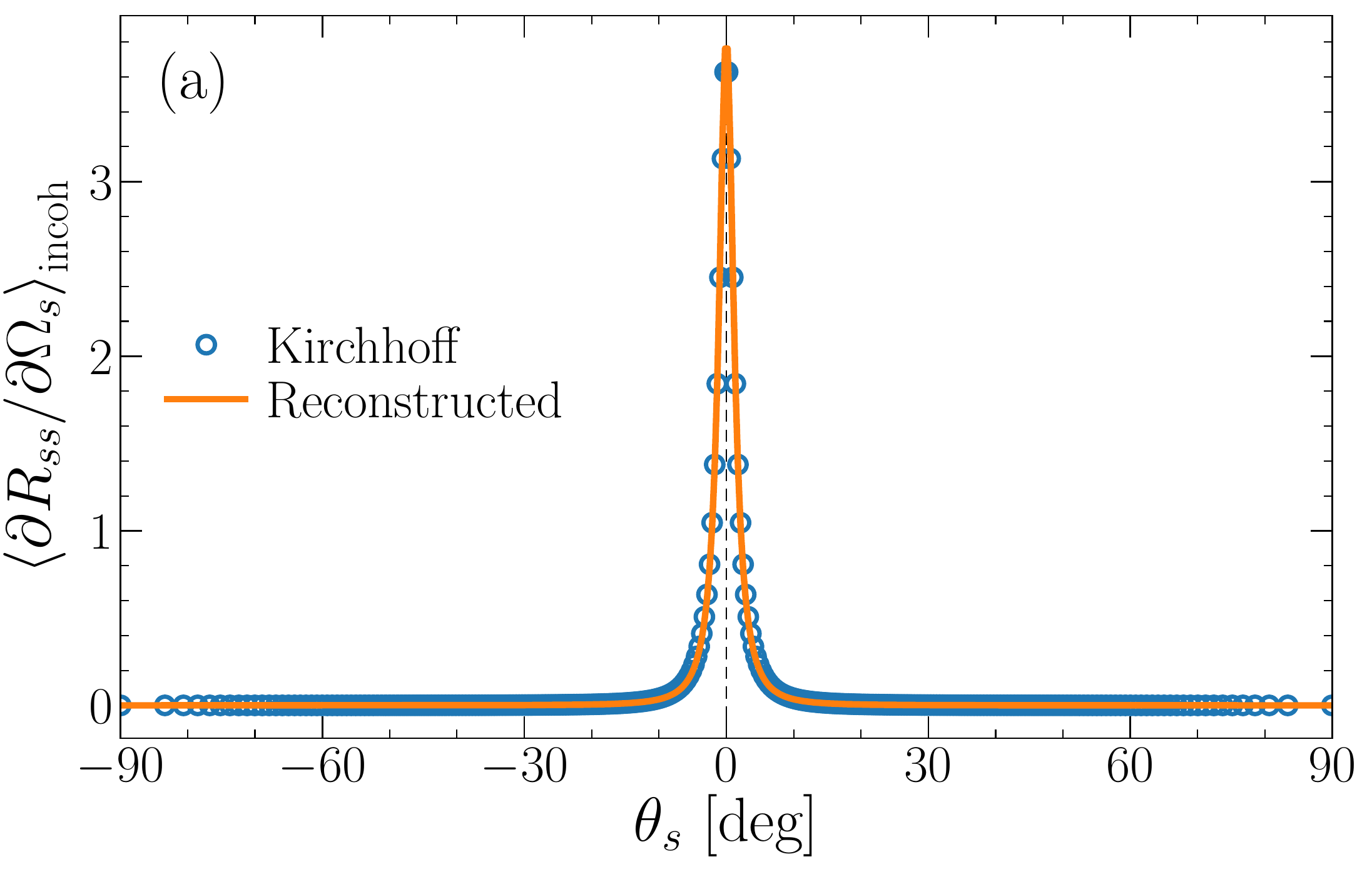}
  \qquad
  \includegraphics[width=0.475\columnwidth]{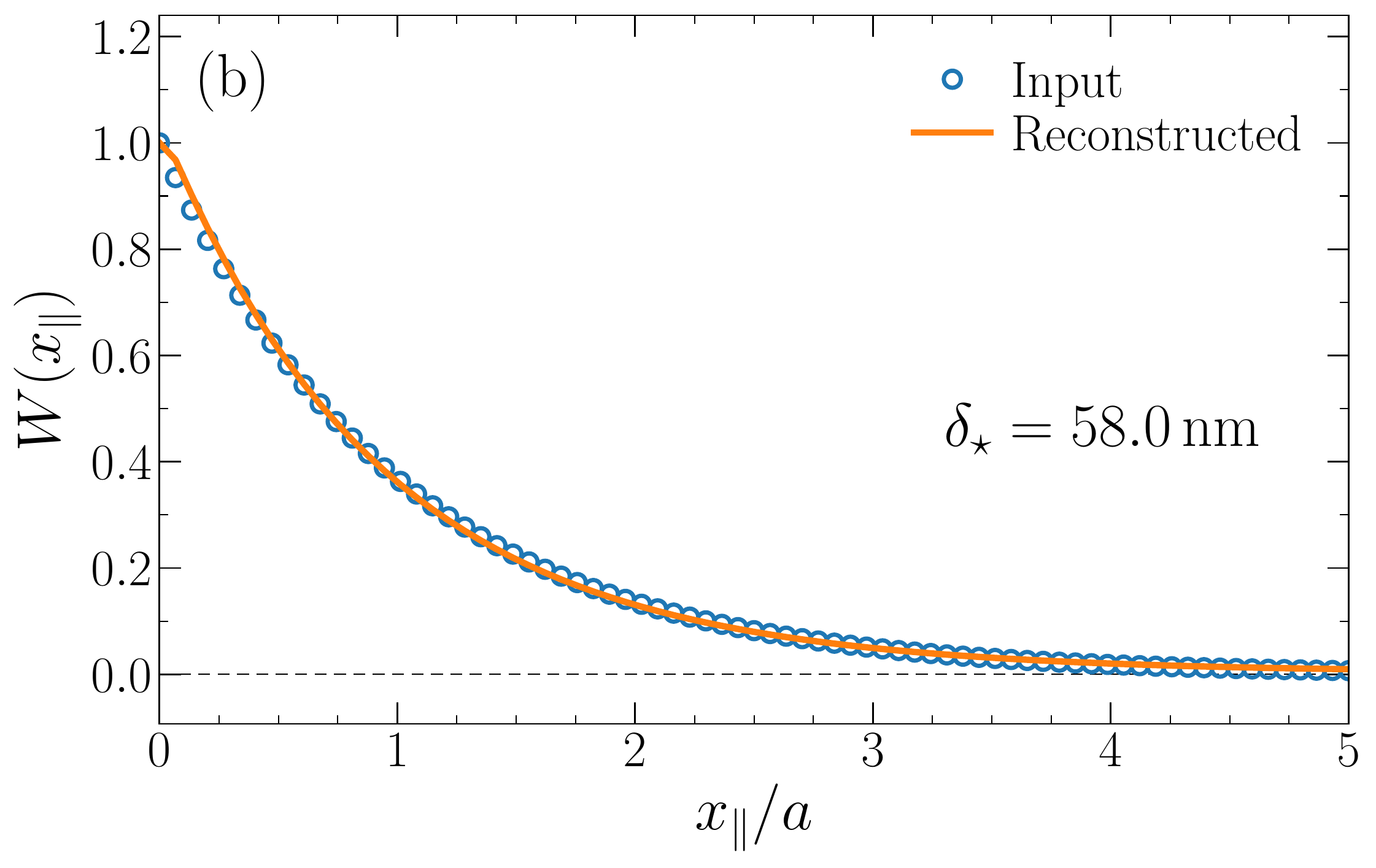}
  \caption{Reconstruction of the normalized-surface-height auto-correlation function  $W(x_\parallel)$ and surface roughness $\delta$ based on in-plane and s-to-s scattering data obtained for a rough \textit{vacuum-glass} system~[$\varepsilon=2.25$] for the polar angle of incidence $\theta_0=\ang{0}$. The surface is characterized by the roughness $\delta=\lambda/10\approx\SI{63.3}{nm}$ and the \textit{exponential} correlation function~\eqref{eq:exponW} defined by $a=6\lambda$. The input scattering data were generated on the basis of the Kirchhoff approximation by the use of Eq.~\eqref{eq:MDRC-approx}. All the remaining parameters and the explanations of the various curves and symbols in each of the two panels are identical to those of Figs.~\protect\ref{fig:Kirchhoff-Ex1}(a)--(b).}
 \label{fig:Kirchhoff-Ex2}
\end{figure}
%

%
\begin{figure}[!tbh]
  \centering
  \includegraphics[width=0.475\columnwidth]{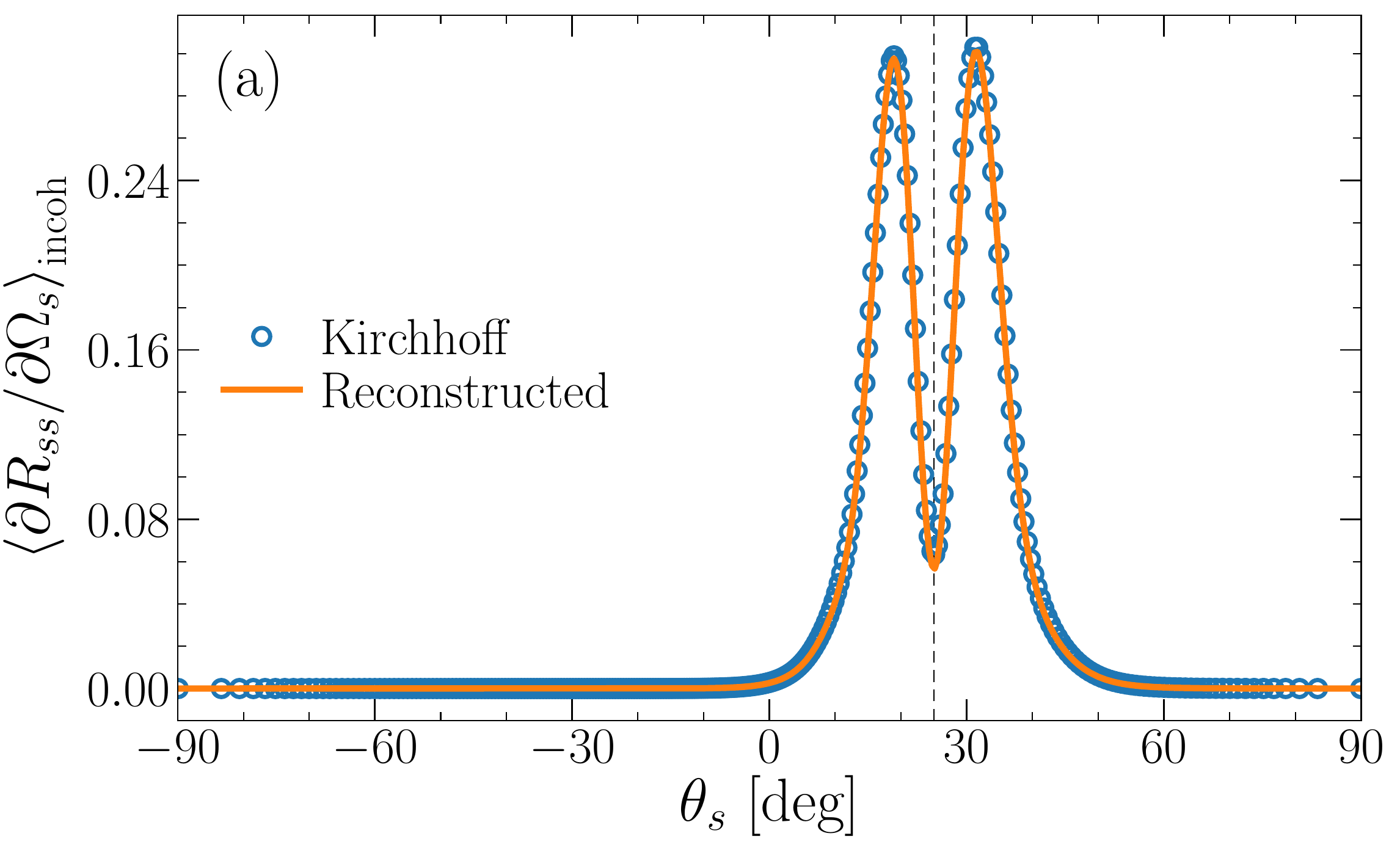}
  \qquad
  \includegraphics[width=0.475\columnwidth]{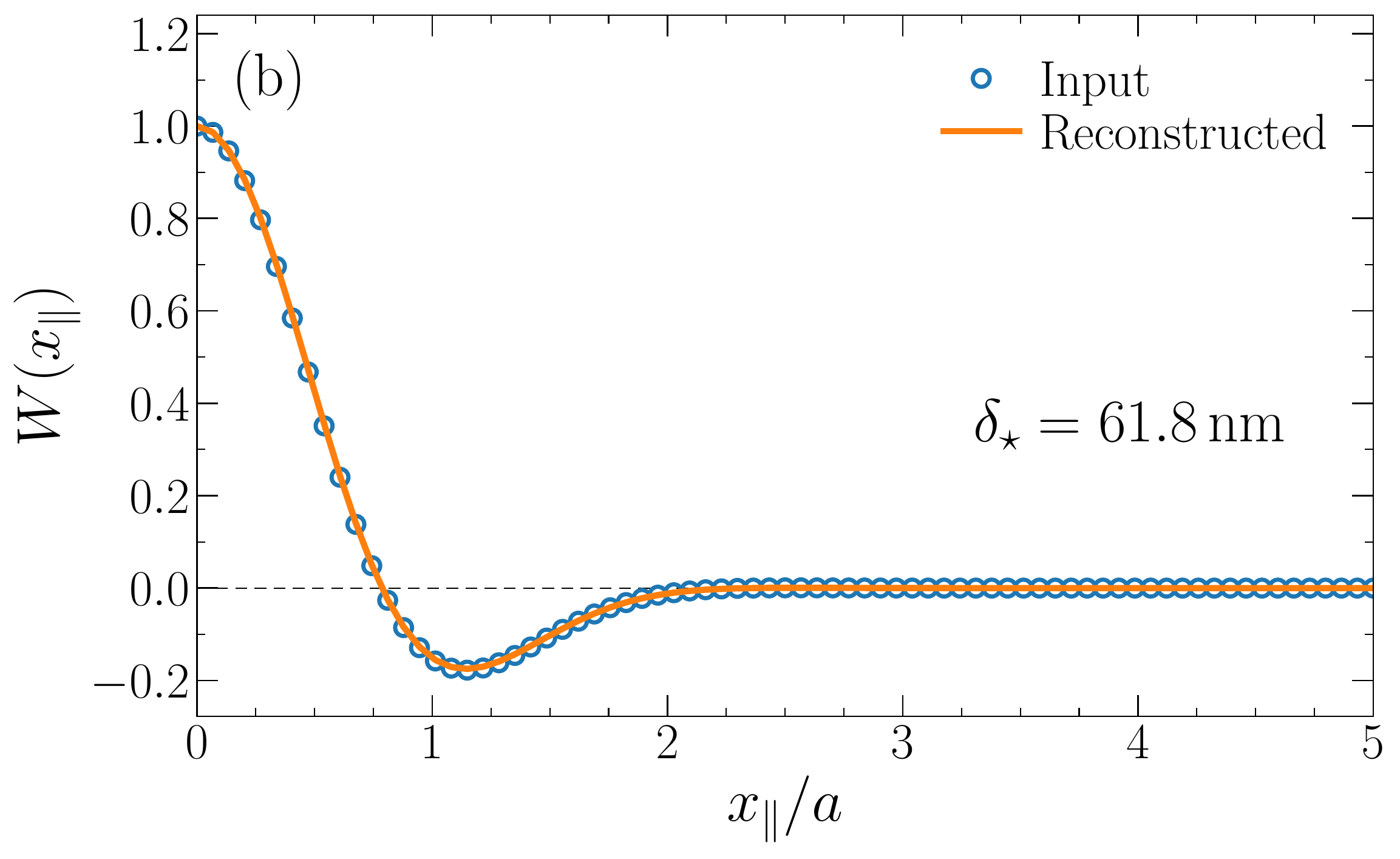}
  \caption{Reconstruction of the normalized-surface-height auto-correlation function  $W(x_\parallel)$ and surface roughness $\delta$ for in-plane and s-to-s input scattering data obtained for a rough \textit{vacuum-glass} system~[$\varepsilon=2.25$] for the polar angle of incidence $\theta_0=\ang{25}$. The surface is characterized by the roughness $\delta=\lambda/10\approx\SI{63.3}{nm}$ and the  \textit{Gaussian-cosine} form~\eqref{eq:cosineW} defined by the of parameters $a=4\lambda$ and $b=a/2$. The input scattering data were generated on the basis of the Kirchhoff approximation by the use of Eq.~\eqref{eq:MDRC-approx}. All the remaining parameters and the explanations of the various curves and symbols in each of the two panels are identical to those of Figs.~\protect\ref{fig:Kirchhoff-Ex1}(a)--(b).}
 \label{fig:Kirchhoff-Ex3}
\end{figure}
%
%

\subsection{Validation of the reconstruction approach}

However, before presenting such inversion results, we will address the question of how accurate the proposed reconstruction approach is, and therefore, how well the statistical properties of the randomly rough surfaces can be obtained on the basis of the scattering data that they produce.  To assist in answering this important question, we will generate scattering data on the basis of the Kirchhoff approximation~[Eq.~\eqref{eq:MDRC-approx}], and take them as input data to the reconstruction approach.  Following this procedure has to produce the assumed input roughness parameters $\delta$ and $W(x_\parallel)$ if no additional approximations are assumed in performing the reconstruction, since it takes the expressions in Eq.~\eqref{eq:MDRC-approx} as the starting point.  As should be apparent from the derivation leading to the expression in  Eqs.~\eqref{eq:statistical-property-estimation} and \eqref{eq:statistical-property-estimation-simplified}, \emph{additional} approximations were introduced in order to  arrive at these expressions.  Therefore, reconstruction based on scattering data generated within the Kirchhoff approximation is expected to reproduce the input surface roughness and correlation function well only when the additional approximations underlying the reconstruction procedure are well satisfied.

The first set of reconstruction results based on Kirchhoff scattering input data are presented in Fig.~\ref{fig:Kirchhoff-Ex1}.  Here the scattering system consists of a rough \emph{vacuum-silver} where the surface roughness is characterized by a correlation function of the \textit{Gaussian} form   
\begin{align}
  W(x_\parallel ) = \exp \left[ -\left( \frac{ x_\parallel}{a}\right)^2 \right],
  \label{eq:gaussW}
\end{align}
where the positive constant $a$ denotes the correlation length.  The wavelength (in vacuum) of the incident light is $\lambda=\SI{632.8}{nm}$, for which the dielectric function of silver is  $\varepsilon=-18.28+\mathrm{i}\,0.48$~\cite{Johnson1972}.  The rms-height of the surface is $\delta=\lambda/7=\SI{90.4}{nm}$, and the polar angle of incidence is $\theta_0=\ang{0}$.  The blue open symbols in Fig.~\ref{fig:Kirchhoff-Ex1}(a) represent the incoherent component of the p-to-p mean DRC $\left<\partial R_{pp}/\partial\Omega_s\right>_{\textrm{incoh}}$ in the plane of incidence, calculated within the Kirchhoff approximation on the basis of the expressions in Eq.~\eqref{eq:MDRC-approx}.  These data are the input data set on which the subsequent reconstruction will be performed.  The correlation length assumed in obtaining the input scattering data is $a=6\lambda$, and the corresponding input Gaussian correlation function is shown as open blue symbols in Fig.~\ref{fig:Kirchhoff-Ex1}(b).  When reconstruction of these input scattering data is performed on the basis of the expressions in Eq.~\eqref{eq:statistical-property-estimation-simplified}, the (reconstructed) correlation function presented as a solid orange line in Fig.~\ref{fig:Kirchhoff-Ex1}(b) is obtained.  It is found to agree rather well with the input correlation function~[blue open symbols in Fig.~\ref{fig:Kirchhoff-Ex1}(b)].  The rms-roughness of the surface obtained during the reconstruction is $\delta_\star=\SI{89.3}{nm}$, which also agrees rather well with the input value $\delta=\SI{90.4}{nm}$.  By assuming the reconstructed statistical properties of the rough surface, that is to use $\delta=\delta_\star$ and the reconstructed correlation function $W(x_\parallel)$ [solid line in Fig.~\ref{fig:Kirchhoff-Ex1}(b)], the incoherent component of the mean DRC can be calculated from Eq.~\eqref{eq:MDRC-approx}.  In this way we obtain the mean DRC curve displayed as a solid orange line in Fig.~\ref{fig:Kirchhoff-Ex1}(a) and labeled ``Reconstructed''.  It is remarked that in order to perform this calculation, the reconstructed correlation function $W(x_\parallel)$, know on a finite set of points along $x_\parallel$, was used to construct a Lagrange interpolation scheme that allowed for the calculation of the reconstructed correlation function for any value of $x_\parallel$ (no greater than the maximum value assumed in the reconstruction).   

Similarly, Fig.~\ref{fig:Kirchhoff-Ex1}(c) presents the input and reconstructed surface height correlation function when the input correlation length is  $a=2\lambda$ for the Gaussian correlated surface, and the scattering data on which the reconstruction is based are generated from Eq.~\eqref{eq:MDRC-approx}.  The reaming parameters are identical to those assumed in obtaining the results presented in Figs.~\ref{fig:Kirchhoff-Ex1}(a)--(b).  The rms-roughness obtained during the reconstruction is $\delta_\star=\SI{80.2}{nm}$.  Compared to the results presented in  Fig.~\ref{fig:Kirchhoff-Ex1}(b) for the longer correlation length [$a=6\lambda$], we find that the reconstruction results become poorer when the correlation length is decreased; the correlation function is acceptable and monotonously decreasing with the argument $x_\parallel$, while the rms-roughness obtained during the reconstruction is clearly underestimated relative to the input value.  This trend is increased when the input correlation length is decreased even further.  Figure~\ref{fig:Kirchhoff-Ex1}(d) shows results for the reconstruction based on scattering data obtained from Eq.~\eqref{eq:MDRC-approx} for a surface characterized by $a=\lambda/2$.  It should be remarked that for the roughness parameters assumed in producing the correlation function in this figure, the validity of the Kirchhoff approximation is expected to be poor~\cite{Thorsos1988,Voronovich2007}.  Anyhow, it is found that the reconstructed correlation function deviates significantly from the input correlation function; that former function is no longer a monotonously decreasing function of its argument and it displays regions of anti-correlation. Furthermore, the error in the reconstructed rms-roughness $\delta_\star=\SI{44.9}{nm}$ starts to become significant.  Based on the results presented in Fig.~\ref{fig:Kirchhoff-Ex1}, it is found that accurate reconstruction results are obtained when the correlation length of the rough surface is sufficiently long; at least this is the case for the Gaussian correlated surfaces illuminated at normal incidence that we considered.

%
\medskip
We now turn to the reconstruction of scattering data obtained for rough surfaces for which the surface height correlation function is of the \textit{exponential} form
\begin{align}
  W( x_\parallel ) = \exp \left( - \frac{ x_\parallel}{a}\right).
  \label{eq:exponW}
\end{align}
Here it will be assumed that the substrate is glass~[$\varepsilon=2.25$], and that s-polarized light of wavelength $\lambda=\SI{632.8}{nm}$ is incident normally~[$\theta_0=\ang{0}$] onto the mean surface. The roughness parameters of the rough surface are assumed to be $\delta=\lambda/10\approx\SI{63.3}{nm}$ and $a=6\lambda$. The input scattering data  generated on the basis of  Eq.~\eqref{eq:MDRC-approx} are presented as open symbols in Fig.~\ref{fig:Kirchhoff-Ex2}(a) and it is observed, as expected, that the mean DRC curve is more centered  around the specular direction. Furthermore, one observes that the overall incoherently reflected light is significantly less intense than what was found for the Gaussianly correlated surface of the same correlation length~[Fig.~\ref{fig:Kirchhoff-Ex1}(a)]. This is mainly caused by the different optical properties of dielectric and metallic substrates. When the reconstruction is based on these input data and Eq.~\eqref{eq:statistical-property-estimation-simplified}, one obtains the results presented as solid orange lines in Fig.~\ref{fig:Kirchhoff-Ex2}; the meaning of the various curves in this figure are the same as in Fig.~\ref{fig:Kirchhoff-Ex1}(a)--(b). The results presented in Fig.~\ref{fig:Kirchhoff-Ex2} do agree rather well with the corresponding quantities assumed in generating the input scattering data. One notes in particular that the rms-roughness obtained by reconstruction, $\delta_\star=\SI{58.0}{nm}$, is less accurate than the corresponding quantity for the Gaussian surface.

%
In our final example with the use of input scattering data generated within the Kirchhoff approximation~[Eq.~\eqref{eq:MDRC-approx}], we assume a correlation function of the \textit{Gaussian-cosine} form~\cite{Rasigni1983}
\begin{align}
  W(x_\parallel ) = \exp \left[ -\left( \frac{ x_\parallel}{a}\right)^2 \right]\cos\left(\frac{ x_\parallel}{b}\right). \label{eq:cosineW}
\end{align}
Here the positive constants $a$ and $b$ are both lengths scales characterizing the correlations of the surface roughness; since there are two such lengths, this form of  $W(x_\parallel )$ represents an example of a two-scale correlation function. Except for the difference in the form of the correlation function and the assumption of another polar angle of incidence $\theta_0=\ang{25}$, the scattering system and all remaining parameters are the same as what was assumed in producing the results presented in Fig.~\ref{fig:Kirchhoff-Ex2}. For the parameters characterizing the correlation function we assume $a=4\lambda$ and $b=a/2$. With these values, the functional form of $W(x_\parallel )$ is presented as open symbols in  Fig.~\ref{fig:Kirchhoff-Ex3}(b), and one observes that a region of anti-correlation exists  (defined by $W(x_\parallel )<0$). The results presented as open symbols in Fig.~\ref{fig:Kirchhoff-Ex3}(a) show that the physical consequence of this anti-correlation is that the incoherent component of the in-plane and co-polarized mean DRC display a pronounced dip around the specular direction. Figure~\ref{fig:Kirchhoff-Ex3} presents as solid lines the results that are obtained when non-iterative reconstruction is performed on the basis of this (Kirchhoff generated) input scattering data set. The rms-roughness obtained by reconstruction is   $\delta_\star=\SI{61.8}{nm}$, and it compares fairly well with the input value $\delta\approx\lambda/10=\SI{63.3}{nm}$. Furthermore, the results presented in Fig.~\ref{fig:Kirchhoff-Ex3}(b) demonstrate explicitly the good agreement between the functional forms of the input and reconstructed correlation functions. In particular, it is stressed that these results were obtained without making any prior assumptions about the functional form of the correlation functions; this demonstrates the power of the approach that we propose. That no assumption has to be made about the functional form of $W(x_\parallel)$ prior to reconstruction, is in sharp contrast to what was done in recent studies based on phase-perturbation theory~\cite{Simonsen2016-06,Simonsen2014-05}.

\medskip
Several conclusions can be drawn on the basis of the results presented in Figs.~\ref{fig:Kirchhoff-Ex1}--\ref{fig:Kirchhoff-Ex3}. First, one finds that the reconstruction works well for sufficiently long correlation lengths, which is also where the Kirchhoff approximation is accurate~\cite{Fung1981,Franco2017}. Hence, it is tempting to draw the conclusion that the reconstruction is reliable if the Kirchhoff approximation is accurate~\footnote{Recall here that it is strictly speaking the stationary phase approximation to the Kirchhoff integrals that is meant here by the term ``Kirchhoff approximation'' called the tangent plane approximation by some authors.}, but additional work is needed before such general conclusion can be drawn. Second, the reconstruction approach that we propose works well for (i) penetrable randomly rough surfaces, both metallic and dielectric; (ii) normal and non-normal incident light;  and (iii) for both p- and s-polarized incident light. Finally, no prior assumptions about the functional form of the correlation function are required, and the approach that we propose is, therefore, \textit{non-parametric}.

\subsection{Computer generated scattering data}

%
\begin{figure}[!tbh]
 \centering 
\includegraphics[width=0.475\columnwidth]{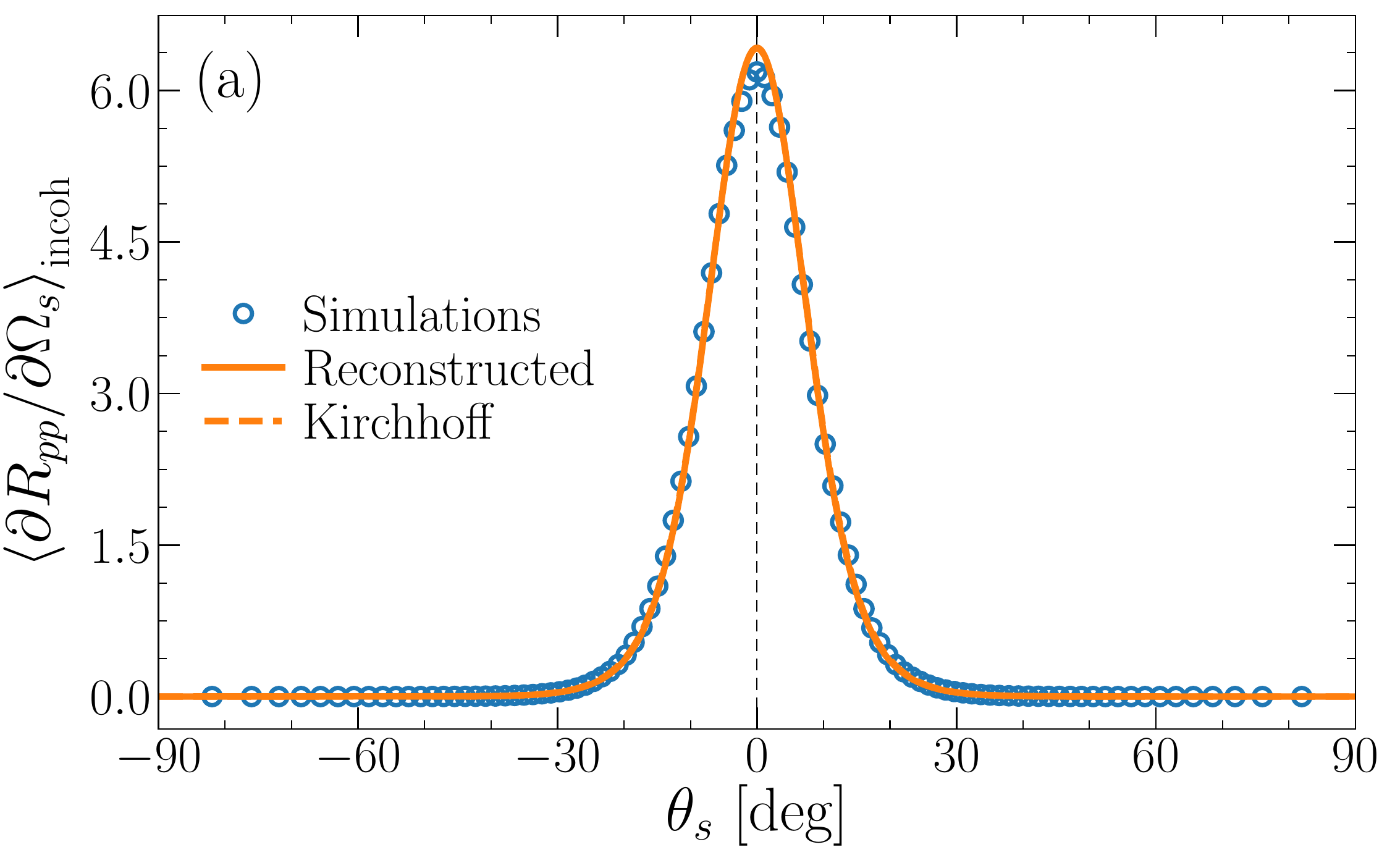}
\quad
\includegraphics[width=0.475\columnwidth]{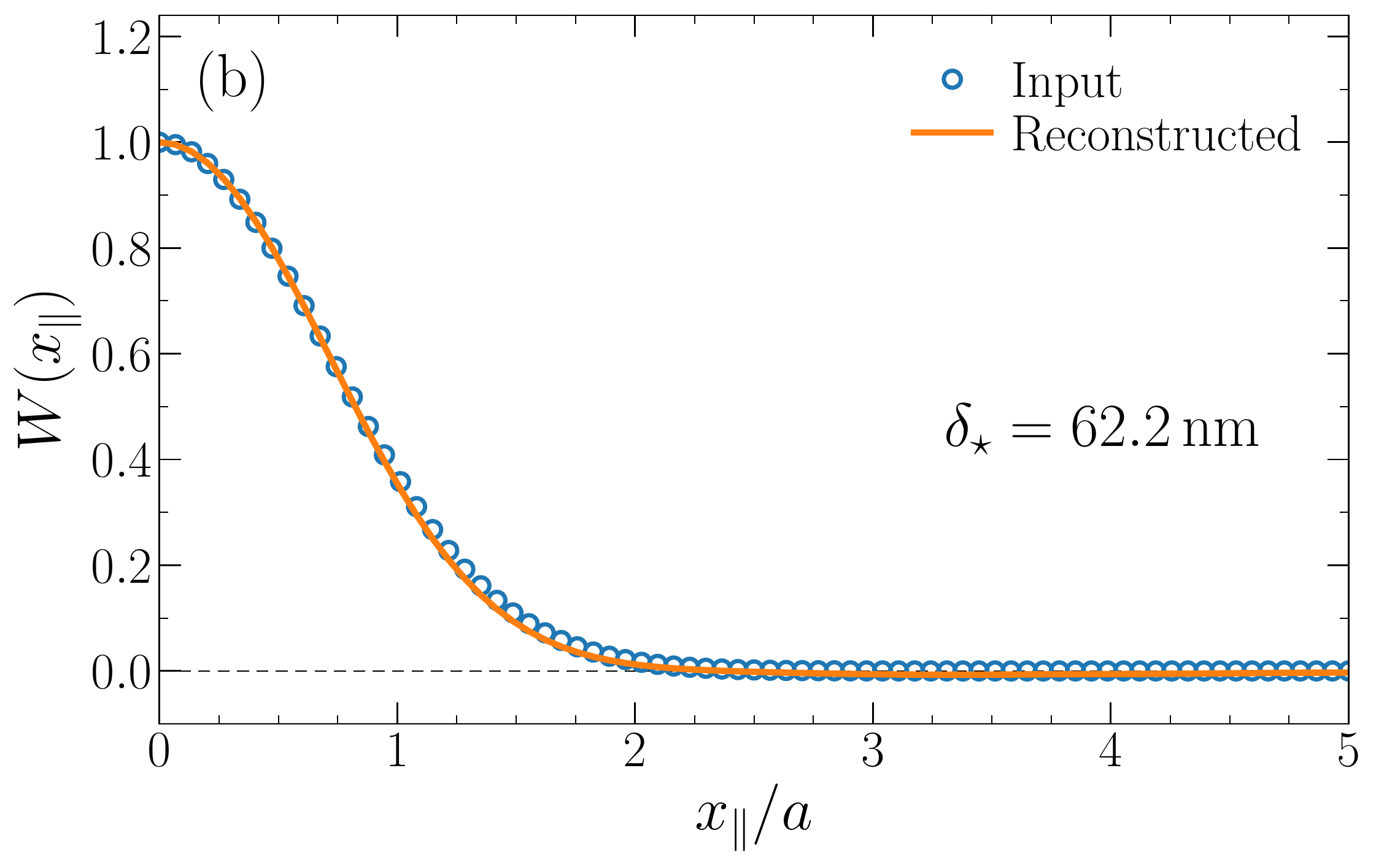}
  \\*[1ex]   
  %
  %
  \includegraphics[width=0.475\columnwidth]{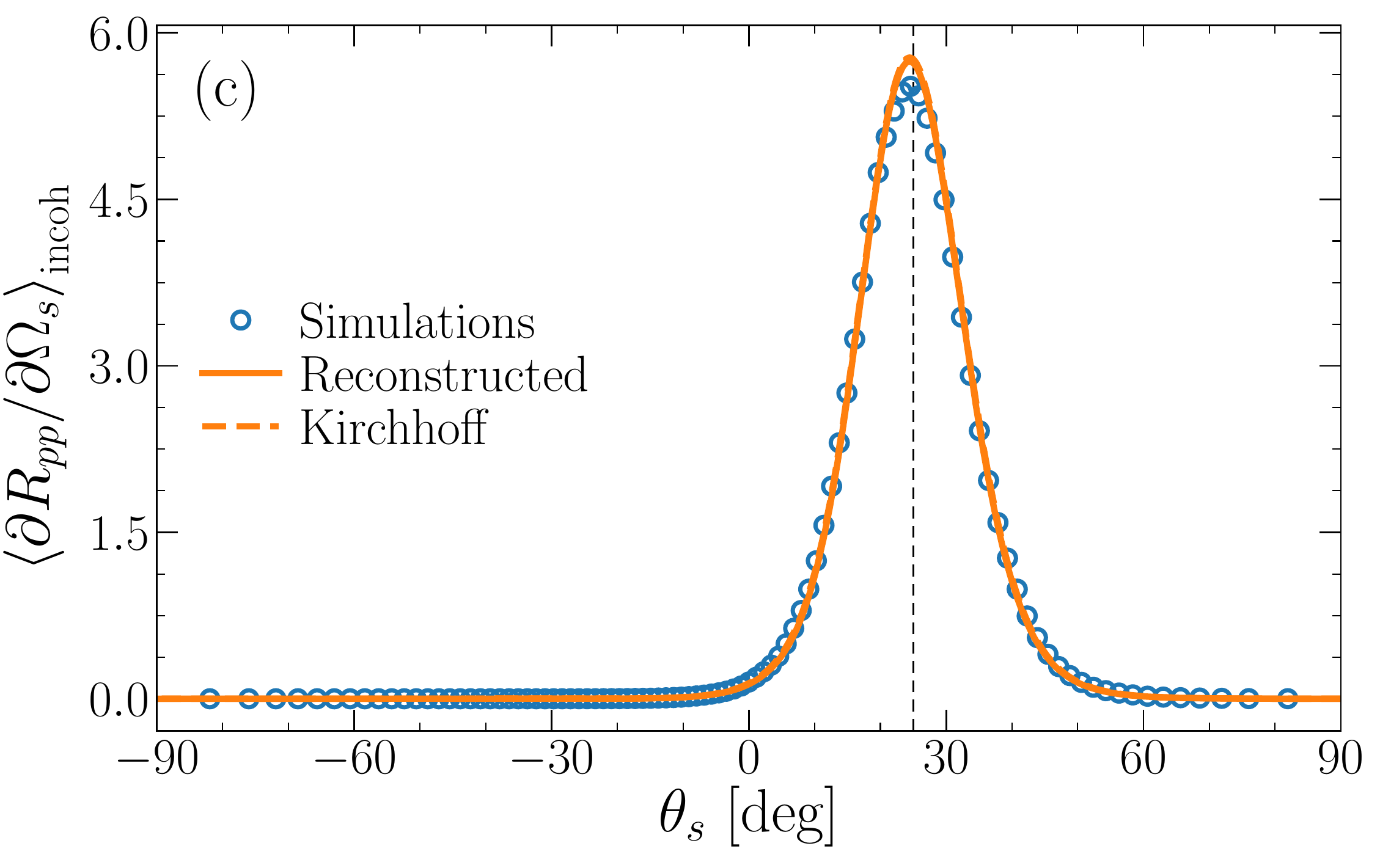}
  \quad
  \includegraphics[width=0.475\columnwidth]{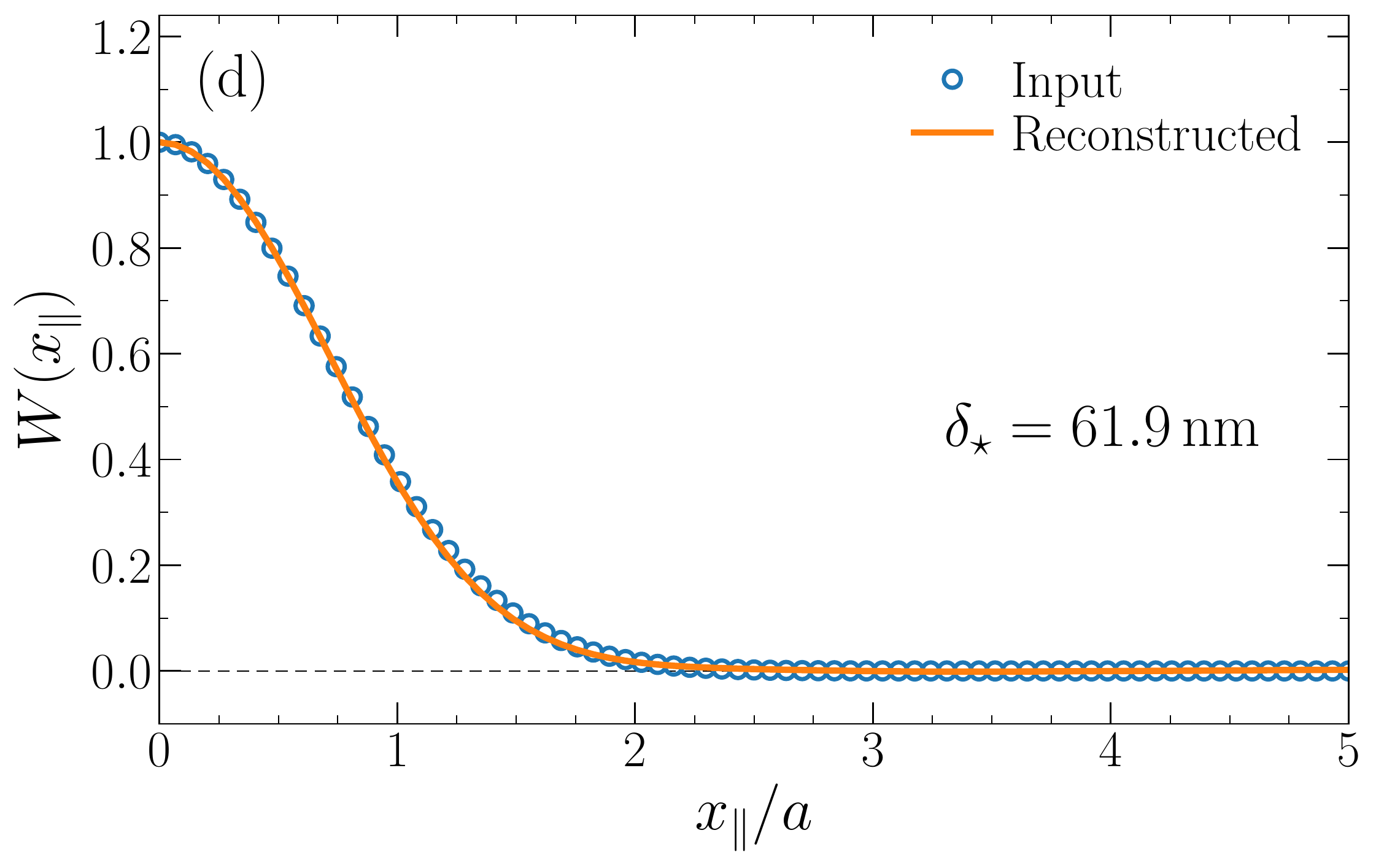} 
  \\*[1ex]   
  %
  \includegraphics[width=0.475\columnwidth]{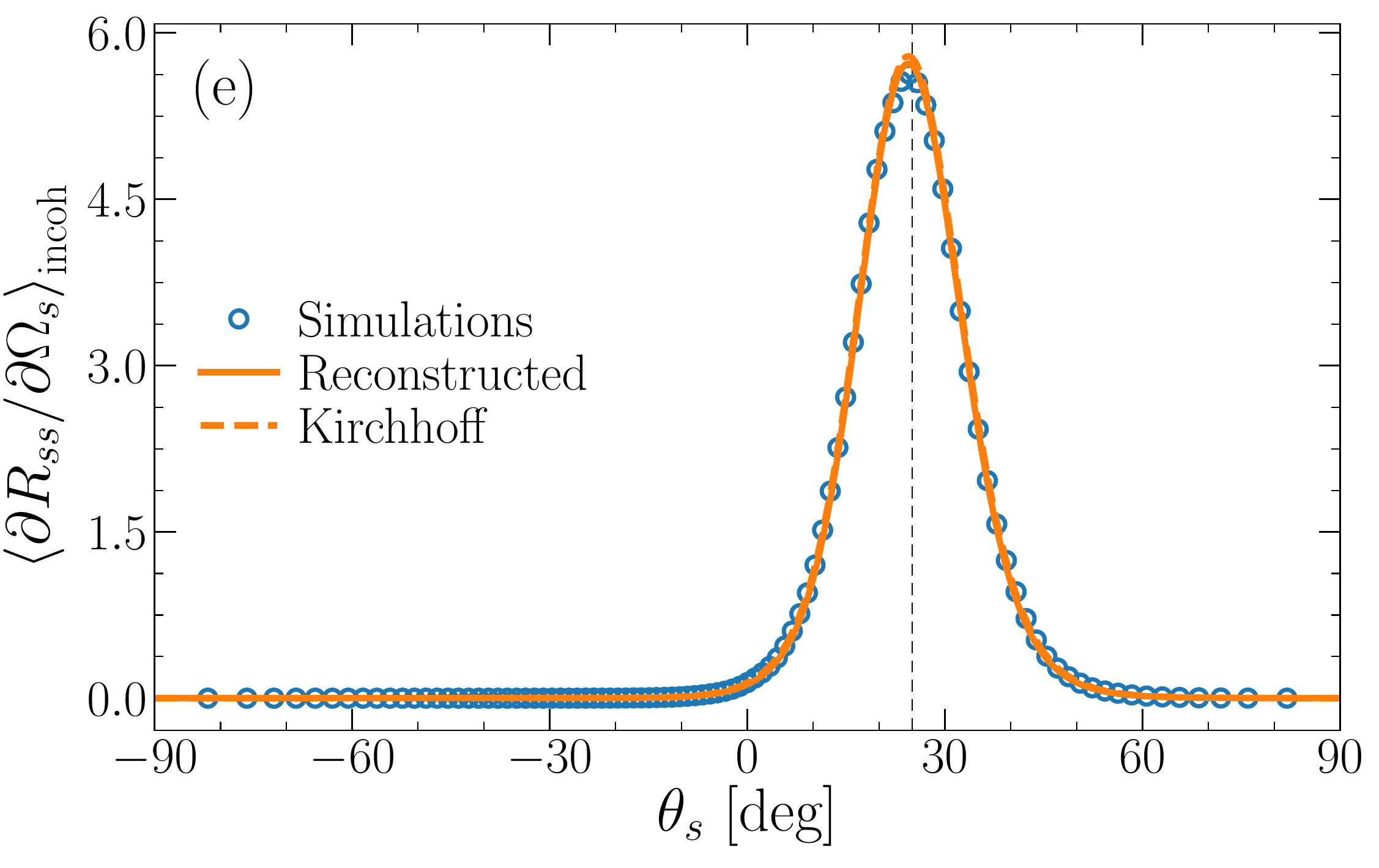}
  \quad
  \includegraphics[width=0.475\columnwidth]{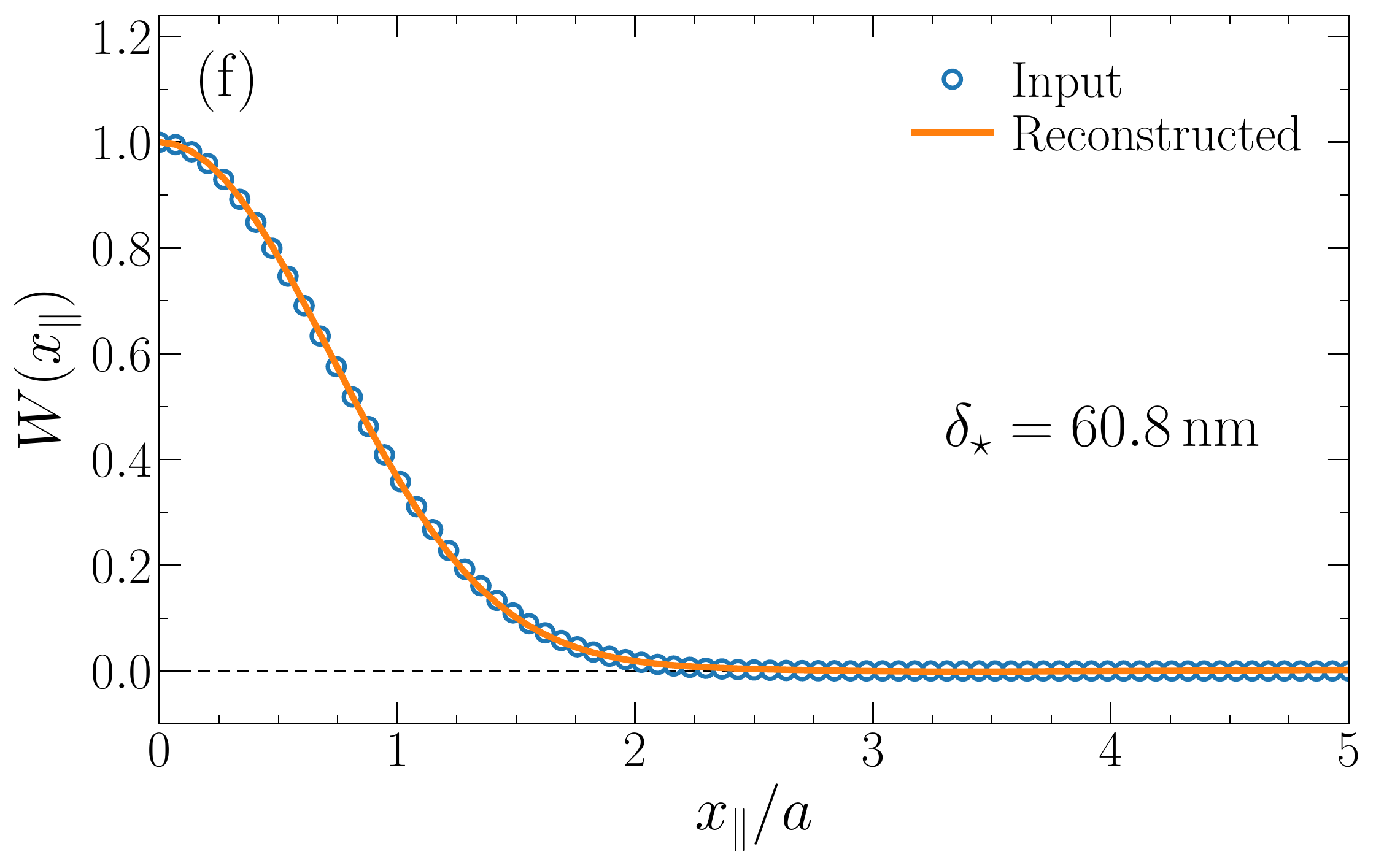} 
  \caption{Reconstruction of the rms-roughness $\delta_\star$ and the correlation function $W(x_\parallel)$ for a rough \textit{vacuum-silver} surface from several sets of in-plane and co-polarized scattering data. The wavelength of the incident light was $\lambda=\SI{632.8}{nm}$ for which value the dielectric constant of silver is $\varepsilon=-18.28+\mathrm{i}\,0.48$~\cite{Johnson1972}. The rough surface is Gaussianly correlated and characterized by $a=2\lambda$ and $\delta=\lambda/7=\SI{90.4}{nm}$. For these parameters, the incoherent component of the mean~DRCs were calculated on the basis of \num{2500} surface realizations by a rigorous (surface integral) computer simulation approach~(see main text for details). Such results are presented as open circles in the left-most panels (the ``input data sets''); they correspond to (a)~the polar angle of incidence $\theta_0=\ang{0}$ and p-polarization; (c) $\theta_0=\ang{26}$ and p-polarization; and (e) $\theta_0=\ang{25}$ and s-polarization and the vertical thin dashed lines indicate the polar angle of incidence.  When the non-iterative reconstruction approach~\eqref{eq:statistical-property-estimation-simplified} is applied to the computer generated scattering data, one obtains the surface roughness $\delta_\star$ and reconstructed correlation function  $W(x_\parallel)$ presented as solid lines in the right column of panels; to facilitate comparison, the form of the correlation function assumed in producing the input data is presented as open circles in the same panels. Moreover, the values for the rms-roughness obtained during the reconstructions are $\delta_\star=\SI{62.2}{nm}$, $\SI{61.9}{nm}$ and $\SI{60.8}{nm}$ from top-to-bottom, respectively. The solid lines in the left column of panels were obtained from Eq.~\eqref{eq:MDRC-approx} when the reconstructed rough surface properties were assumed. Similarly, the dashed lines in the same panels were obtained in the same manner when the ``input'' properties of the rough surface were assumed.}    
\label{fig:Simulations-Silver-R006}
\end{figure}
%
%


\begin{figure}[!tbh]
  \centering
  \includegraphics[width=0.475\columnwidth]{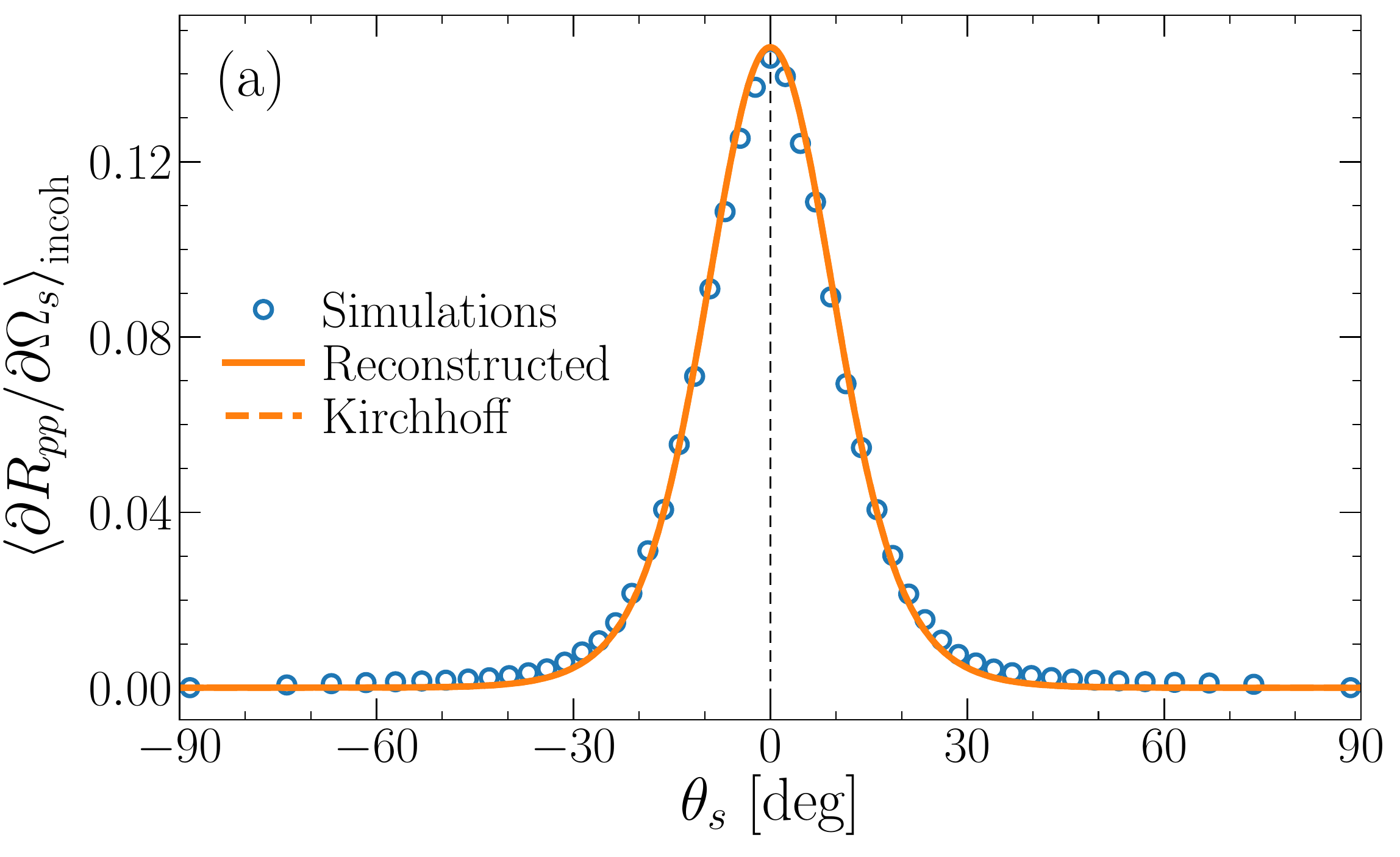}
  \qquad
  \includegraphics[width=0.475\columnwidth]{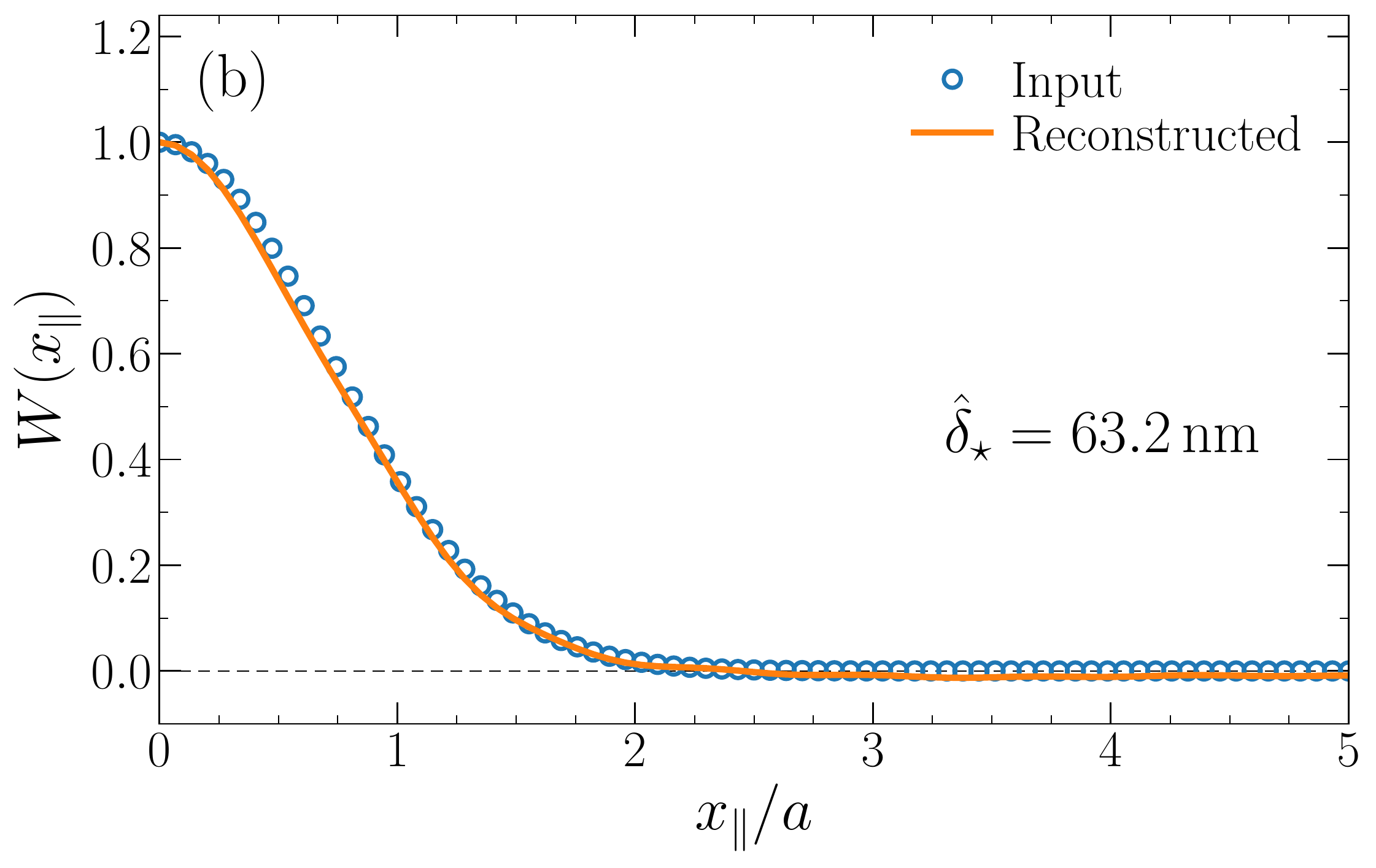}
  \\*[1ex]
  %
    \includegraphics[width=0.475\columnwidth]{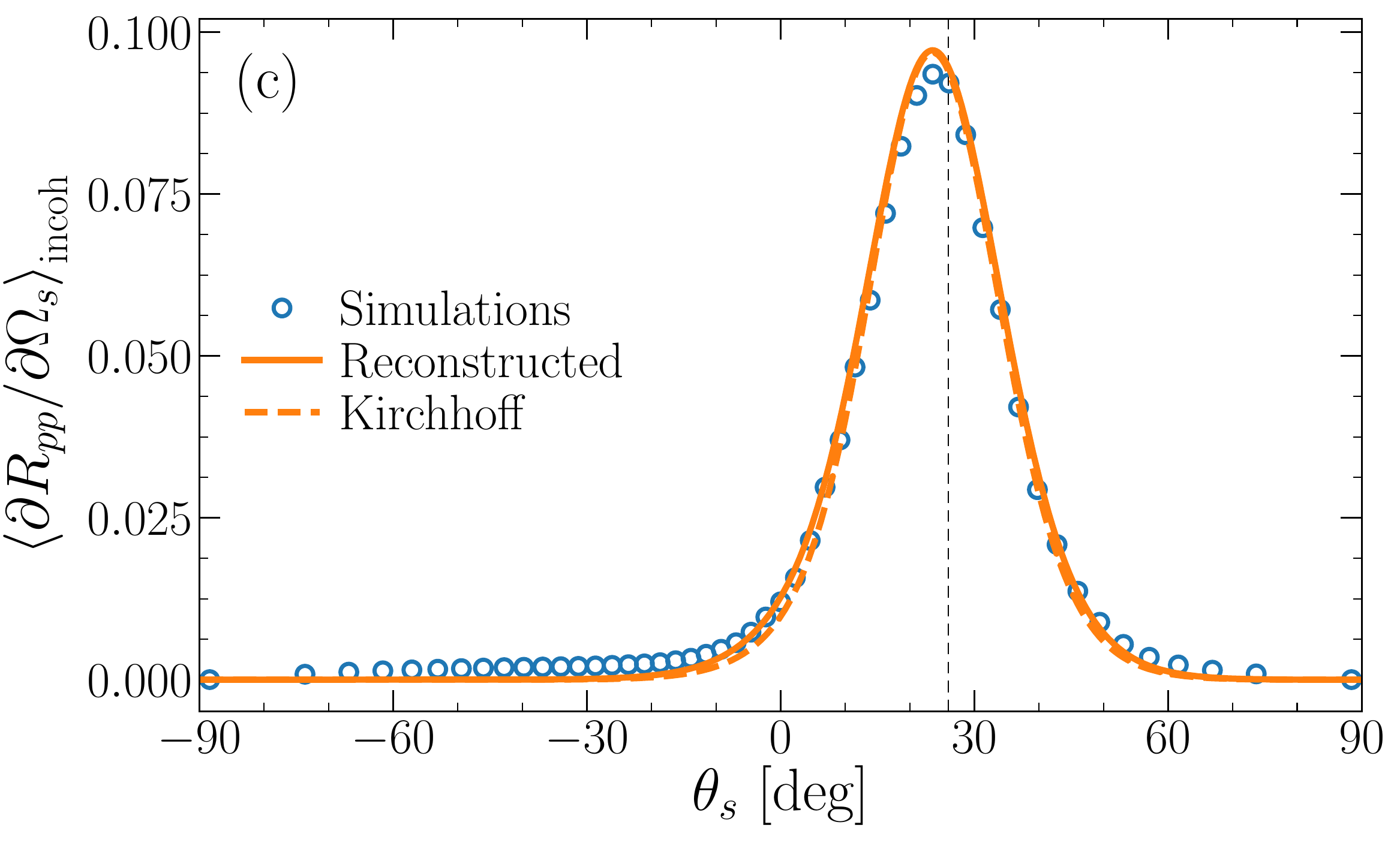}
  \qquad
  \includegraphics[width=0.475\columnwidth]{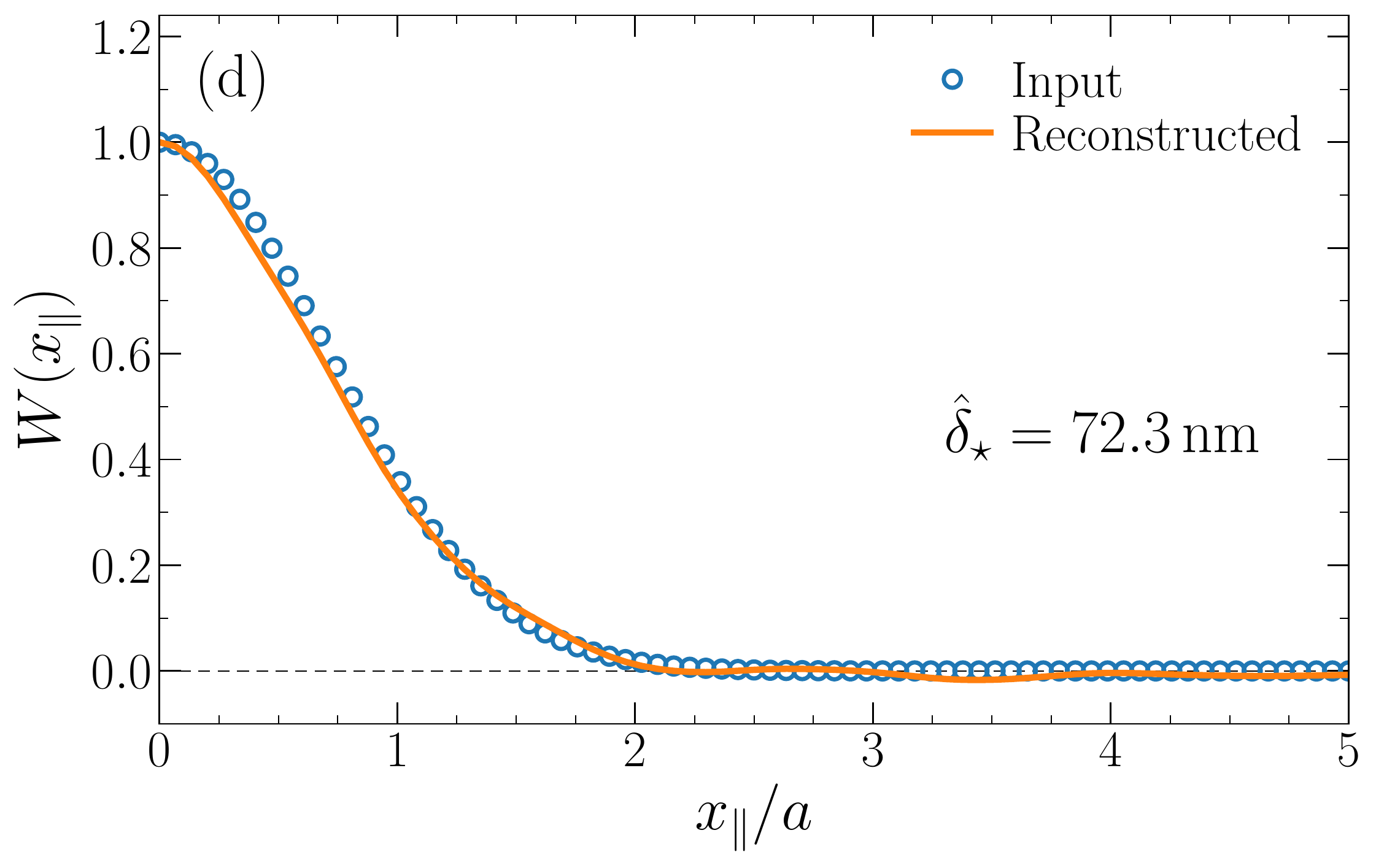}
  \\*[1ex]   
  %
    \includegraphics[width=0.475\columnwidth]{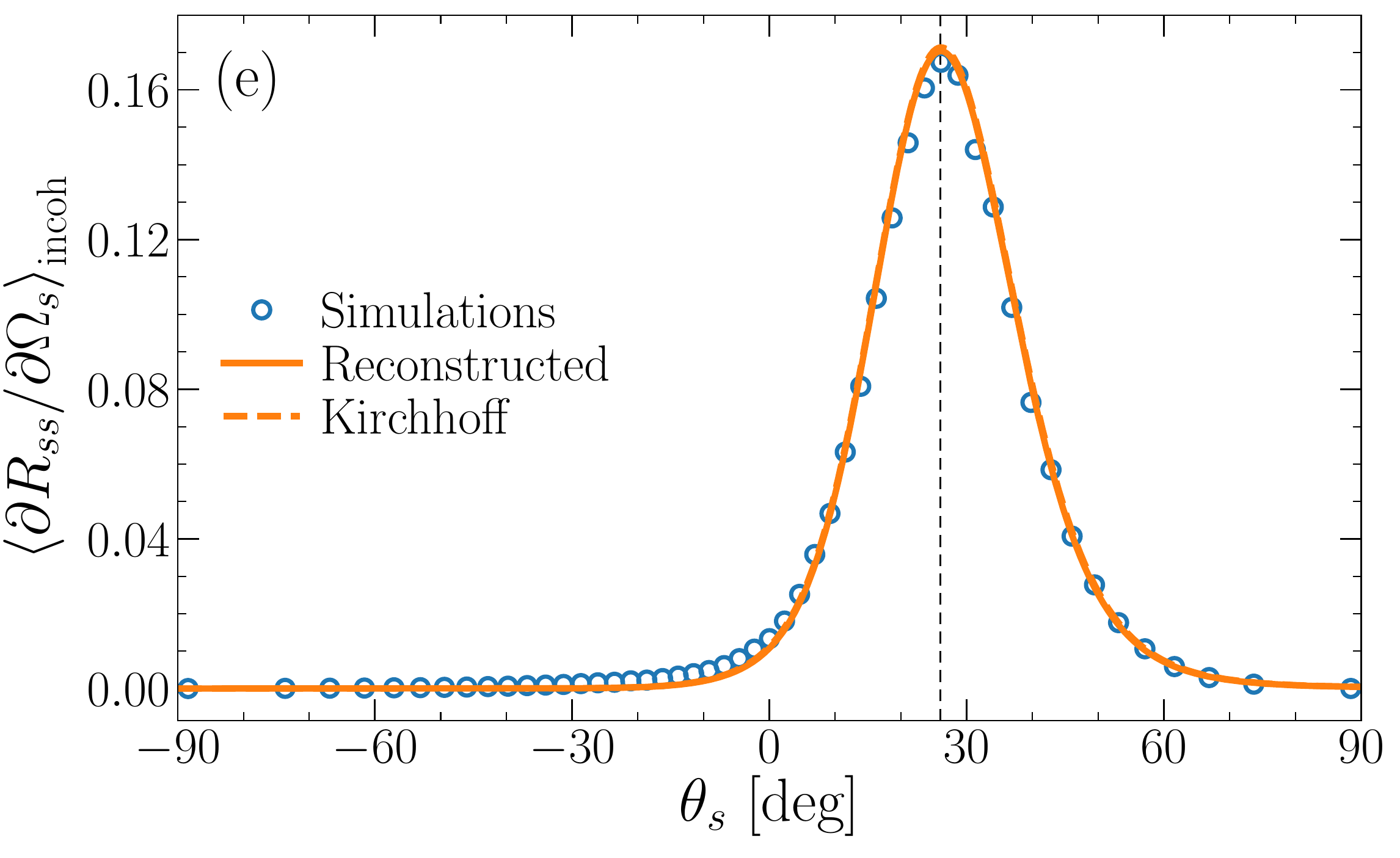}
  \qquad
  \includegraphics[width=0.475\columnwidth]{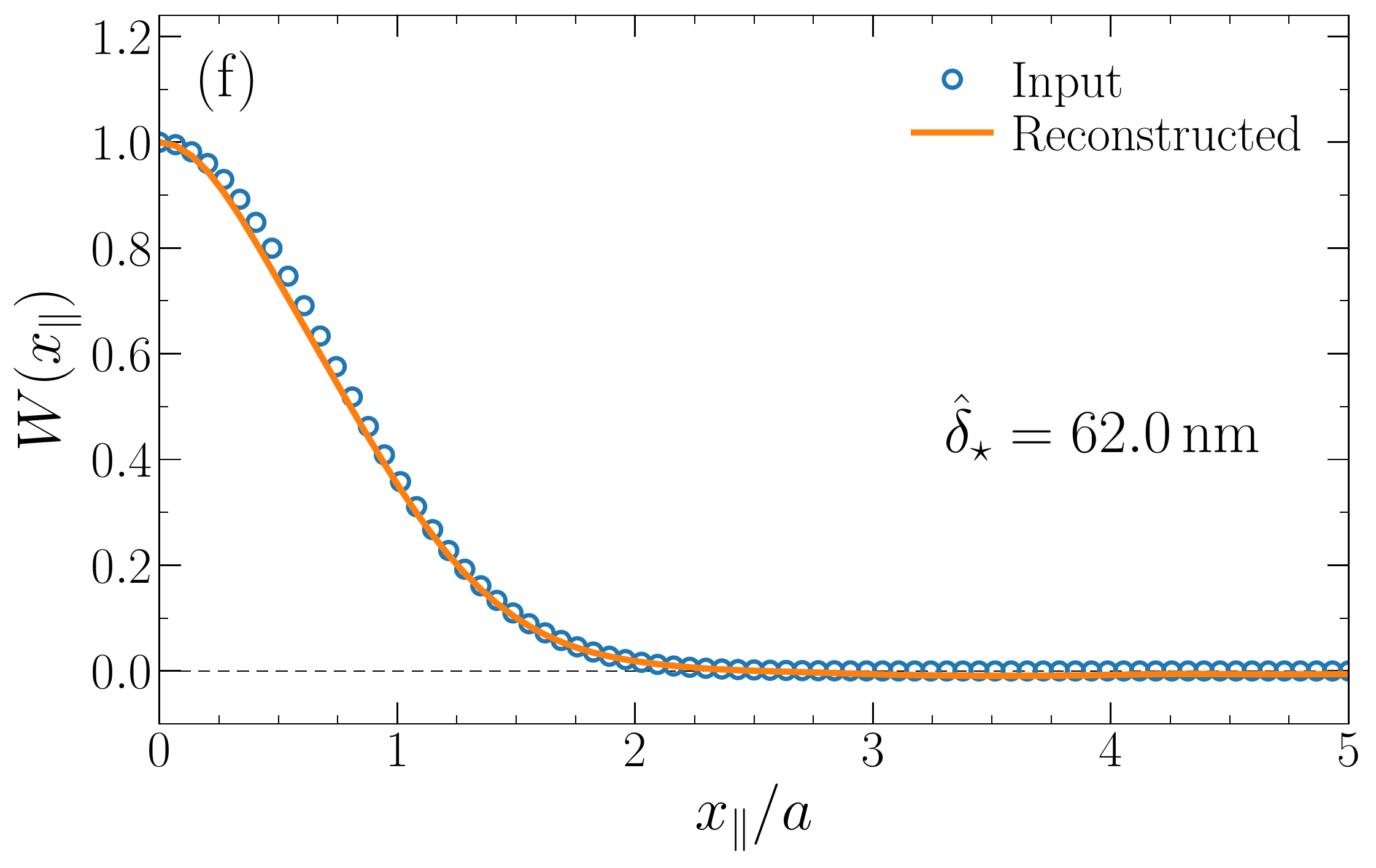}

  \caption{Reconstruction of the rms-roughness $\hat{\delta}_\star$ and the correlation function $W(x_\parallel)$  from the in-plane angular dependence of mean DRCs curves produced by a non-perturbative numerical solution of the reduced Rayleigh equations. The rough \textit{vacuum-glass} surface is illuminated by linearly polarized light of wavelength $\lambda= \SI{0.6328}{\micro\meter}$ [$\varepsilon=2.25$], and the polar angles of incidence were (a,b) $\theta=\ang{0}$ or (c--f)~$\theta=\ang{26}$. The rough surface is characterized by a Gaussian correlation function $W(x_\parallel)$, and the parameters  $\delta=\lambda/10\approx\SI{63.3}{\nano\meter}$ and $a=1.5\lambda= \SI{0.9492}{\micro\meter}$. The reconstruction was performed by the iterative approach from Eq.~\eqref{eq:statistical-property-estimation} and it produced the surface roughness values $\hat{\delta}_\star=\SI{63.2}{nm}$, $\SI{72.3}{nm}$, and $\SI{62.0}{nm}$~(top-to-bottom). The remaining description of this figure is identical to  that of Fig.~\protect\ref{fig:Simulations-Silver-R006}.
  }
  \label{fig:Simulations-Glass}  
\end{figure}

After having established that the proposed reconstruction approach can produce reliable and self-consistent results when applied to (approximate) scattering data obtained for certain classes of rough surfaces, we now turn to the reconstruction of computer generated scattering data that include both single and multiple scattering effects. Such input scattering data are obtained on the basis of several different \textit{rigorous} computer simulation approaches~\cite{Simonsen2009-9,Simonsen2009-1,Simonsen2011-05,Simonsen2004-3} when the roughness parameters were predefined. In this way, we obtain in-plane scattering data for p- and s-polarized incident light that is scattered from either metallic~(silver) or dielectric~(glass) randomly rough surfaces.

\smallskip
The first set of computer simulation generated scattering data were obtained for a two-dimensional randomly rough silver surface illuminated from the vacuum by p- or s-polarized light of wavelength $\lambda= \SI{0.6328}{\micro\meter}$, characterized by an rms-roughness $\delta=\lambda/10\approx\SI{63.3}{\nano\meter}$ and a Gaussian correlation function~\eqref{eq:gaussW} of correlation length  $a=2\lambda= \SI{1.26}{\micro\meter}$. The input scattering data are presented as open symbols in Figs.~\ref{fig:Simulations-Silver-R006}(a), \ref{fig:Simulations-Silver-R006}(c) and ~\ref{fig:Simulations-Silver-R006}(e), correspond to p-to-p scattering for a polar angle of incidence $\theta_0=\ang{0}$, and to p-to-p and s-to-s scattering, both for a polar angle of incidence $\theta_0=\ang{25}$, respectively. These results were obtained by a rigorous computer simulation approach based on the surface integral (or boundary element) equations in the M\"uller form~\cite{Simonsen2009-9}, which can be derived from the Stratton-Chu equations~\cite[p. 674]{Book:Kong2005}. For each surface realization, this approach consists of numerically solving a coupled set of four inhomogeneous integral equations for the (four) unknown tangential components of the total electric and magnetic fields on the surface~\cite{Simonsen2009-9}. The solution of this system of equations predicts the total electromagnetic fields on the surface from which the electric and magnetic fields everywhere in the scattering geometry, that is both above and below the rough surface, can be calculated by the Franz formulas~\cite[pp.674--675]{Book:Kong2005} and used to obtain the reflection amplitudes from which the DRCs can be calculated~\cite{Simonsen2009-9,Simonsen2009-1}. The simulation results for the in-plane angular dependence of the mean~DRC presented in Fig.~\ref{fig:Simulations-Silver-R006} were obtained by averaging the results of \num{2500} surface realizations. In passing, we note that these simulation results are rigorous and do, for instance, include the potential excitation and de-excitation of surface waves like surface plasmon polaritons~(SPPs), which can influence strongly the reflection from rough metal surface~\cite{Simonsen2004-3}.  

Based on the computer generated in-plane p-to-p scattering data for normal incidence presented in Fig.~\ref{fig:Simulations-Silver-R006}(a), the non-iterative reconstruction approach was performed on the basis of Eq.~\eqref{eq:statistical-property-estimation-simplified} to produce the rms-roughness $\delta_\star=\SI{62.2}{nm}$ and the reconstructed correlation function presented as a solid line in Fig.~\ref{fig:Simulations-Silver-R006}(b). When the results obtained during the  reconstruction are compared to the (input) surface roughness $\delta \approx\SI{63.3}{nm}$ and the correlation function assumed in generating the input scattering data~[open symbols in Fig.~\ref{fig:Simulations-Silver-R006}(b)], rather good agreement is found between these two sets of data. Hence, we conclude that the reconstruction is producing reliable results, at least, this is the case for the scattering system and roughness parameters that we assumed.  Moreover, when the reconstructed correlation function and surface roughness value $\delta_\star$ are used to calculate the mean DRC from Eq.~\eqref{eq:MDRC-approx}, one obtains the solid line in Fig.~\ref{fig:Simulations-Silver-R006}(a) which agrees quite well (as expected) with the computer generated input data presented as open symbols in the same figure. In performing this calculation, the reconstructed correlation function needs to be known for any spatial argument $x_\parallel$, while during the reconstruction, it is only calculated at a finite set of points. To this end, an interpolation procedure is applied to the set of points defining the reconstructed correlation function, while this function is assumed to vanish for arguments larger than those provided during the reconstruction. In this way, the solid line in Fig.~\ref{fig:Simulations-Silver-R006}(a) was obtained. For reasons of comparison, we in Fig.~\ref{fig:Simulations-Silver-R006}(a) also present the mean DRC~\eqref{eq:MDRC-approx} as a dashed line for the surface roughness and surface height correlation function $W(x_\parallel)$ assumed in producing the input scattering data; here the predictions from Eq.~\eqref{eq:MDRC-approx} using input or reconstructed surface statistics are so similar that the resulting two data sets are hard to distinguish. If instead the reconstruction was performed on the basis of normally incident s-to-s scattering data obtained for the same scattering system, the results that we obtain during the reconstruction (results not shown) will only show minor differences relative to the corresponding p-to-p reconstruction results. The main difference between these two sets of reconstruction results is that the reconstructed surface roughness (for s-polarized light) now is  $\delta_\star=\SI{60.5}{nm}$ which is slightly less than what is obtained when basing the reconstruction on p-polarized scattering data. The percentage relative errors of the reconstructed rms-roughness values relative to the input value are $\SI{1.8}{\percent}$ and $\SI{4.4}{\percent}$ for p- and s-polarized normally incident light, respectively.

%
%
For the same scattering system but a polar angle of incidence $\theta_0=\ang{25}$,  Figs.~\ref{fig:Simulations-Silver-R006}(c)--(d)  and Figs.~\ref{fig:Simulations-Silver-R006}(e)--(f) present the input data (as open symbols) for p-to-p and s-to-s in-plane scattering, and the results that can be obtained when reconstructions are based on them, respectively;  all the remaining parameters are identical to those assumed in obtaining the results in Figs.~\ref{fig:Simulations-Silver-R006}(a)--(b). The results in Figs.~\ref{fig:Simulations-Silver-R006}(c)--(f) demonstrate that also for non-normally incident p- and s-polarized light, rather good agreement between the input and reconstructed roughness parameters can be achieved. In particular, this applies to the surface height correlation functions which are accurately reconstructed for both linear polarizations of the incident light. The values obtained for the surface roughness during the reconstruction were $\delta_\star=\SI{61.9}{nm}$ and $\delta_\star=\SI{60.8}{nm}$ for p- and s-polarized light, respectively. These values correspond to a percentage relative error of $\SI{2.2}{\percent}$ and $\SI{4.0}{\percent}$, respectively.

It should be remarked that for the input scattering data presented in Fig.~\ref{fig:Simulations-Silver-R006}, only marginal changes are found in the results of the reconstruction by basing it on the iterative approach~\eqref{eq:statistical-property-estimation}, as compared to using the non-iterative approach~\eqref{eq:statistical-property-estimation-simplified} to produce the results presented in this figure. It should also be mentioned that for the numerous reconstruction tests that we have performed based on scattering data produced by computer simulations, we have found that the accuracy in the reconstruction of the correlation functions typically is the same, or better, than the accuracy in the reconstruction of the surface roughness.  

We have also performed reconstruction of scattering data for metallic scattering systems that are similar to the one assumed in obtaining the results of Fig.~\ref{fig:Simulations-Silver-R006}, but characterized by longer correlation lengths (with the remaining parameters being unchanged). Doing so generally leads to reconstruction results that agree even better with the surface statistics assumed to produce the input scattering data. For instance, the reconstruction of scattering data that are similar to those in Fig.~\ref{fig:Simulations-Silver-R006} but obtained by computer simulations under the assumption that $a=3\lambda$, resulted in reconstructed surface roughness of $\delta_\star=\SI{62.7}{nm}$ and $\delta_\star=\SI{62.6}{nm}$ when $\theta_0=\ang{0}$ and $\theta_0=\ang{25}$, respectively. Both of these values are quite close to the surface roughness $\delta\approx\SI{63.3}{nm}$ assumed in generating the input scattering data.

\medskip
We now turn to the reconstruction of surface statistics based on scattering data collected during the reflection of light from rough \emph{dielectric} surfaces. To this end, we assume that linearly polarized light of wavelength $\lambda=\SI{0.6328}{\micro\meter}$ is incident from the vacuum onto a glass substrate~[$\varepsilon=2.25$] bounded by a randomly rough surface that is characterized by surface roughness $\delta=\lambda/10\approx\SI{63.3}{\nano\meter}$, and a Gaussian correlation function of correlation length $a=1.5\lambda= \SI{0.9492}{\micro\meter}$. For this scattering system, the input scattering data are presented as open symbols in the left-side panels of Fig.~\ref{fig:Simulations-Glass} and they correspond to $\theta_0=\ang{0}$ and p-polarization~[Fig.~\ref{fig:Simulations-Glass}(a)]; to $\theta_0=\ang{26}$ and p-polarization~[Fig.~\ref{fig:Simulations-Glass}(c)]; and finally to $\theta_0=\ang{26}$ and s-polarization~[Fig.~\ref{fig:Simulations-Glass}(e)]. These results were obtained by a purely numerical non-perturbative solution of the reduced Rayleigh equation, which is a single inhomogeneous integral equation satisfied for the scattering amplitudes~\cite{Simonsen2011-05,Simonsen2004-3}. The simulation results for the in-plane angular dependence of the co-polarized mean~DRCs, presented in the left-side panels of Fig.~\ref{fig:Simulations-Glass} as open symbols, were obtained by averaging the results of \num{2500} surface realizations.

For each of these input scattering data sets, the iterative reconstruction approach based on Eq.~\eqref{eq:statistical-property-estimation} was performed. In this way, we obtained the reconstructed correlation functions presented as solid lines in the right-side panels of Fig.~\ref{fig:Simulations-Glass}, and they all show good quantitative agreement with the input correlation functions assumed in producing the input scattering data. The surface roughness values produced by the iterative reconstruction procedure are $\hat{\delta}_\star=\SI{63.2}{nm}$~($\theta_0=\ang{0}$ and p-polarization); $\hat{\delta}_\star=\SI{72.3}{nm}$~($\theta_0=\ang{26}$ and p-polarization); and finally $\hat{\delta}_\star=\SI{62.0}{nm}$~($\theta_0=\ang{26}$ and s-polarization) that correspond to relative errors of $\SI{0.2}{\percent}$; $\SI{14.2}{\percent}$ and $\SI{2.1}{\percent}$, respectively. Even with the correlation functions being reconstructed accurately for all of the input data sets that we assumed, the accuracy of the reconstructed surface roughness values  varies much more between the data sets that we considered. In the case of the dielectric surface, the most pronounced discrepancy between input and reconstructed surface roughness values was found for p-polarization and $\theta=\ang{26}$, while in the case of the metal surface [Fig.~\ref{fig:Simulations-Silver-R006}], the same was found for the s-polarization and $\theta=\ang{25}$, but the relative error was significantly less in this latter case. The reason for this difference between the metallic and dielectric cases we are not sure of, but we speculate that it might be related to the validity of the Kirchhoff approximation on which our reconstruction approach is based, which is more appropriate for the metallic than the dielectric scattering system. 

%
%

%
\begin{figure}[!tbh]
  \centering
  \includegraphics[width=0.475\columnwidth,height=0.275\columnwidth]{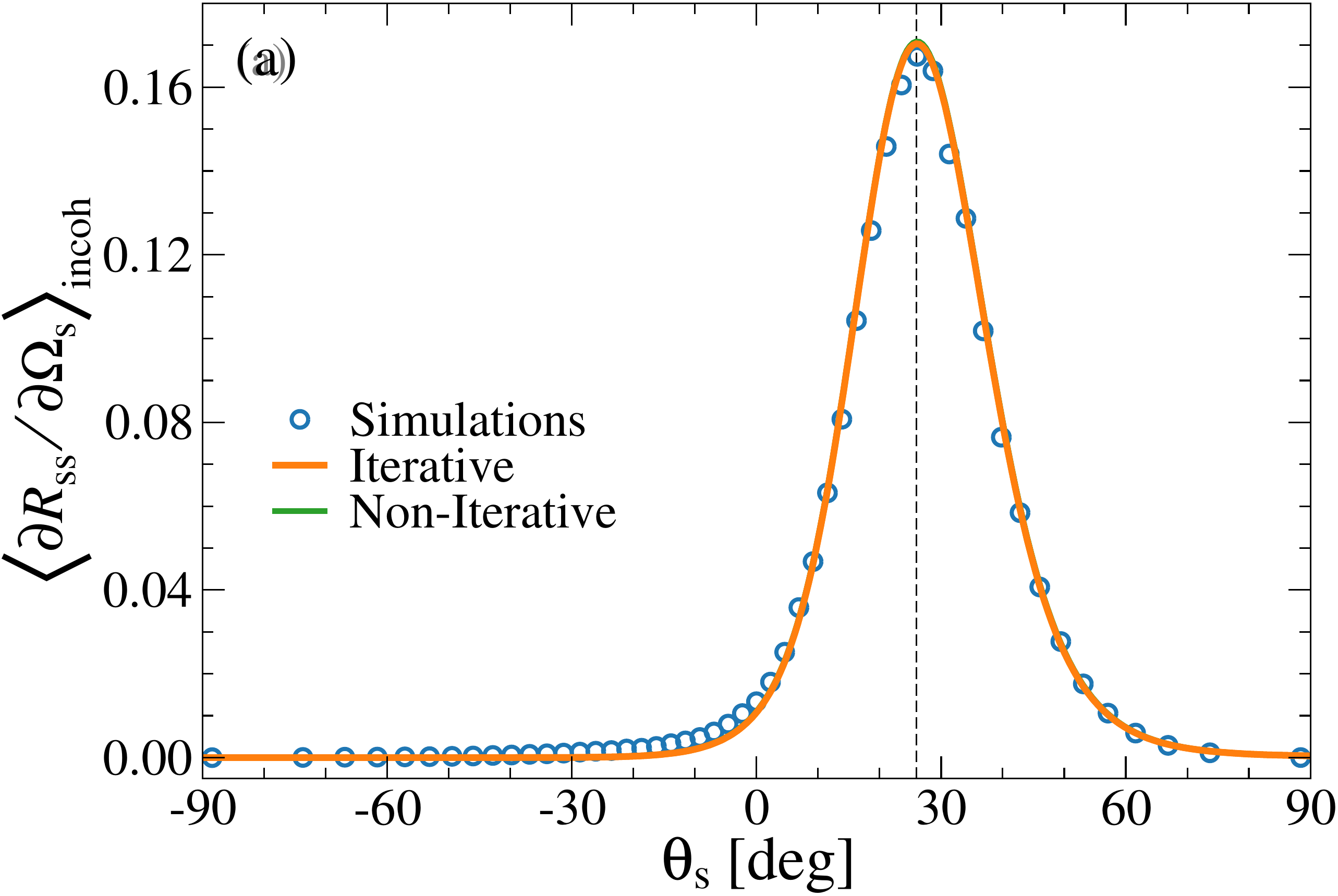}
  \qquad
  \includegraphics[width=0.475\columnwidth,height=0.275\columnwidth]{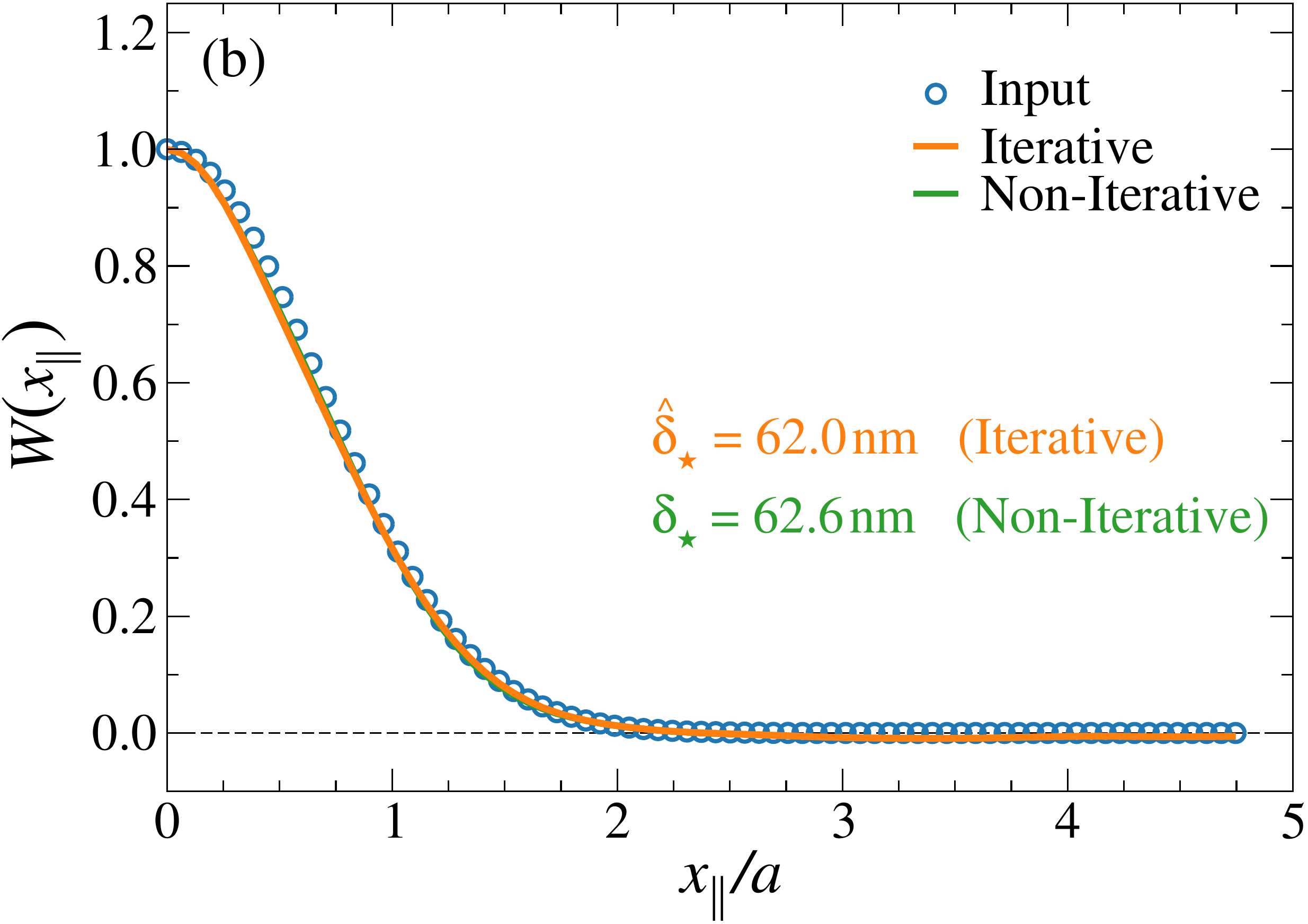}  
  \\*[1ex]   
  %
  \includegraphics[width=0.475\columnwidth,height=0.275\columnwidth]{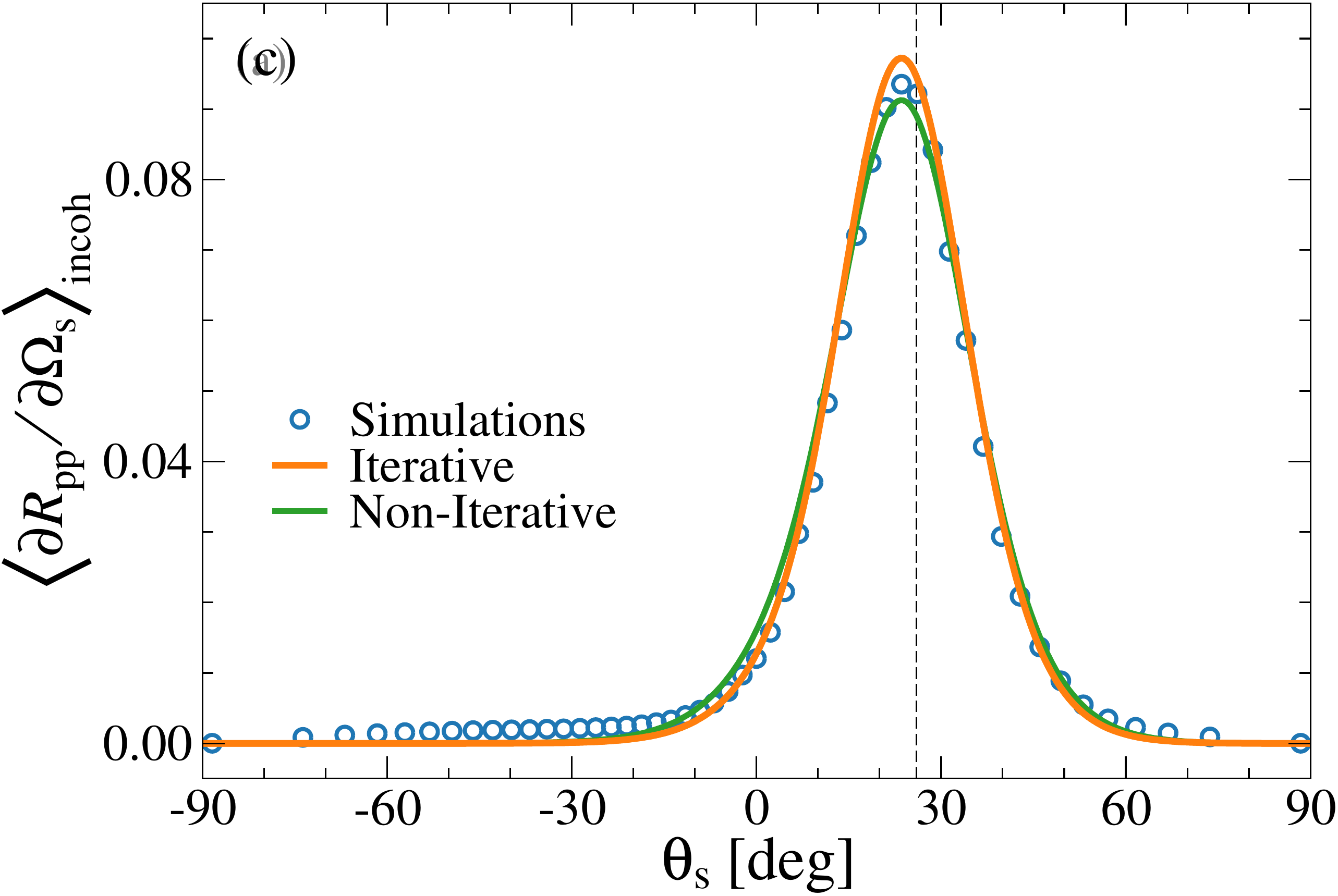}
  \qquad
  \includegraphics[width=0.475\columnwidth,height=0.275\columnwidth]{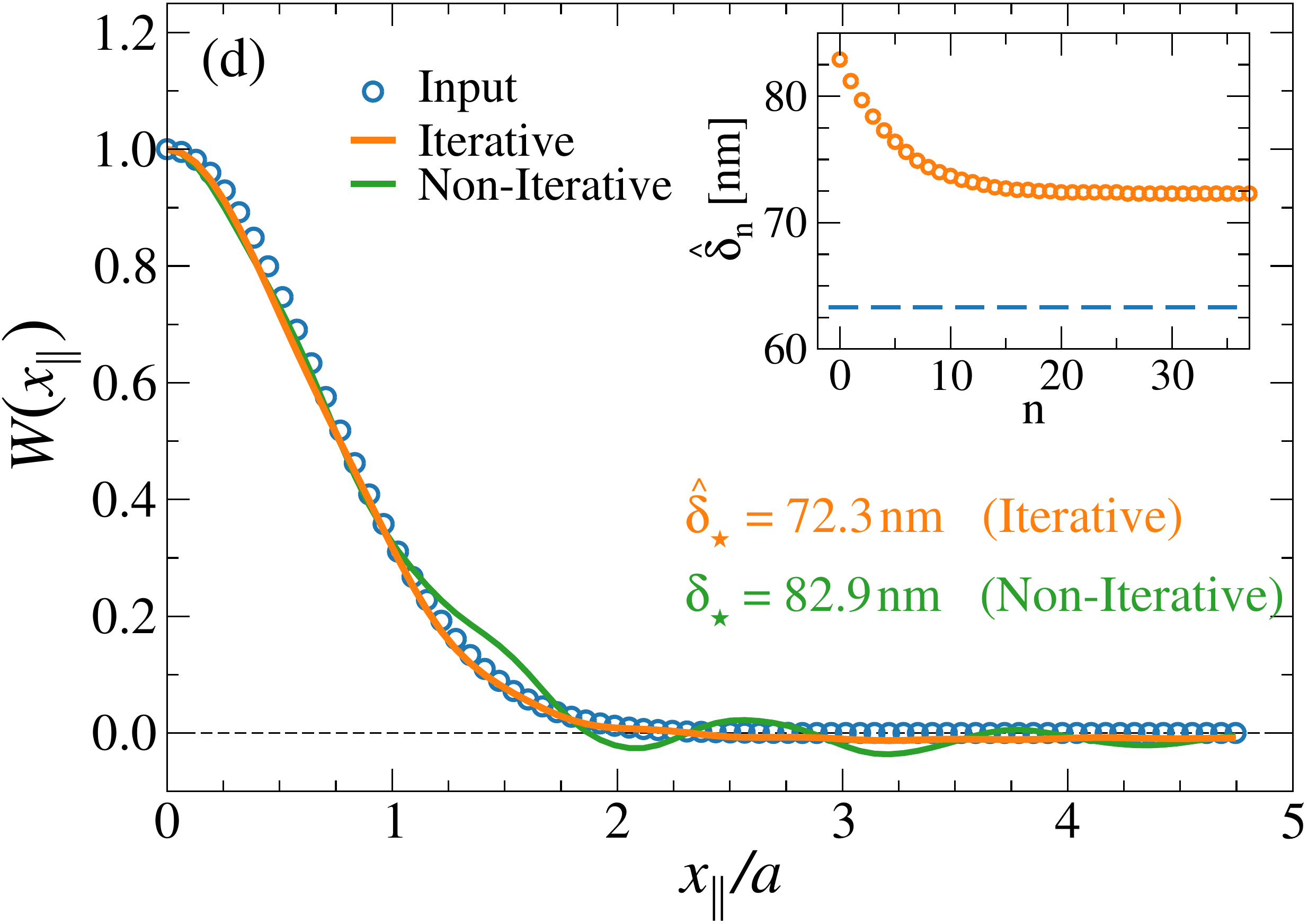}
  \caption{The same as Figs.~\ref{fig:Simulations-Glass}(c)--(f) but now presenting \textit{both} the iterative and non-iterative reconstruction results when applied to the same input scattering data. Note that the order of the panels for p- and s-polarized light have been interchanged relative to the order used in Figs.~\ref{fig:Simulations-Glass}(c)--(f). The inset in Fig.~\ref{fig:Iterative_vs_NonIterative}(d) presents for p-polarized light the reconstructed surface roughness $\hat{\delta}_n$ at iteration level $n$ and how it converges  from the initial value $\delta_\star=\SI{82.9}{nm}$ (non-iterative result) towards $\hat{\delta}_\star=\SI{72.3}{nm}$~(iterative result) after about \num{25} iterations.}
  \label{fig:Iterative_vs_NonIterative}
\end{figure}
%
%

\smallskip
The reconstruction for the silver scattering system~[Fig.~\ref{fig:Simulations-Silver-R006}] was performed on the basis of the \textit{non-iterative} approach, while the reconstruction that we did for the glass scattering system~[Fig.~\ref{fig:Simulations-Glass}] used the \textit{iterative} approach. Furthermore, for the silver scattering system discussed previously, we argued that only marginal differences were obtained by the use of non-iterative or iterative reconstruction. We will now demonstrate that under certain conditions, this may not be the case for the glass scattering system assumed in obtaining the results presented in Fig.~\ref{fig:Simulations-Glass}. Figure~\ref{fig:Iterative_vs_NonIterative} compares results for the glass scattering system for s- and p-polarization with a polar angle of incidence of $\theta_0=\ang{26}$, when the reconstruction is performed on the basis of either the iterative or non-iterative reconstruction approaches, that is, on the basis of the expressions in Eqs.~\eqref{eq:statistical-property-estimation} or \eqref{eq:statistical-property-estimation-simplified}. From the results presented in  this figure, several interesting observations should be made. First, when the incident light is s-polarized, the correlation functions obtained by reconstruction with the non-iterative and iterative approaches are essentially the same; at least, on the scale of Fig.~\ref{fig:Iterative_vs_NonIterative}(b) it is hard to distinguish between these two sets of results. Moreover, the reconstructed surface roughness values obtained this way are rather similar, and the numerical values are $\delta_\star=\SI{62.6}{nm}$~(non-iterative) and $\hat{\delta}_\star=\SI{62.0}{nm}$~(iterative). Interestingly, it is the value obtained by the non-iterative approach that is the most accurate of the two reconstructed roughness values. With so similar results obtained by the two reconstruction approaches it comes as little surprise that also the in-plane mean DRC curves that are calculated from them on the basis of Eq.~\eqref{eq:MDRC-approx} are rather similar~[Fig.~\ref{fig:Iterative_vs_NonIterative}(a)].

We now turn to what happens when the incident light is p-polarized. Figures~\ref{fig:Iterative_vs_NonIterative}(c)--(d) present the results obtained by the two reconstruction approaches, and the apparent differences between the results are readily observed. The reconstructed correlation function obtained by the non-iterative approach displays some spurious oscillations~[Fig.~\ref{fig:Iterative_vs_NonIterative}(d); green solid line] that are not present in the input correlation function that decays monotonously towards zero for sufficiently large arguments. On the other hand, when the correlation function is obtained on the basis of the iterative reconstruction approach, it represents a good approximation to the input correlation function; only in the tail of the correlation function can one observe some rather weak oscillations which are almost difficult to notice. The surface roughness values obtained by reconstruction for p-polarized light are $\delta_\star=\SI{82.9}{nm}$ and $\hat{\delta}_\star=\SI{72.3}{nm}$, and of the two, it is the value obtained by the iterative approach that is closest to the input value (even if it is not particularly accurate). The inset to Fig.~\ref{fig:Iterative_vs_NonIterative}(d) presents the convergence of the iterative approach for the rms-roughness of the surface, $\hat{\delta}_n$, as a function of the iteration level $n$. When the results for the reconstructed surface statistics are used to calculate the mean DRC, the two solid lines in  Fig.~\ref{fig:Iterative_vs_NonIterative}(c) are obtained. One observes that the mean DRC curve obtained by iterate reconstruction~(orange line) agrees better with the input mean DRC curve~(opens symbols) than the non-iterative reconstruction result~(green line). The in-plane angular dependence of the mean DRC produced in the latter case is broader than both the input data and the iterative mean DRC data sets, and it has also a smaller amplitude around the specular direction (vertical dashed line) than both of the other two data sets.   

It is now tempting to ask why the reconstruction for the glass system, in particular the value of the rms-roughness,  is of better quality when basing it on s-polarized than on p-polarized input scattering data? To give a  definite answer to this question is outside the scope of this study. However, we speculate that the reason for the observed difference in reconstruction quality is due to the Kirchhoff approximation more poorly approximates the in-plane angular dependence of the input mean DRC; this is seen readily from the results in Figs.~\ref{fig:Simulations-Glass}(c) and \ref{fig:Simulations-Glass}(e). Further research is needed to give a definite answer to this question.

We have also performed reconstruction of computer generated scattering data for dielectric and metallic scattering systems for which the correlation lengths were longer than those used in obtaining the results Figs.~\ref{fig:Simulations-Silver-R006}--\ref{fig:Iterative_vs_NonIterative}. These results are not presented since they are rather similar to what we have already presented. The main difference that we found is that the quality of the reconstruction is increasing with increasing correlation length of the rough surface. This is consistent, and therefore expected, since the Kirchhoff approximation is more accurate for longer correlation lengths~\cite{Voronovich2007,Book:Voronovich1999}.

\smallskip
Before leaving the computer generated scattering data examples, we will comment on the time required to performed the reconstruction.
Our reconstruction approach is non-parametric and based on the expressions in Eqs.~\eqref{eq:statistical-property-estimation-simplified} or \eqref{eq:statistical-property-estimation}. To be performed, it requires the numerical evaluation of a one-dimensional integral and a few standard mathematical functions for each value of $x_\parallel$ for which one wants to reconstruct the correlation function $W(x_\parallel)$. Such numerical calculations are quite fast. For instance, each of the reconstructions to obtain the full correlation function reported in Figs.~\ref{fig:Kirchhoff-Ex1}--\ref{fig:Iterative_vs_NonIterative} took only a few seconds on a modest laptop. This is in contrast to approaches that are based on a least-square procedure, for instance, the approaches presented in Refs.~\cite{Chakrabarti2013a,Simonsen2014-03,Simonsen2014-05,Simonsen2016-06,Simonsen2018-06}, which would take significantly longer to converge towards a result.

\subsection{Experimental scattering data}

\begin{figure}[!tbh]
  \centering
  %
  \includegraphics[width=0.475\columnwidth,height=0.275\columnwidth]{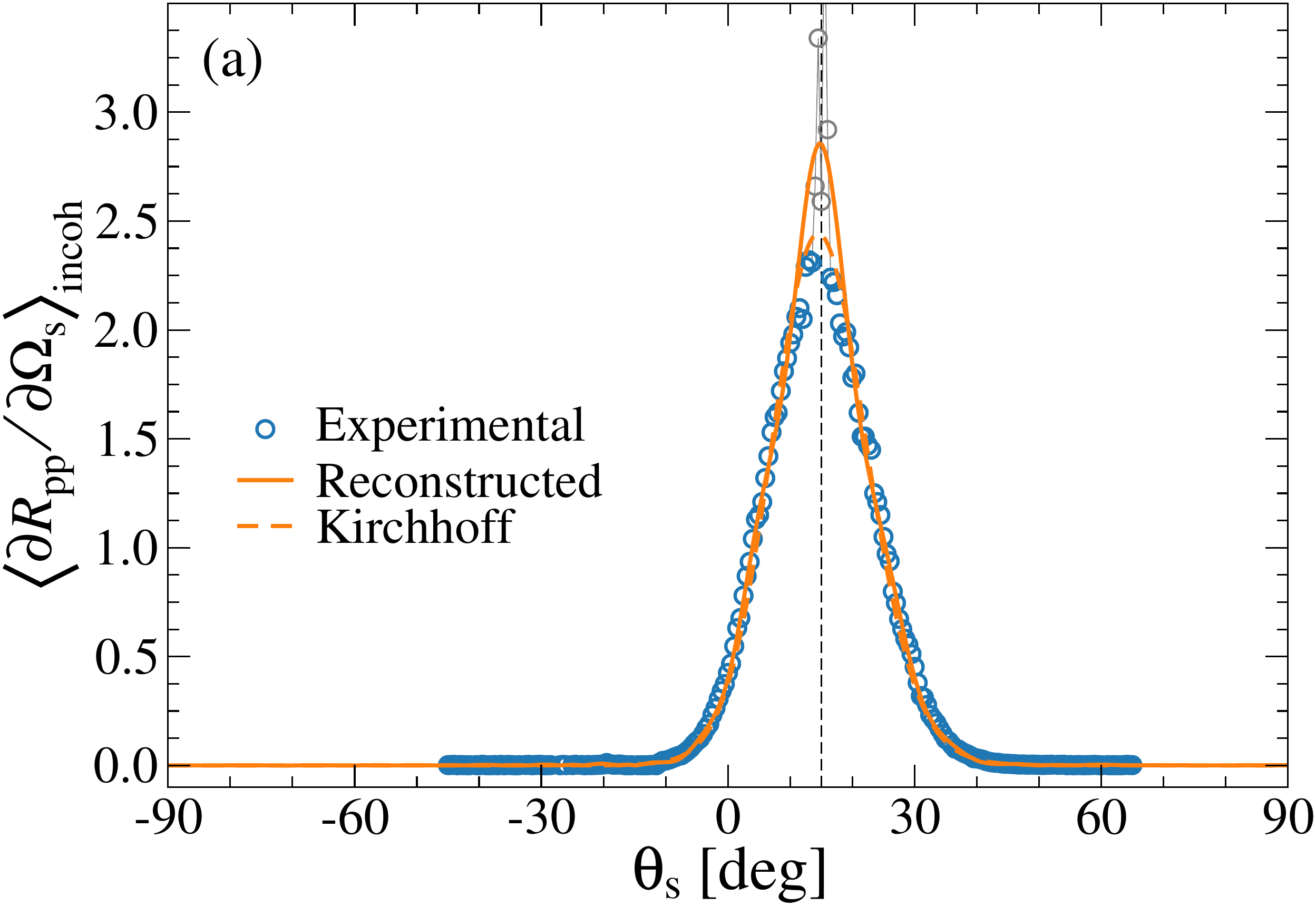}
  \qquad
  \includegraphics[width=0.475\columnwidth,height=0.275\columnwidth]{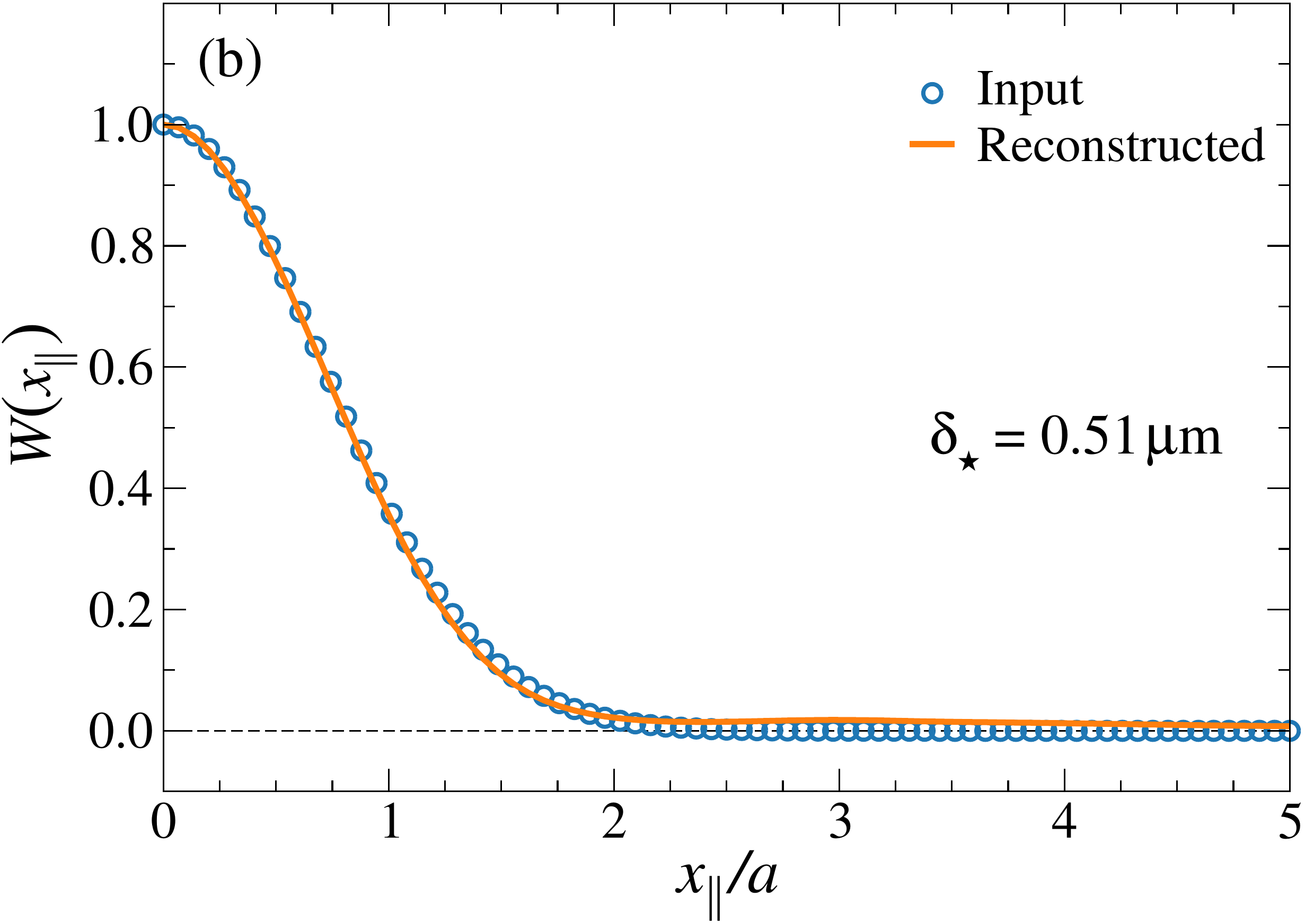}
  \\*[3ex]  
  %
    \includegraphics[width=0.475\columnwidth,height=0.275\columnwidth]{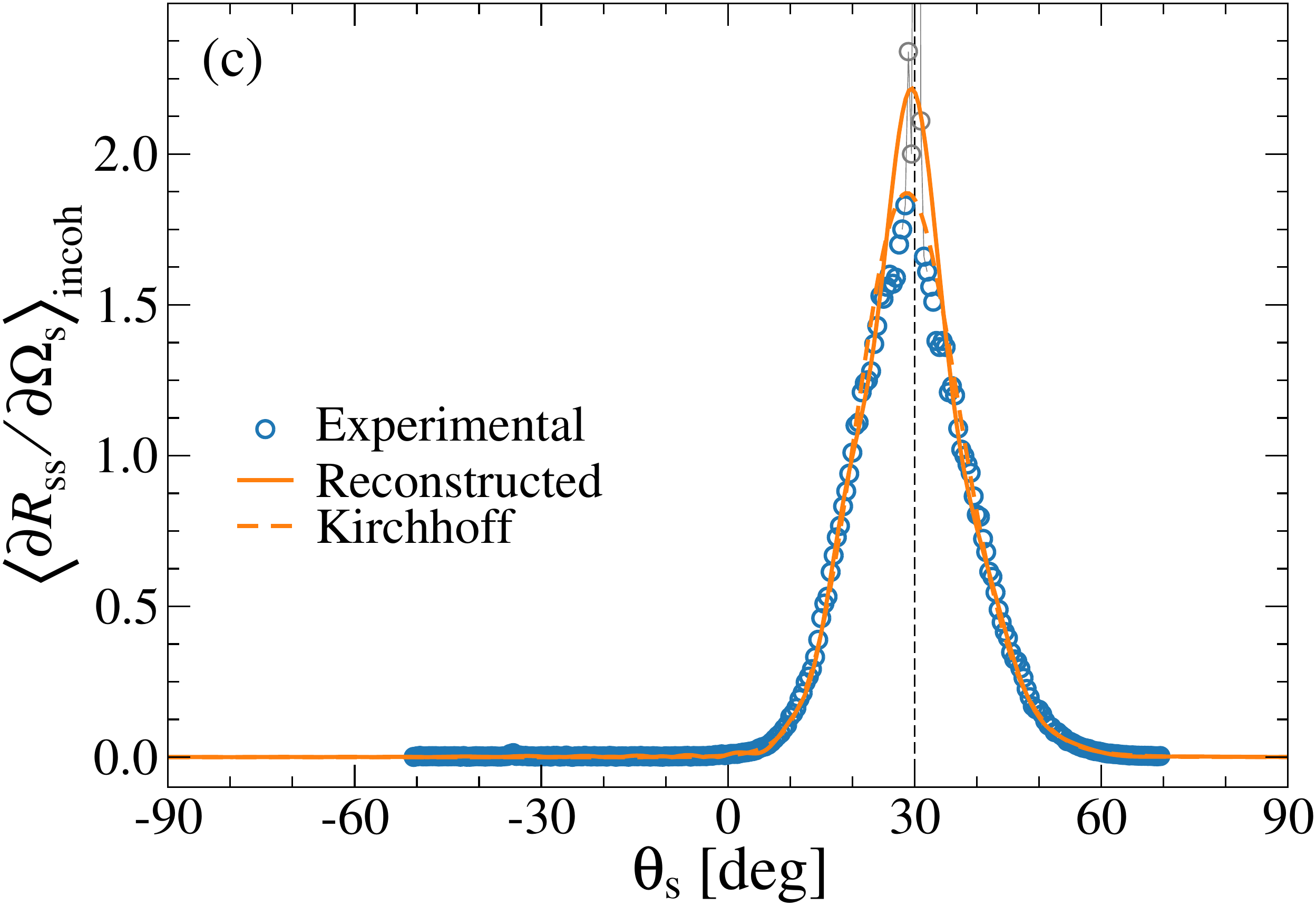}
  \qquad
  \includegraphics[width=0.475\columnwidth,height=0.275\columnwidth]{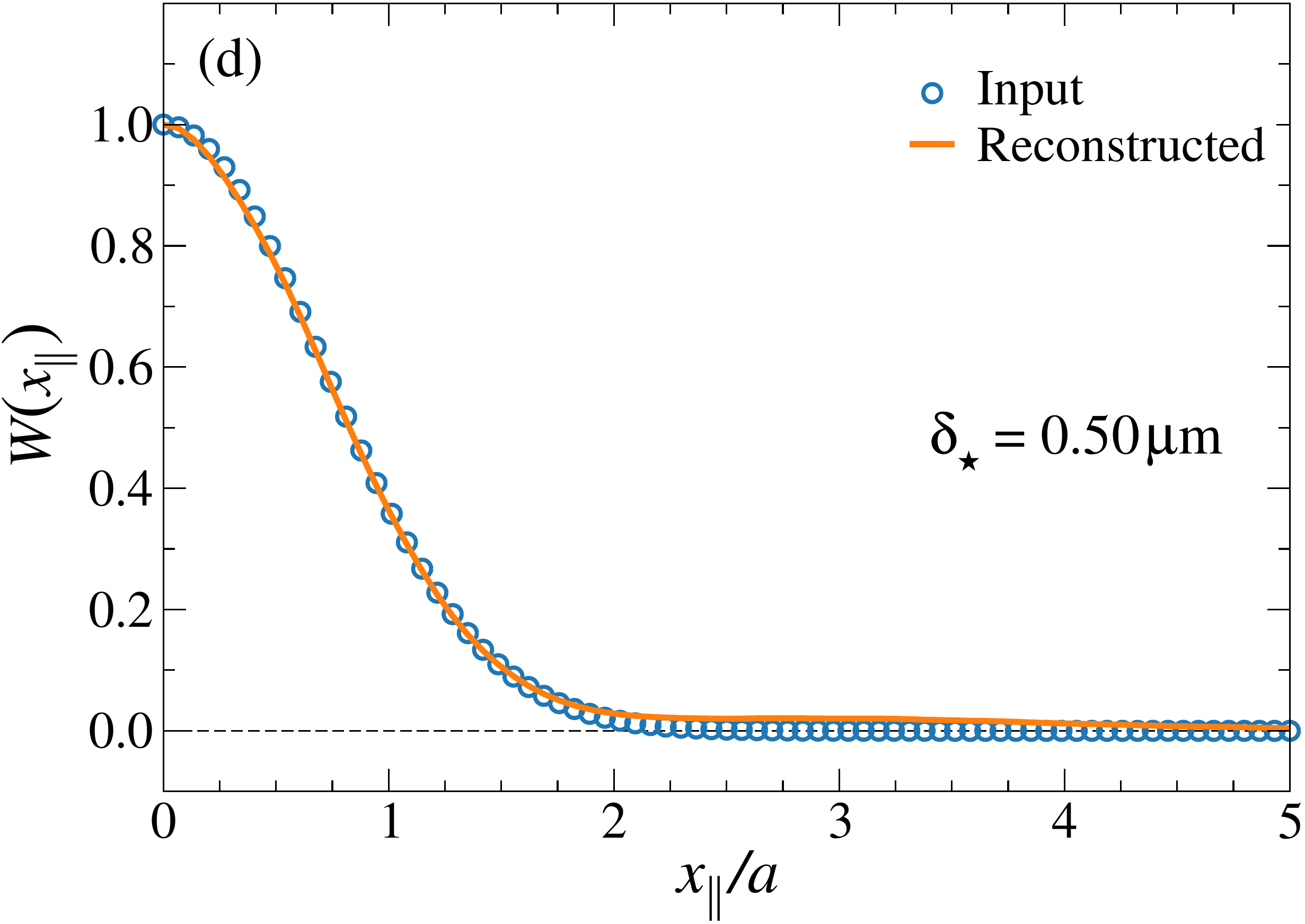}
  \caption{The (non-iterative) reconstruction of the surface-height auto-correlation function $W(x_\parallel)$ and the surface roughness $\delta_\star$ from light scattering data \textit{measured} for gold sample~$0061$ reported in Figs.~10 and 11 of  Ref.~\onlinecite{NavarreteAlcala2009}. Experimentally it was determined that the rough surface was Gaussian correlated and characterized by the rms-roughness  $\delta=\SI[separate-uncertainty = true]{0.50 \pm 0.01}{\micro\meter}$ and the correlation length $a=\SI[separate-uncertainty = true]{19.0 \pm 1.7}{\micro\meter}$.
    The measured in-plane angular dependence of the co-polarized DRCs for light of wavelength  $\lambda= \SI{10.6}{\micro\meter}$ are reproduced from  Ref.~\onlinecite{NavarreteAlcala2009} and shown as open symbols in Fig.~\ref{fig:experimental-Sample0061}(a) for p-polarized light incident at an angle $\theta_0=\ang{15}$, and in Fig.~\ref{fig:experimental-Sample0061}(c) for $\theta_0=\ang{30}$ and s-polarized light. Only the blue data points were included in the reconstruction; the gray data points were neglected as they have contributions from specular reflections (coherent component of the mean DRCs). The thin vertical dashed lines represent the specular positions.
    When our reconstruction approach~\eqref{eq:statistical-property-estimation-simplified} is applied to these measured scattering data sets we obtain the results presented in Figs.~\ref{fig:experimental-Sample0061}(b)~[$\theta_0=\ang{15}$] and \ref{fig:experimental-Sample0061}(d)~[$\theta_0=\ang{30}$]; the values for the reconstructed rms-roughness were found to be $\delta_\star=\SI{0.51}{\micro\meter}$ and $\delta_\star=\SI{0.50}{\micro\meter}$, respectively. In these panels the solid lines are the reconstructed correlation functions while the symbols represent the correlation function obtained by the morphological study of the sample~[see Ref.~\onlinecite{Navarrete2002}, Fig.~2]. The dielectric constant of gold assumed in performing the reconstruction is $\varepsilon(\omega)=-2489.77+\textrm{i}2817.36$~\cite{Book:Palik1997}. 
The remaining description of this figure is identical to that of Figs.~\protect\ref{fig:Simulations-Silver-R006}(c)--(f).}
  \label{fig:experimental-Sample0061} 
\end{figure}

\begin{figure}[!tbh]
  \centering
  %
  \includegraphics[width=0.475\columnwidth,height=0.275\columnwidth]{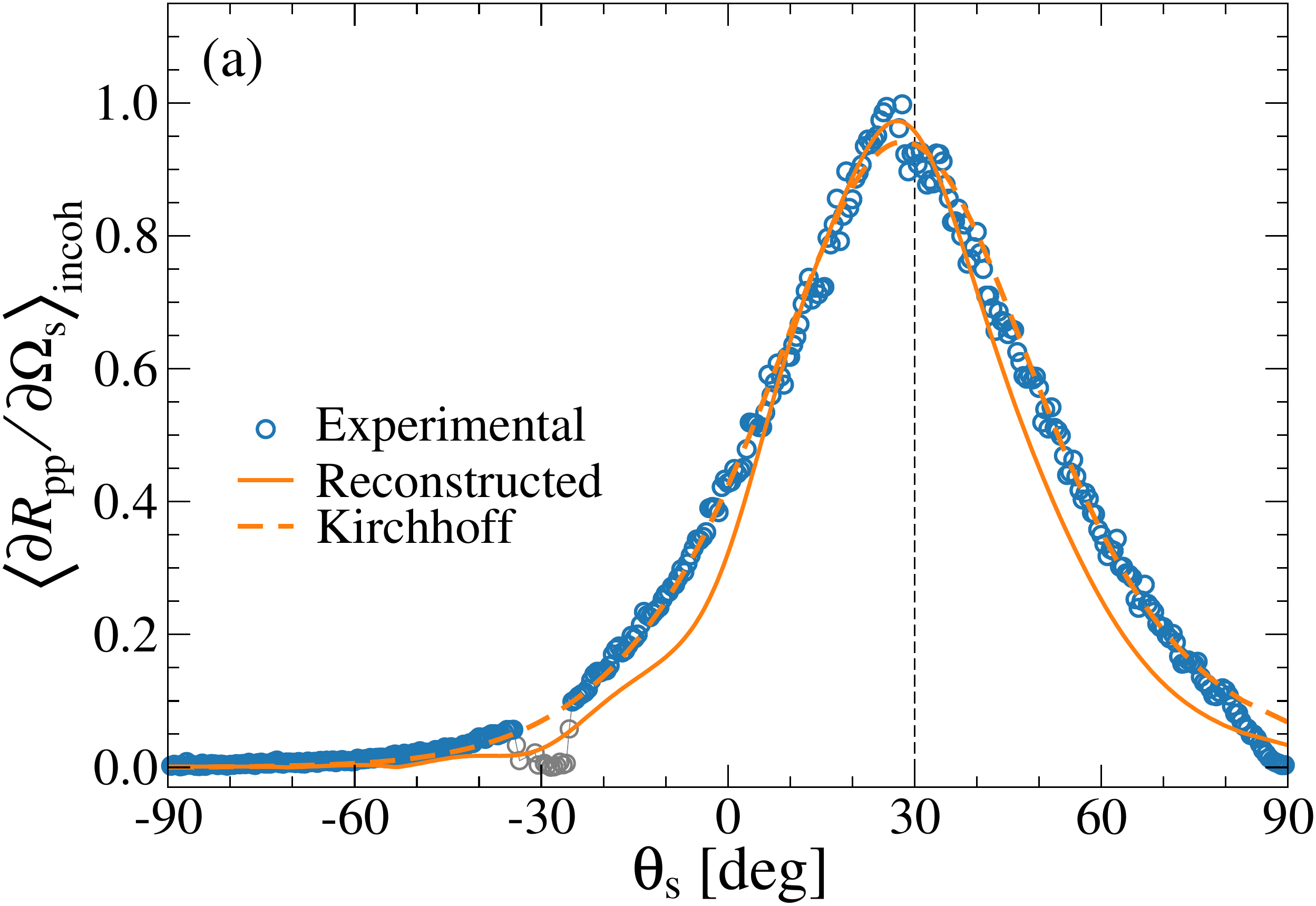}
  \qquad
  \includegraphics[width=0.475\columnwidth,height=0.275\columnwidth]{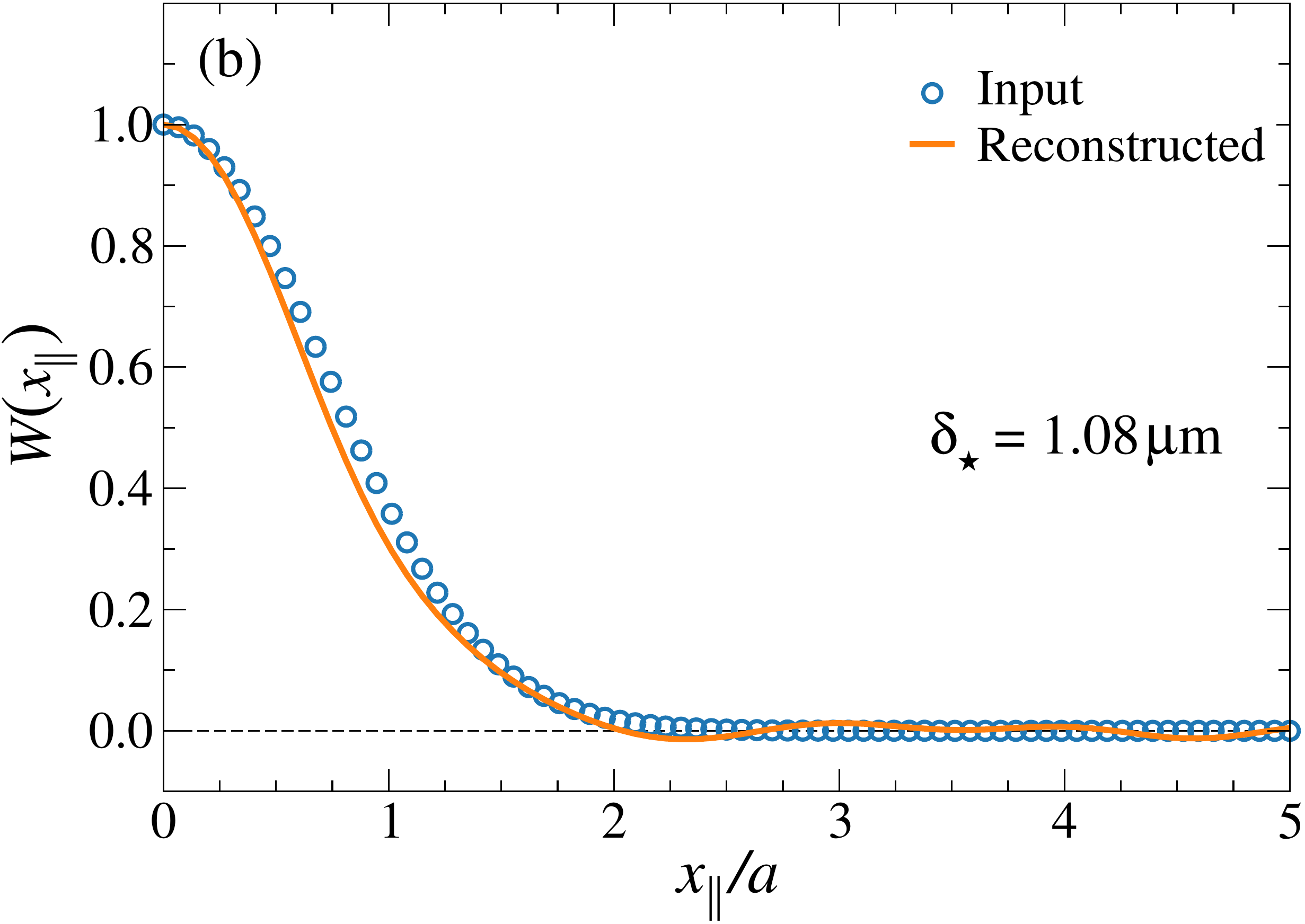}
  \\*[3ex]  
  %
    \includegraphics[width=0.475\columnwidth,height=0.275\columnwidth]{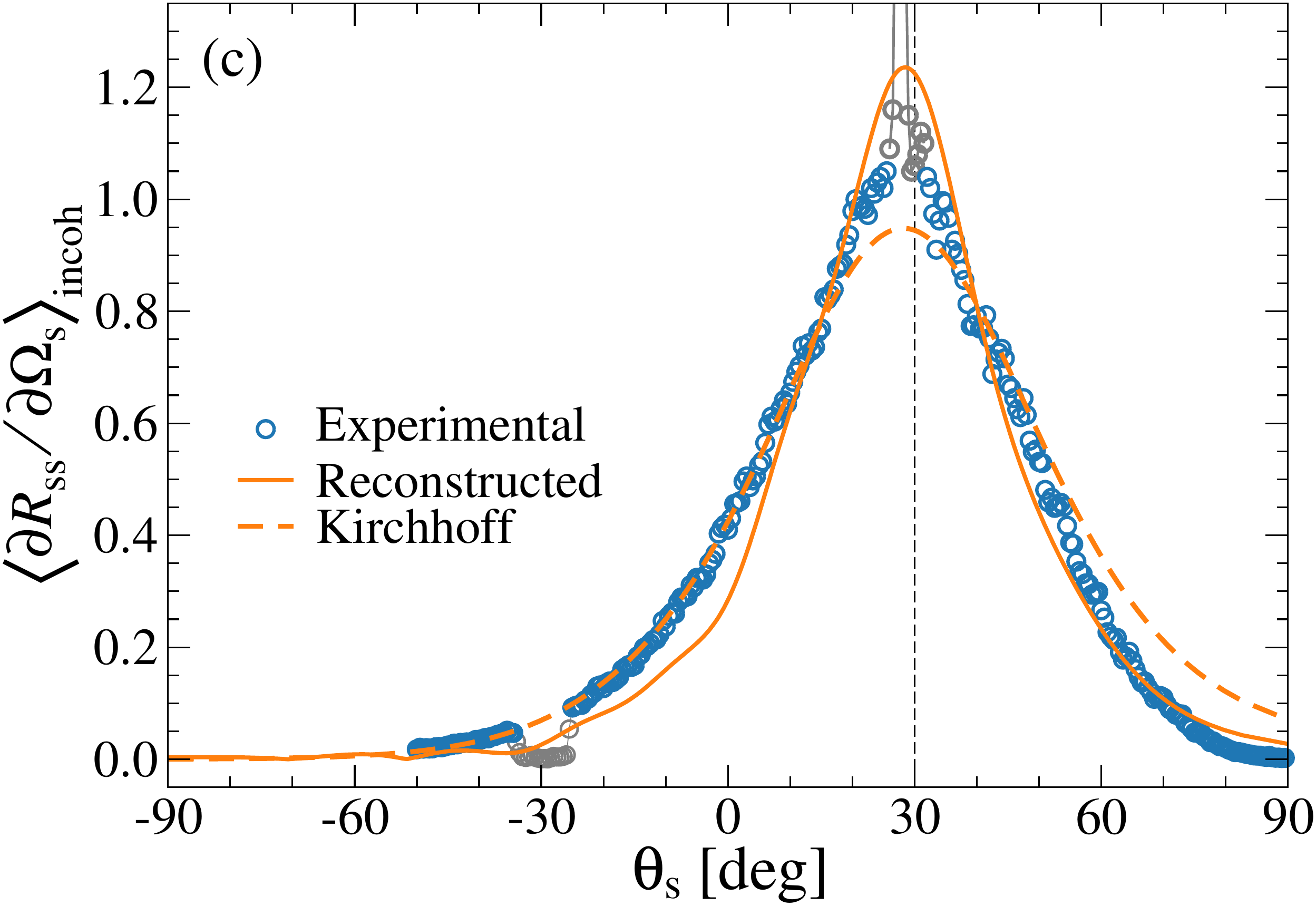}
  \qquad
  \includegraphics[width=0.475\columnwidth,height=0.275\columnwidth]{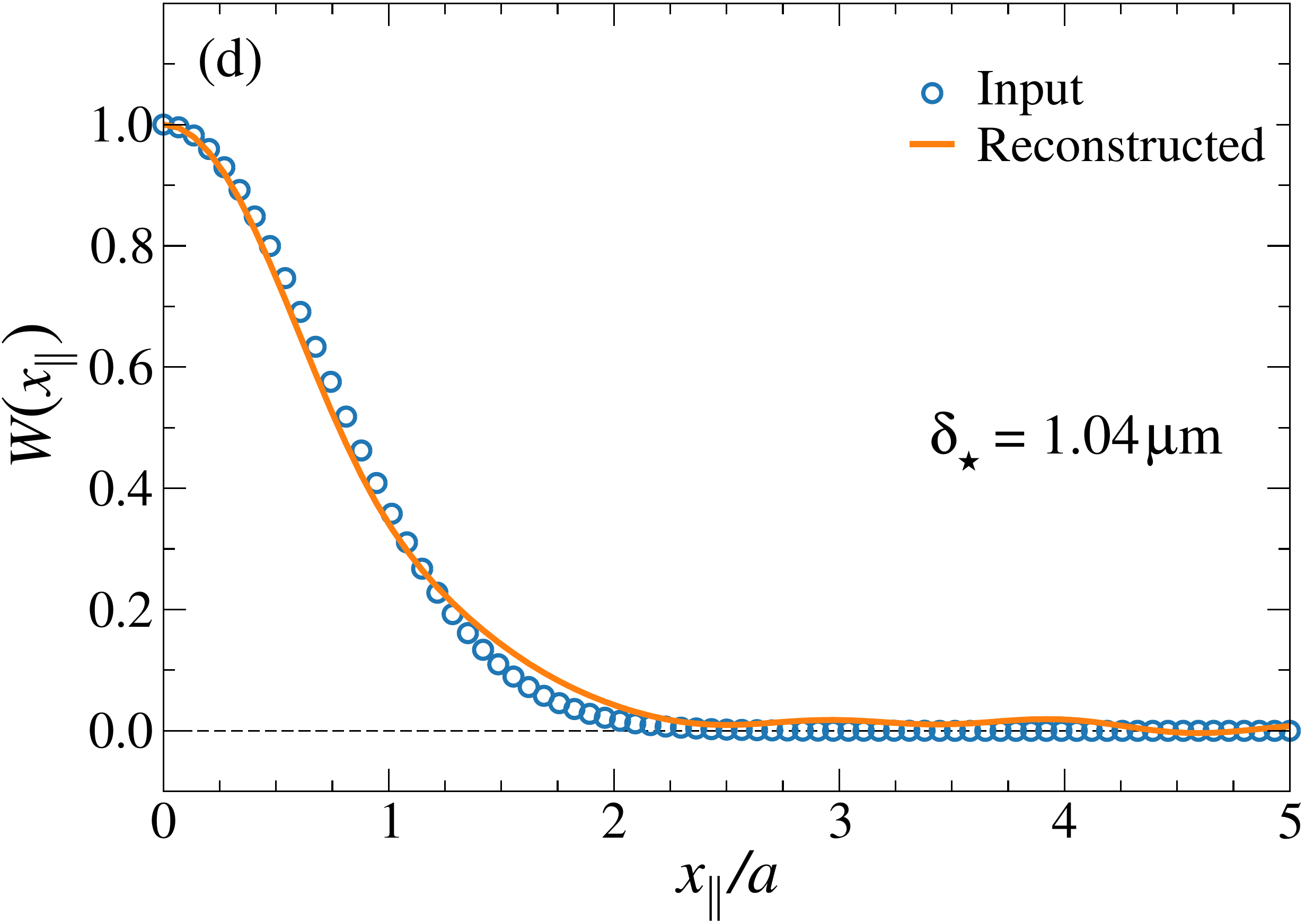}
  \caption{The reconstruction of the surface-height auto-correlation function $W(x_\parallel)$ and the surface roughness $\delta_\star$ from light scattering data \textit{measured} for gold sample~$7047$ reported in Figs.~8 and 9 in Ref~\onlinecite{NavarreteAlcala2009}. The morphological characterization of this sample revealed that the correlation function was well approximated by a Gaussian form, and the  rms-roughness and correlation length were determined to be
$\delta=\SI[separate-uncertainty = true]{1.6\pm 0.05}{\micro\meter}$ and $a=\SI[separate-uncertainty = true]{9.5 \pm 1.3}{\micro\meter}$, respectively.
The remaining description of this figure is identical to that of Fig.~\ref{fig:experimental-Sample0061}, except that here the polar angles of incidence is $\theta_0=\ang{30}$ for both polarizations. Moreover, also the data points around the backscattering direction $\theta_s=-\theta_0$ for which measurements are not possible, were neglected in the reconstruction.}
  \label{fig:experimental-Sample7047} 
\end{figure}

We are now prepared to apply our reconstruction approach to scattering data measured in experiments. To this end, we will consider some of the scattering data collected by Navarrete {\textit et al.}~\cite{NavarreteAlcala2009,Navarrete2002} for the coherent and incoherent contributions to the differential reflection coefficient for linearly polarized electromagnetic waves scattered by two-dimensional Gaussian correlated randomly rough gold surfaces. The samples used in these experiments were fabricated using photoresist-coated glass plates that were exposed to several uncorrelated speckle patterns of Gaussian statistics~\cite{Gray1978,
ODonnell1987,Navarrete2002}.  Next the sample was exposed to a developer, and finally the rough surfaces obtained in this way were coated by thin layers of gold. The gold layers were sufficiently thick to ensured that the scattering of light from the coated surfaces could be treated as equivalent to the scattering from semi-infinite gold substrates. Each of the gold surfaces produced in this manner were given a morphological characterization to ensure that the rough surfaces approximately possessed the desired statistical properties~\cite{Navarrete2002}.

For each of the samples that Navarrete and coworkers produced, they measured the intensity of the scattered field as a function of the angle of incidence, and its corresponding reflectivity was determined analytically as the normalized strength of its coherent (specular) component. The diffuse (or incoherent) components of the light scattered from the surfaces were measured in the plane of incidence as a  function of the polar angle of scattering for several polar angles of incidence.  When preforming these measurements, the specular intensity was blocked to avoid saturation of the detector. Subsequently, the contribution to the differential reflection coefficients from the light scattered incoherently by the rough surfaces was calculated from the measurements; see Refs.~\onlinecite{NavarreteAlcala2009,Navarrete2002} for additional details. In this way, the experimental input scattering data that we will consider were obtained. 

The open symbols in the left-side panels of Figs.~\ref{fig:experimental-Sample0061} and \ref{fig:experimental-Sample7047} reproduce the results for the in-plane angular dependent measurements for the co-polarized component of the mean DRC for sample~$0061$ and ~$7047$, respectively, from Figs.~8--11 of Ref.~\onlinecite{NavarreteAlcala2009}. Since our reconstruction approach is based on scattering data for the light that has been scattered diffusely by the rough surface, each of these measured data sets were divided into several groups;  the first group of data, denoted by blue open circles in the left-side panels of Figs.~\ref{fig:experimental-Sample0061} and 
\ref{fig:experimental-Sample7047}, constitute the incoherent components of the mean DRC (diffusely scattered light) --- these scattering data we will subsequently base our reconstruction on. The gray open circles located around the specular direction (vertical dashed black lines) in the same figures represent the coherent components (specularly scattered light), and this classification was done manually; as a visual guidance these circular symbols have been connected by thin lines. Finally, the gray data points around the back-scattering direction $\theta_s=-\theta_0$, where measurements can not be performed were neglected since they represents artifacts~[see Figs.~\ref{fig:experimental-Sample7047}(a) and \ref{fig:experimental-Sample7047}(c)]. The measurements reported in Figs.~\ref{fig:experimental-Sample0061} and \ref{fig:experimental-Sample7047} were performed for the wavelength (in vacuum) of  $\lambda=\SI{10.6}{\micro\meter}$ for which the dielectric constant of gold is $\varepsilon(\omega)=-2489.77+\textrm{i}2817.36$~\cite{Book:Palik1997}. The morphological characterization of these samples revealed that both were consistently described by Gaussian statistics.  Sample~$0061$ was found to be characterized by the rms-roughness $\delta=\SI[separate-uncertainty = true]{0.50 \pm 0.01}{\micro\meter}$, and a correlation function of the Gaussian form \eqref{eq:gaussW} with correlation length $a=\SI[separate-uncertainty = true]{19.0 \pm 1.7}{\micro\meter}$; sample~$7047$ was found to have roughness parameters $\delta=\SI[separate-uncertainty = true]{1.6\pm 0.05}{\micro\meter}$ and $a=\SI[separate-uncertainty = true]{9.5 \pm 1.3}{\micro\meter}$~\cite{NavarreteAlcala2009,Navarrete2002}.

%
\smallskip
Our first set of reconstruction results performed on the basis of experimental data will be based on scattering measurements for sample~$0061$. The open blue symbols in Fig.~\ref{fig:experimental-Sample0061}(a) represent the measured angular dependence of the incoherent, in-plane, p-to-p scattering for a polar angle of incidence of $\theta_0=\ang{15}$. For the same sample, the open blue symbols in Fig.~\ref{fig:experimental-Sample0061}(c) present similar measurement results but for s-polarization and a polar angle of incidence $\theta_0=\ang{30}$. When non-iterative reconstruction is performed on the basis of these two measured data sets, the results for $W(x_\parallel)$ presented in Figs.~\ref{fig:experimental-Sample0061}(b) and \ref{fig:experimental-Sample0061}(d) are obtained for the polar angles of incidence  $\theta_0=\ang{15}$ and  $\theta_0=\ang{30}$, respectively. In these figures the solid orange lines represent the reconstructed correlation functions $W(x_\parallel)$, while the open blue symbols correspond to the Gaussian correlation function~\eqref{eq:gaussW} evaluated for the value of the correlation length $a$, obtained during the morphological characterization of sample~$0061$, that is, for the value $a=\SI{19.0}{\micro\meter}$~\cite{Navarrete2002}. From the results presented  in Fig.~\ref{fig:experimental-Sample0061}(b) and \ref{fig:experimental-Sample0061}(d), one observes a strikingly good agreement between the reconstructed and input correlation functions $W(x_\parallel)$ for both angles of incidence and for both polarizations. Moreover, the two reconstructed $W(x_\parallel)$-functions  are internally consistent (approximately the same), as they should be for measurements performed on the same sample. The values for the surface roughness are found to be $\delta_\star=\SI{0.51}{\micro\meter}$~[$\theta_0=\ang{15}$]  and $\delta_\star=\SI{0.50}{\micro\meter}$~[$\theta_0=\ang{30}$].  Both results are in excellent agreement with the roughness $\delta=\SI[separate-uncertainty = true]{0.50 \pm 0.01}{\micro\meter}$ determined from the measured topography data.  When the results for the statistical properties of the rough surface obtained during the reconstruction are used to calculate the mean DRC curves in the plane of incidence, the (orange) solid lines in Figs.~\ref{fig:experimental-Sample0061}(a) and \ref{fig:experimental-Sample0061}(c) are obtained. On the basis of the  empirical roughness parameters, the in-plane mean DRCs were calculated  and the results are presented as (orange) dashed lines in these figures. Mainly around the specular direction one finds some minor discrepancies between these two sets of results. We have found that these discrepancies, even when the reconstructed and input roughness parameters are rather similar, are caused by the reconstructed correlation functions being slightly larger then the input correlation function for $x_\parallel/a>2$. Finally, we mention that using the iterative reconstruction approach, instead of the non-iterative approach used to obtain the results in Fig.~\ref{fig:experimental-Sample0061}, did produced almost the same reconstruction results. For instance, the values obtained for the surface roughness using the iterative reconstruction approach deviated by no more than \SI{0.3}{\percent} from the corresponding values reported in Fig.~\ref{fig:experimental-Sample0061}.

The results presented in Fig.~\ref{fig:experimental-Sample0061} clearly testify to the robustness, accuracy and power of the reconstruction approach that we proposed. For instance, it demonstrates that even for experimental data which unavoidably contains statistical fluctuations, the approach can produce robust and accurate results for the statistical properties of the randomly rough surface that was used to obtain the input scattering data. We stress that our approach has no adjustable parameters and that, for instance, the form obtained for the correlation function is a result of the input scattering data and the reconstruction approach alone, and not from assuming a certain form for $W(x_\parallel)$ prior to performing the reconstruction. For many naturally occurring and/or man-made surfaces this is a significant advantage since often very little is known in advance about the form of the surface height correlation function of such samples. 

%
%
\smallskip
We now turn to sample~$7047$ and to the in-plane, co-polarized scattering that was measured for it for the polar angle of incidence   $\theta_0=\ang{30}$ and $p$-polarized~[Fig.~\ref{fig:experimental-Sample7047}(a)] or $s$-polarized light~[Fig.~\ref{fig:experimental-Sample7047}(c)]. To obtain reliable results from our reconstruction approach based on scattering data obtained for this sample is expected to be significantly more challenging than what was the case for sample~$0061$. The reason for this is that the validity of the Kirchhoff approximation, on which the reconstruction approach is based, starts to become questionable at a polar angle of incidence $\theta_0=\ang{30}$ since $a/\lambda<1$~\cite{Millet2004,Fung1981,Franco2017}. We recall that the Kirchhoff approximation is valid when $2kr_c\cos^3\theta_0\gg 1$, where $k=2\pi/\lambda$ and $r_c$ is the integrated radius of curvature~\cite{Millet2004}. Even after this reservation,  we proceeded by applying our Kichhoff-based reconstruction approach to these scattering data. Figures~\ref{fig:experimental-Sample7047}(b) and \ref{fig:experimental-Sample7047}(d) present as (orange) solid lines the results of the reconstruction that was performed in an manner that is completely analogous to how the reconstruction results in Fig.~\ref{fig:experimental-Sample0061} were obtained. Also for this sample, the iterative and non-iterative reconstruction approaches produced essentially the same results. One observes from the results in these figures that there still is a reasonable good agreement between the input and reconstructed correlation functions. Both the form of the correlation function and the length scale $a$ characterizing their decays are fairly well reproduced in the reconstruction; this result is rather encouraging and definitely better than expected in view of the questionable  validity of the Kichhoff approximation. The fact that we still obtain a relatively adequate reconstruction result for $W(x_\parallel)$ we suspect is caused by the Kirchhoff approximation producing results that are not too far from the measured input scattering data; this is shown by the dashed lines in Figs.~\ref{fig:experimental-Sample7047}(a) and \ref{fig:experimental-Sample7047}(c). When it comes to the reconstructed surface roughness values, we obtain  $\delta_\star=\SI{1.08}{\micro\meter}$ and $\delta_\star=\SI{1.04}{\micro\meter}$ for p- and s-polarized incident light, respectively. These values represent significant underestimation of the value $\delta=\SI[separate-uncertainty = true]{1.6\pm 0.05}{\micro\meter}$ obtained by analyzing the morphology of the sample. We note that applying the iterative reconstruction approach did not improve the accuracy of the reconstructed surface roughness values (or the correlation function). The trend that the correlation function is more accurately reproduced than the value of the surface roughness, is something that we have seen in numerous cases when applying our reconstruction approach to computer generated scattering data. We speculate that the reason for this behavior is that the shape of the in-plane mean DRC curve seems to be less sensitive to the value of the surface roughness than the form of the correlation function. For instance, from Fig.~\ref{fig:experimental-Sample7047}(b) we observe that the input and reconstructed correlation function are rather similar, while the results in Fig.~\ref{fig:experimental-Sample7047}(a) show two mean DRC curves (solid and dashed lines) that are not that different even if the surface roughness values assumed when producing them are rather different. Motivated by this finding, we assumed the correlation function obtained during the reconstruction and shown as solid lines in the right-side panels of Fig.~\ref{fig:experimental-Sample7047}, and obtained the surface roughness by a least-square optimization procedure of the input mean DRC, and the mean DRC calculated from Eq.~\eqref{eq:MDRC-approx} for a given value of $\delta_\star$. In this way we obtained surface roughness values that did not deviate more than \SI{4}{\percent} from the input value $\delta$~(results not shown). Since we do not a priori know the correlation function, this latter approach is somewhat questionable, and we therefore not in general recommend it, even if it produced  encouraging results for the samples that we applied it to.


\section{Conclusions and Outlook}
\label{sec:Conclusions}

An approach is introduced for the non-parametric reconstruction of the surface height correlation function and the rms-roughness of penetrable two-dimensional randomly rough surfaces based on the angular dependence of the co-polarized light scattered by the surface in the plane of incidence. The reconstruction approach is based on an expression from electromagnetic scattering theory derived within the Kirchhoff approximation. It is stressed that unlike many other inversion methods, the form of the correlation function is not assumed prior to  the reconstruction (non-parametric); in fact, our approach has no adjustable parameters at all. We apply our method to in-plane, co-polarized scattering data obtained for rough dielectric and metallic surfaces illuminated by p- or s-polarized light. Such scattering data were obtained either by rigorous computer simulations, or in experimental measurements performed on well-characterized randomly rough surfaces. For a wide range of surface morphologies, the reconstruction performed in this way revealed good agreement between the surface height correlation function and the rms-roughness of the rough surface assumed in obtaining the (``input'') scattering data, and the correlation function and the rms-roughness obtained by reconstruction. The region of  validity of the reconstruction approach was judged by applying it to input scattering data calculated on the basis of the Kirchhoff approximation. In this way, it was determined that it works well when the local slopes of the surface are not too large; this result is similar, but more restrictive, than the region of validity of the Kirchhoff approximation itself on which the inversion scheme is based.

\begin{acknowledgments}
  We  are indebted to all the authors of Refs.~\onlinecite{NavarreteAlcala2009,Navarrete2002}~(Navarrete, Chaikina, M\'endez, Leskova and Maradudin) for kindly granting us permission to use their measured scattering data that they reported in their publications and to share them with us in digital form. DB and VPS acknowledge the Research Council of Norway through its Center of Excellence Funding Scheme, project number 262644, PoreLab.  The research of I.S. was supported in part by the French National Research Agency~(ANR) under contract ANR-15-CHIN-0003.  VPS and IS would like to thank Dr.~A.A.~Maradudin for his keen observations and fruitful discussions on the topic of this publication. 
\end{acknowledgments}


\appendix 

\section{Computations details}
\label{Appendix:A}

Upon the substitution of the scattering amplitude~\eqref{eq:R-Kichhoff-a} into Eq.~\eqref{eq:mdrc-definition}, an expression for the mean~DRC in the form~\eqref{eq:MDRC-approx-a} is obtained  where $\Phi_{\alpha\beta}(\pvec{q}|\pvec{k})$ is defined by Eq.~\eqref{eq:Phi} and  
\begin{align}
  L( \pvec{q}| \pvec{k} )_{\textrm{incoh}}
  &=
    \frac{1}{(2\pi)^2S}
    \left[
    \left< \left| r(\pvec{q} |\pvec{k} ) \right|^2\right>
      -
      \Big| \Big< r(\pvec{q} |\pvec{k} ) \Big> \Big|^2
    \right].
    \label{app-eq:L-incoh}
\end{align}
Here $r(\pvec{q} |\pvec{k} )$ is given by the expression in Eq.~\eqref{eq:R-Kichhoff-b} and the expression for the mean DRC is valid within the stationary phase approximation to the Kirchhoff integrals. In this Appendix, we will explicitly calculate the averages over the ensemble of realizations of the surface profile function $\zeta(\pvec{x})$ that the right-hand side of Eq.~\eqref{app-eq:L-incoh} contains.

With the use of the results from Eqs.~\eqref{eq:R-Kichhoff-b} and \eqref{eq:Lambda}, a direct calculation leads to
\begin{align}
  \left< \left| r(\pvec{q} |\pvec{k} ) \right|^2\right>
  &=
    \int \dint[2]{x_\parallel}\,  \dint[2]{x_\parallel'}
    \exp
    \left[
    -\imu \left( \pvec{q}-\pvec{k} \right) \cdot ( \pvec{x}-\pvec{x}' ) 
    \right]
    \left<
    \exp
    \left[
    -\imu \Lambda(\pvec{q}|\pvec{k})
    \left(
    \zeta(\pvec{x}) - \zeta(\pvec{x}')
    \right)
    \right]
    \right>.
    \label{app-eq:2}
\end{align}
We have here used that the ensemble average $\left< \cdot \right>$ only is concerned with the surface profile function and, hence, can be moved inside the integral. It can be calculated explicitly since the surface profile function is assumed to be a stationary, isotropic, Gaussian random process. Under these assumptions, the (two-point) joint probability density function~(pdf) for finding a surface height $z=\zeta(\pvec{x})$ at in-plane coordinate $\pvec{x}$, and at the same time, the height $z'=\zeta(\pvec{x}')$ at $\pvec{x}'$, takes the bivariate Gaussian form~\cite[p.~19]{Book:Ogilvy1991}
\begin{align}
  P_2(z,z' ; |\pvec{x}-\pvec{x}'|)
  &=
    \frac{ 1 }{ 2\pi \delta^2 \left[ 1 - W^2(|\pvec{x}-\pvec{x}'|) \right]^\frac{1}{2} }
    \exp
    \left[
    -
    \frac{
      z^2 + z'^2 -2 zz' W(|\pvec{x}-\pvec{x}'|)
    }{
     2\delta^2 \left[ 1 - W^2(|\pvec{x}-\pvec{x}'|) \right]
    }
    \right],
    \label{eq:joint-pdf}
\end{align}
where the (isotropic) surface roughness is characterized by the surface height correlation function $W(x_\parallel)$ and the rms-roughness $\delta$ [see Eq.~\eqref{eq:1}]. The reason that this joint pdf $P_2(\cdot)$ only depends on the in-plane vector difference $\pvec{x}-\pvec{x}'$, and not separately on the values of the vectors $\pvec{x}$ and $\pvec{x}'$, is a consequence of the stationary of the surface. The isotropy of the surface dictates that the joint pdf only depends on the length of $\pvec{x}-\pvec{x}'$  and not on its direction. On the contrary, if the surface were anisotropic, then the surface height correlation function will depend on this vector difference and one has to replace $W(|\pvec{x}-\pvec{x}'|)$ in Eq.~\eqref{eq:joint-pdf} by $W(\pvec{x}-\pvec{x}')$; however, this situation will not be considered further here.

The ensemble average that appears in Eq.~\eqref{app-eq:2} is defined via the joint pdf $P_2(\cdot)$ as 
\begin{align}
  \left<
    \exp
    \left[
    -\imu \Lambda(\pvec{q}|\pvec{k})
    \left(
    \zeta(\pvec{x}) - \zeta(\pvec{x}')
    \right)
    \right]
  \right>
  &\equiv
    \int_{-\infty}^\infty \dint{z}     \int_{-\infty}^\infty \dint{z'}
    P_2(z,z' ; |\pvec{x}-\pvec{x}'|)
    \exp
    \left[
    -\imu \Lambda(\pvec{q}|\pvec{k})
    \left(
    z - z'
    \right)
    \right].
    \label{app-eq:4}
\end{align}
With the explicit Gaussian form~\eqref{eq:joint-pdf} for the joint pdf, the integrals on the right-hand side of Eq.~\eqref{app-eq:2} can be calculated explicitly since both  integrals are in the forms of Gaussian integrals. In this way, a lengthy but in principle straightforward calculation (using integral 3.323.2 from Ref.~\onlinecite{Book:Gradshteyn2007}) results in
\begin{align}
  \left<
    \exp
    \left[
    -\imu \Lambda(\pvec{q}|\pvec{k})
    \left(
    \zeta(\pvec{x}) - \zeta(\pvec{x}')
    \right)
    \right]
  \right>
  &=
    \exp
    \left\{
    - \delta^2 \Lambda^2(\pvec{q}|\pvec{k})
    \left[
    1 -
    W(| \pvec{x}-\pvec{x}'|)
    \right]
    \right\}.
    \label{app-eq:5}
\end{align}
Next, this result is introduced into Eq.~\eqref{app-eq:2} and one obtains after making the change of variable $\pvec{u}=\pvec{x}-\pvec{x}'$ 
\begin{align}
  \left< \left| r(\pvec{q} |\pvec{k} ) \right|^2\right>
  &=
    \int \dint[2]{x_\parallel}\, \dint[2]{u_\parallel}
    \exp
    \left[
    -\imu \left( \pvec{q}-\pvec{k} \right) \cdot \pvec{u} 
    \right]
    \exp
    \left\{
    - \delta^2 \Lambda^2(\pvec{q}|\pvec{k})
    \left[
    1 -
    W(| \pvec{u} |)
    \right]
    \right\}
    \nonumber
  \\
  %
  %
  &=
    S
    \int_0^\infty \dint{u_\parallel} \,
    u_\parallel
    \int_{-\pi}^{\pi} \dint{\phi_{u_\parallel}} \,
    \exp\left[ -\imu \left|\pvec{q}-\pvec{k} \right| u_\parallel \cos\phi_{u_\parallel} \right]
    \exp
    \left\{
    - \delta^2 \Lambda^2(\pvec{q}|\pvec{k})
    \left[
    1 -
    W( u_\parallel)
    \right]
    \right\}
    \nonumber
  \\ 
  %
  %
  %
  %
  &=
    2\pi S
    \exp
    \left[
    - \delta^2 \Lambda^2(\pvec{q}|\pvec{k})
    \right]
    \int_0^\infty \dint{u_\parallel} \,
    u_\parallel \textrm{J}_0\!\left( |\pvec{q}-\pvec{k}| u_\parallel \right)
    \exp
    \left[
    \delta^2 \Lambda^2(\pvec{q}|\pvec{k})
    W( u_\parallel)
    \right].
    \label{app-eq:6}
\end{align}
In obtaining this result, it was used that the $\pvec{x}$-integration produces the surface area $S$ since the integrand is independent of this variable, polar coordinates $(u_\parallel,\phi_{u_\parallel})$ were introduced for the $\pvec{u}$-integration with $\phi_{u_\parallel}$ defined as the angle between the vectors $\pvec{u}$ and $\pvec{q}-\pvec{k}$, and the Bessel function of the first kind and order zero $\textrm{J}_0(\cdot)$ appears as a result of the angular integration by the use of the identity~\cite[pp.~228--231]{Book:Temme1996}
\begin{align}
  \textrm{J}_0\!\left( z \right)
  &=
    \frac{1}{2\pi}
    \int_{-\pi}^{\pi} \dint{\phi}
    \exp
    \left[
    - \imu z \cos \phi
    \right].
        \label{app-eq:J0-integral}
\end{align}

\medskip
In a similar fashion, with the use of Eq.~\eqref{eq:R-Kichhoff-b} and  the result
\begin{align}
  \left<
    \exp
    \left[
    -\imu \Lambda(\pvec{q}|\pvec{k})
    \zeta(\pvec{x}) 
    \right]
  \right>
  &\equiv
    \int_{-\infty}^\infty \dint{z}     
    P_1(z)
    \exp
    \left[
    -\imu \Lambda(\pvec{q}|\pvec{k}) z 
    \right]
    \nonumber
  \\
  &=  
    \exp
    \left[
     -\frac{\delta^2}{2} \Lambda^2(\pvec{q}|\pvec{k})
    \right],
        \label{app-eq:8}
\end{align}
where the surface height distribution is predicted to have the Gaussian form 
\begin{align}
  P_1(z)
  &\equiv
    \int_{-\infty}^\infty \dint{z'}         
    P_2(z,z' ; |\pvec{x}-\pvec{x}'|)
    =
    \frac{1}{\sqrt{2\pi}\delta}
    \exp
    \left[
    - \frac{z^2}{2\delta^2}
    \right],
        \label{app-eq:9}
\end{align}
it is obtained that 
\begin{align}
  \left< r(\pvec{q} |\pvec{k} ) \right>
  &=
    \int \dint[2]{x_\parallel}
    \exp
    \left[
    -\imu \left( \pvec{q}-\pvec{k} \right) \cdot \pvec{x}  
    \right]
    \exp
    \left[
     -\frac{\delta^2}{2} \Lambda^2(\pvec{q}|\pvec{k})
    \right].
    \label{app-eq:10}
\end{align}
The integral in Eq.~\eqref{app-eq:8} is also a Gaussian integral and it is evaluated explicitly in the same manner as the integrals in Eq.~\eqref{app-eq:4} were calculated (see Ref.~\onlinecite{Book:Gradshteyn2007}). From this result, one obtains 
\begin{align}
  \left| \left< r(\pvec{q} |\pvec{k} ) \right> \right|^2
  &=
    2\pi S
    \exp
    \left[
    - \delta^2 \Lambda^2(\pvec{q}|\pvec{k})
    \right]
    \int_0^\infty \dint{u_\parallel} \,
    u_\parallel \textrm{J}_0\!\left( |\pvec{q}-\pvec{k}| u_\parallel \right),
    \label{app-eq:11}
\end{align}
after performing a calculation which closely resembles how the expression for $\langle | r(\pvec{q} |\pvec{k} ) |^2 \rangle$  was obtained.

\medskip
The substitution of the results from Eqs.~\eqref{app-eq:6} and \eqref{app-eq:11} into Eq.~\eqref{app-eq:L-incoh} yields Eq.~\eqref{eq:L}, which is the final form for $L( \pvec{q}| \pvec{k} )_{\textrm{incoh}}$.

%
\bibliographystyle{apsrev4-1}
\bibliography{paper2019-04,$HOME/Archive/Papers/BIBLIOGRAPHY,$HOME/Archive/Papers/BOOKS,$HOME/Adm/CV/PubList/Simonsen-Publications}

\end{document}